\documentclass[preprint,12pt]{elsarticle}
\usepackage{amsmath}
\usepackage{amssymb}
\usepackage{graphicx}
\usepackage{xspace}
\usepackage{hyperref}
\hypersetup{colorlinks=true,linkcolor=black,citecolor=black,urlcolor=black,filecolor=black,allcolors=black}
\AtBeginDocument{\hypersetup{linkcolor=black,citecolor=black,urlcolor=black,filecolor=black,allcolors=black}}
\usepackage{tabularx}
\usepackage{booktabs}
\usepackage{placeins}
\usepackage{array}
\usepackage{tikz}
\usetikzlibrary{arrows.meta}
\newcommand{\vitriflow}{\texttt{Vitriflow}\xspace}
\newcommand{\code}[1]{\texttt{#1}}

\journal{Computational Materials Science}

\begin{document}

\begin{frontmatter}

\title{\texttt{Vitriflow}: calibrated amorphous structure ensembles from melt--quench simulations}

\author[inst1,inst2]{Jonathon Cottom}
\author[inst1,inst2]{Robin Delhomme}
\author[inst1,inst2]{Emilia Olsson\corref{cor1}}
\ead{k.i.e.olsson@uva.nl}

\affiliation[inst1]{organization={Institute for Theoretical Physics, University of Amsterdam},
            addressline={Science Park 904},
            city={Amsterdam},
            postcode={1098 XH},
            country={the Netherlands}}
\affiliation[inst2]{organization={Advanced Research Center for Nanolithography},
            addressline={Science Park 106},
            city={Amsterdam},
            postcode={1098 XG},
            country={the Netherlands}}
\cortext[cor1]{Corresponding author.}

\makeatletter
\def\phase@silabel#1#2{\expandafter\gdef\csname r@#1\endcsname{{#2}{}{}{Doc-Start}{}}}
\phase@silabel{eq:si-descriptor-map}{S6}
\phase@silabel{eq:si-ring-fraction}{S14}
\phase@silabel{eq:si-preflight-settings}{S15}
\phase@silabel{eq:si-core-splice}{S24}
\phase@silabel{eq:si-transition-indicators}{S28}
\phase@silabel{eq:si-high-temperature}{S38}
\phase@silabel{eq:si-disordering-time}{S39}
\phase@silabel{eq:si-hold-time}{S42}
\phase@silabel{eq:si-linear-quench-steps}{S43}
\phase@silabel{eq:si-size-selection}{S49}
\phase@silabel{eq:si-screen-vector}{S50}
\phase@silabel{eq:si-generic-crystal-screen}{S60}
\phase@silabel{eq:si-prefix-mean}{S61}
\phase@silabel{eq:si-ks-distance}{S74}
\phase@silabel{eq:si-stage-separation}{S75}
\phase@silabel{eq:si-freeze-temperature}{S76}
\phase@silabel{eq:si-sio2-defect-fraction}{S77}
\phase@silabel{eq:si-sm2o3-screen}{S92}
\phase@silabel{fig:si-nve-timestep}{S1}
\phase@silabel{fig:si-vdos-timestep}{S2}
\phase@silabel{fig:si-cal-sio2}{S3}
\phase@silabel{fig:si-cal-si3n4}{S4}
\phase@silabel{fig:si-cal-sm2o3}{S5}
\phase@silabel{fig:si-sio2-uncertainty-bonds}{S6}
\phase@silabel{fig:si-sio2-uncertainty-angles}{S6}
\phase@silabel{fig:si-sio2-uncertainty-rings-density}{S6}
\phase@silabel{fig:si-si3n4-uncertainty-pairs}{S7}
\phase@silabel{fig:si-si3n4-uncertainty-angles-coordination}{S7}
\phase@silabel{fig:si-si3n4-uncertainty-rings-density}{S7}
\phase@silabel{fig:si-sm2o3-uncertainty-pairs}{S8}
\phase@silabel{fig:si-sm2o3-uncertainty-angles-coordination}{S8}
\phase@silabel{fig:si-sm2o3-uncertainty-scalars}{S8}
\phase@silabel{fig:si-sio2-structural}{S9}
\phase@silabel{fig:si-sio2-screening}{S10}
\phase@silabel{fig:si-sio2-screen-sensitivity}{S11}
\phase@silabel{fig:si-si3n4-structural}{S12}
\phase@silabel{fig:si-si3n4-scalars}{S13}
\phase@silabel{fig:si-si3n4-audit-sensitivity}{S14}
\phase@silabel{fig:si-si3n4-boundary}{S15}
\phase@silabel{fig:si-sm2o3-structural}{S16}
\phase@silabel{fig:si-sm2o3-screening}{S17}
\phase@silabel{fig:si-sm2o3-classifier-sensitivity}{S18}
\phase@silabel{fig:si-sm2o3-component-response}{S19}
\phase@silabel{fig:si-declaration-consequences}{S20}
\phase@silabel{tab:autotuned-run-settings}{S1}
\phase@silabel{tab:si-family-uncertainty-thresholds}{S2}
\phase@silabel{tab:si-descriptor-uncertainty-thresholds}{S3}
\let\phase@silabel\relax
\makeatother
\begin{abstract} 
Many structure-property relations in amorphous materials are encoded not only in the mean structure, but in the prevalence, correlations and spatial organisation of minority environments across length scales. Yet atomistic models are commonly validated against bulk averages and a nominal number of structures, without establishing whether the finite population resolves the motifs invoked to explain material behaviour. We present an auditable framework, implemented in the \vitriflow{} software package, that connects calibrated melt-quench generation to explicit populations and quantity-specific uncertainty. This approach is demonstrated on three distinct benchmark systems. In a-SiO$_2$, strain is followed from intra-tetrahedral deformation through compressed Si--O--Si bridges to primitive 3-ring topology, showing that the lower angular tail is an organised medium-range population rather than undifferentiated variation about the network mean. In matched a-Si$_3$N$_4$ refinements, DFT redistributes density and first-neighbour length scale, sharpens the force-field first shell and repairs marginal contacts, while the two DFT descendants retain essentially the same first-neighbour topology. In a-Sm$_2$O$_3$, the amount and connectivity of partially ordered domains are quantified as they emerge from the disordered parent population, while the mixed coordination characteristic of the amorphous oxide remains robust. Mean structure, distribution tails, inherited topology and connected order are therefore distinct materials variables with distinct uncertainties. There is no universal ensemble size: the \vitriflow{} package records what was generated, which population was analysed and what that finite population establishes, providing a reproducible route from amorphous structure datasets to quantitatively supported structure-property hypotheses.
\end{abstract}

\begin{keyword}
amorphous materials \sep melt--quench \sep molecular dynamics \sep reproducibility \sep protocol calibration \sep ensemble uncertainty
\end{keyword}

\end{frontmatter}

\section{Introduction}

Amorphous materials underpin a wide range of dielectric, optical, electronic, and protective thin-film technologies.\cite{Wilk2001HighK,Robertson2006HighK,Kamiya2010TAOS,Nomura2004TransparentTFT,Xiang2022SiNPhotonics,Yen2003SiNProtective,Lee2017ThinFilmSolar,Leskela2003RareEarthOxides,Dakhel2004Sm2O3ThinFilms} Their key observables--density, coordination statistics, medium-range topology, defect populations, and the properties derived from them--are structural and cannot be specified by a small unit cell and a set of symmetry labels.\cite{Zachariasen1932Glass,Elliott1991MRO,Wooten1985WWW,Franzblau1991Rings} Atomistic simulation is therefore often required to construct structural models that can be connected directly to diffraction, spectroscopy, transport, interfacial ion migration, and mechanical response.\cite{Fischer2006Diffraction,Salmon1994FSDP,Keen1999SilicaTotalScattering,Pedone2009SilicaGlassSim,Cottom2019TiNSiO2} At the same time, the absence of long-range periodicity makes amorphous materials intrinsically difficult to model, and a persistent gap separates laboratory structures from the configurations accessible on atomistic time and length scales.\cite{Christie2023AmorphousHPC,Madanchi2025FutureAmorphous,Liu2022GlassReview,Massobrio2015DisorderedMaterials,Wright1993Comparison}

Among available atomistic routes, melt--quench molecular dynamics is one of the most broadly applicable methods for constructing amorphous models with empirical interatomic potentials, \textit{ab initio} electronic structure forces, or machine-learned interatomic potentials.\cite{Massobrio2015DisorderedMaterials,Car1985CPMD,VanBeest1990BKS,Behler2007NNP,Bartok2010GAP,Behler2016MLP,Unke2021MLFF,Deringer2018RealisticASi} For network glasses such as a-SiO$_2$, it is a plausible proxy for vitrification;\cite{Vollmayr1996CoolingSilica,Horbach1999SilicaMelt,Lane2015CoolingSilica} for non-glass-forming systems, it is better understood as a controlled route to disordered structural models than as a literal representation of synthesis.\cite{Christie2023AmorphousHPC,Madanchi2025FutureAmorphous,Liu2022GlassReview} The nominal protocol is simple: heat above the temperature at which initial-packing memory is lost, equilibrate the liquid, cool to the target temperature, and relax. In practice, however, the final amorphous state is path dependent. In silica, cooling rate and system size alter density, coordination, bond-angle distributions, ring statistics, and intermediate-range structural signatures such as the first sharp diffraction peak.\cite{Vollmayr1996CoolingSilica,Lane2015CoolingSilica,Nakano1994FiniteSizeSilica,vonAlfthan2003SilicaPotentials,Horbach1999SilicaMelt,VanBeest1990BKS,Uchino2005FSDP,Salmon1994FSDP,Sundararaman2018SHIK} In amorphous Si$_3$N$_4$ and amorphous Si, both defect content and network topology depend on preparation route and on the chosen interaction model.\cite{Ippolito2011ASi3N4,Dasmahapatra2018ASi3N4,deBritoMota1998SiN,Marian2000MG2,Stich1991ASiAIMD,Kluge1987ASiQuench,Ishimaru1997ASiRapidQuench,Wooten1985WWW,Deringer2018RealisticASi,Li2019CoolingRateASi,Santos2019ASiQuenching}

The ensemble character of amorphous materials is long established\cite{Zachariasen1932Glass}; the practical question is the uncertainty carried by one large cell, a few large cells, or a finite population of independent cell-level sampling units, under a declared population definition and descriptor resolution. Melt temperature, time spent at high temperature, cooling rate, system size, numerical controls, screening criteria, and statistical tolerances are often chosen by convention or computational convenience, while established checks for cooling-rate sensitivity, finite-size effects, liquid stability, ensemble variation, thermostat/barostat behaviour, and neighbour-list handling are seldom linked into one scientific decision chain.\cite{Vollmayr1996CoolingSilica,Lane2015CoolingSilica,Li2019CoolingRateASi,Santos2019ASiQuenching,Nose1984Thermostat,Hoover1985Canonical,Martyna1994MTK,Basconi2013ThermostatArtifacts,Harvey1998FlyingIceCube,Braun2018VelocityRescaling,Kim2023NeighborList,Thompson2020TRUE}

Workflow and provenance frameworks such as AiiDA, FireWorks, atomate, signac, and pyiron have substantially improved reproducible execution by standardising automation, recording data lineage, and managing heterogeneous simulation backends.\cite{Pizzi2016AiiDA,Uhrin2021AiiDAWorkflows,Jain2015FireWorks,Mathew2017Atomate,Adorf2018Signac,Janssen2019Pyiron} Reproducible execution does not, however, establish whether a finite amorphous population resolves the structural quantity used to support a materials conclusion.\cite{Thompson2020TRUE} The missing link is between how structures are generated, which structures or sites define the analysis population, and the uncertainty carried by the resulting estimate. This distinction is increasingly important when amorphous structures are used to train, fine-tune, or benchmark machine-learned interatomic potentials, including foundation models, or when generative models construct candidate populations: a consistently generated dataset may still under-resolve the distribution of environments within the chosen descriptor space.\cite{Batatia2025Foundation,Kasoar2025MLPEG,Yang2026DM2,Finkler2026AMDEN}

In this work, we present an auditable framework that links a declared materials question to population construction and quantity-specific uncertainty. It accommodates both prospective generation and retrospective analysis. For a new population, candidate numerical settings can be screened for admissibility and melt--quench choices calibrated within a user-declared protocol family before the resulting population is assessed against a resolution requirement defined by the materials question. For an existing population, the achieved quantity-specific uncertainty and corresponding resolution are reported. The \vitriflow{} software package implements the framework and records the resulting decision chain, distinguishing the available source population from the declared analysis population. It does not replace expert judgement or prescribe a universal protocol, descriptor set, tolerance, or population size; instead, it makes those choices and the scope of the conclusions they support explicit.

We demonstrate this approach on three chemically distinct amorphous systems. The selected oxides and nitride reflect recurring modelling questions across functional amorphous materials, where local motifs and network topology are linked to mechanical evolution, oxygen-ion migration, charge trapping, hydrogen response, oxidation, and charge-induced network transformation.\cite{Cottom2019TiNSiO2,Patel2021ColumnBoundaries,Huckmann2024ChargeTrapping,Cottom2024HydrogenSiN,Huckmann2026DryOxidation,Cottom2025ForgedCharge,Deringer2020BatteryML,Konstantinou2019GSTMidgap,Konstantinou2022GSTChargeTrapping} In a-SiO$_2$, strain is followed from intra-tetrahedral deformation through compressed Si--O--Si bridges to primitive 3-ring topology, testing whether the lower angular tail forms an organised medium-range population rather than undifferentiated variation about the network mean.\cite{Zachariasen1932Glass,Vollmayr1996CoolingSilica,Sundararaman2018SHIK} In matched a-Si$_3$N$_4$ refinements, the MG2 parent population and its PBE and HSE06 descendants are used to separate changes in density and first-neighbour structure from topology inherited from the common parent.\cite{Marian2000MG2,Ippolito2011ASi3N4,Dasmahapatra2018ASi3N4} In a-Sm$_2$O$_3$, the prevalence and connectivity of partially ordered regions emerging within a disordered population are quantified, while the robustness of the mixed coordination characteristic of the amorphous oxide is assessed.\cite{Olsson2019AmorphousSm2O3,Leskela2003RareEarthOxides,Dakhel2004Sm2O3ThinFilms} Together, these cases show that mean structure, distribution tails, inherited topology, and connected order are distinct materials variables with distinct uncertainty requirements: a population adequate for one quantity need not be adequate for another.

\section{Methodological framework}

\begin{figure*}[t]
\centering
\includegraphics[width=\textwidth]{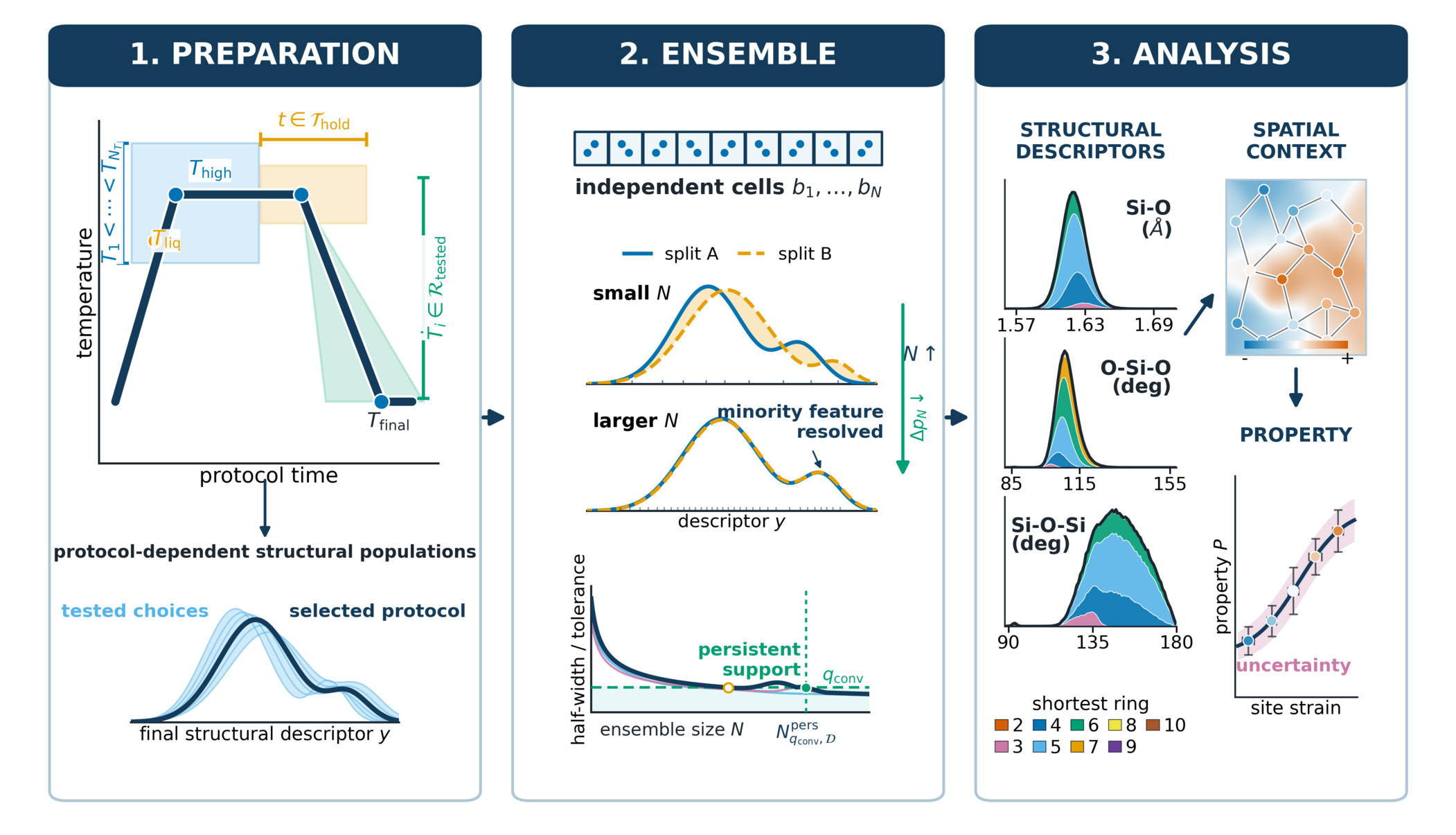}
\caption{Decision chain implemented by the \vitriflow{} software package. (1) Preparation: candidate numerical settings and a user-declared protocol family are evaluated, yielding selected settings within the tested domain. (2) Run: independent cells are generated, with the ensemble assessed against a resolution specified by the materials question. (3) Analysis: the source population is screened to define the analysis population, and quantity-specific uncertainty is evaluated. Externally calibrated settings may enter at stage~2, while an existing population may enter directly at stage~3. In the schematic, $M$ is the number of temperature-ladder points, $\mathcal R_{\mathrm{tested}}$ is the tested cooling-rate set, $n$ is the independent-sampling-unit prefix size, $D_{\mathrm{KS}}$ is the Kolmogorov--Smirnov distance, and $n^{\mathrm{pers}}_{q_{\mathrm{conv}},y}$ is the persistent crossing size for descriptor $y$.}
\label{fig:workflow}
\end{figure*}

The framework begins with the materials question: which structural quantity is to be estimated, which population it describes, and what precision is required to support the intended conclusion. The \vitriflow{} software package implements the resulting decision chain shown in Fig.~\ref{fig:workflow}: numerical preflight, protocol calibration within a user-declared family, population generation where required, structural screening, and quantity-specific statistical analysis. Externally validated settings may enter at the production stage, while an existing population may enter directly at analysis; in the latter case, \vitriflow{} reports the uncertainty achieved by that finite population rather than imposing a population size.

\subsection{Material problems and descriptor basis}

Amorphous SiO$_2$, generated from a randomly initialised 192-atom source melt containing 64 Si and 128 O atoms in a cubic cell of side length 21~\AA{} and treated with the SHIK-1 potential\cite{Sundararaman2018SHIK}, tests tetrahedral-network fidelity by separating defect-free silica from oxygen-bridge-defective structures. Amorphous Si$_3$N$_4$ tests controlled multi-fidelity refinement: MG2 provides the empirical melt--quench source population, and PBE and HSE06 provide cell-optimised descendants of that same population\cite{Marian2000MG2}. Amorphous Sm$_2$O$_3$ tests crystal-like screening in an ionic mixed-coordination rare-earth oxide, where fixed-coordination filtering would remove physically relevant local disorder.

For a generated structure or trajectory segment $x$, the material-specific descriptor map is
\begin{equation}
\Phi(x;\mathcal{D})=\left(d_1(x),d_2(x),\ldots,d_{n_{\mathcal D}}(x)\right),
\label{eq:descriptor-map}
\end{equation}
where $\mathcal D$ is the declared descriptor set, $d_j$ is descriptor $j$, and $n_{\mathcal D}$ is the number of descriptors in that set. In these benchmarks, the roles include density, diffusion inferred from mean-squared displacement, radial-distribution peak descriptors, first-shell coordination, bond-angle distributions, ring or topological descriptors, and material-specific local-order metrics. The roles are distinct: descriptors may be used for numerical preflight, protocol calibration, screening, finite-population support, or reporting. They need not be identical across materials or stages and are modular so that they can be adapted to the materials question. The descriptor map and common structural definitions are given in Supplementary Eqs.~\ref{eq:si-descriptor-map}--\ref{eq:si-ring-fraction}.

\begin{table*}[hbt!]
\centering
\caption{Selected numerical, protocol, and structural-screen settings used in the three demonstrations. The selected production settings are also reported in Supplementary Table~\ref{tab:autotuned-run-settings}; full candidate domains, acceptance criteria, candidate scores, and machine-readable decision records are reported in the archived JSON files. Here, $n_{\rm rep}$ is the calibration-replica count; $t_{\rm snap}$ is the source-snapshot spacing after melt equilibration; $r_{\rm in}^{\rm req}$ and $r_{\rm out}^{\rm req}$ delimit the requested global $C^2$ core-splice interval; $\mathcal A$ is the ordered tuple of allowed central:neighbour coordinations; $r_c$ is the corresponding first-shell screen cutoff; and $\boldsymbol{\eta}$ is the ordered, dimensionless vector of fractional screen thresholds. Their orders are Si:O/O:Si, Si:N/N:Si, and $(\eta_{\rm cryst},\eta_{\rm cluster})$ for a-SiO$_2$, a-Si$_3$N$_4$, and a-Sm$_2$O$_3$, respectively. The a-Si$_3$N$_4$ entries are the branch-specific audit cutoffs in MG2/PBE/HSE06 order. The action $\epsilon_s=1$ excludes a screen-failing cell, whereas $\epsilon_s=0$ retains it with an audit label.}
\label{tab:decision-inputs}
\small
\setlength{\tabcolsep}{6pt}
\renewcommand{\arraystretch}{1.12}
\begin{tabularx}{\textwidth}{@{}
>{\raggedright\arraybackslash}p{0.22\textwidth}
>{\centering\arraybackslash}X
>{\centering\arraybackslash}X
>{\centering\arraybackslash}X@{}}
\toprule
Quantity
& a-SiO$_2$
& a-Si$_3$N$_4$
& a-Sm$_2$O$_3$ \\
\midrule

\multicolumn{4}{@{}l}{\textit{Numerical preflight}} \\

$\Delta t$ (fs)
& $0.5$
& $1.0$
& $0.5$ \\

$\tau_T$ (ps)
& 0.05
& 0.50
& 0.10 \\

$\tau_P$ (ps)
& 0.5
& 5.0
& 1.0 \\

$s_{\mathrm{skin}}$ (\AA)
& $0.25$
& $0.50$
& $1.00$ \\

$r_{\rm in}^{\rm req}\,;\,r_{\rm out}^{\rm req}$ (\AA)
& --
& --
& $1.100/1.375$ \\

\addlinespace[2pt]
\multicolumn{4}{@{}l}{\textit{Protocol calibration}} \\

$T_{\rm liq}\,;\,T_{\rm high}$ (K)
& $4500/5000$
& $3500/4000$
& $3500/4000$ \\

$n_{\rm rep}$ (count); $t_{\rm melt}^{\star}$ (ps)
& 20;\,50.0
& $20\,;\,107.8$
& $20\,;\,44.0$ \\

$t_{\rm snap}$ (ps)
& 10
& --
& -- \\

$\dot T^{\star}$ (K~ps$^{-1}$)
& 1.0
& 1.0
& 50.0 \\

$N$ (atoms)
& $192$
& $280$
& $350$ \\

\addlinespace[2pt]
\multicolumn{4}{@{}l}{\textit{Structural screen}} \\

$\mathcal A$ (coordination); $r_c$ (\AA)
& $(4,2)\,;\,2.07$
& \shortstack{$(4,3)\,;$\\$2.23/2.31/2.29$}
& -- \\

$\boldsymbol{\eta}$
& $(0,0)$
& $(0.15,0.20)$
& $(0.18,0.12)$ \\

$\epsilon_s$
& $1$
& $0$
& $1$ \\

\bottomrule
\end{tabularx}
\end{table*}

\subsection{Numerical preflight}

Before protocol calibration, the \vitriflow{} package applies a numerical preflight to each user-declared candidate
\[
\theta=(\Delta t,\tau_T,\tau_P,s_{\mathrm{skin}},\mathcal C_{\mathrm{core}}),
\]
where $\Delta t$ is the timestep, $\tau_T$ and $\tau_P$ are the thermostat and barostat damping times, $s_{\mathrm{skin}}$ is the neighbour-list skin distance, and $\mathcal C_{\mathrm{core}}$ is an optional tabulated $C^2$ core-splice specification. The preflight rejects gross execution failures and candidates that violate the configured temperature-, pressure-, or volume-control bounds over the declared probe temperatures. The admissibility conditions, ranking score, and Buckingham--ZBL replacement splice are defined in Supplementary Eqs.~\ref{eq:si-preflight-settings}--\ref{eq:si-core-splice}.

The selected setting $\theta^\star$ is the lowest-scoring admissible candidate in the tested domain, with the more conservative barostat damping retained when scores are comparable. This establishes admissibility only within the declared numerical domain; it is not a universal certification of the interaction model, integrator, or timestep. The production timesteps were therefore validated independently using microcanonical (NVE) energy-conservation scans and the number of integration steps in the period associated with the 99th-percentile vibrational-density-of-states wavenumber, as shown in Supplementary Figs.~\ref{fig:si-nve-timestep} and \ref{fig:si-vdos-timestep}. The selected settings are listed in Table~\ref{tab:decision-inputs} and Supplementary Table~\ref{tab:autotuned-run-settings}; complete candidate domains, acceptance criteria, scores, and outcomes are retained in the archived JSON records.

\subsection{Protocol calibration}

For an admissible numerical setting $\theta^\star$, the thermal protocol is
\begin{equation}
p=(T_{\mathrm{high}},t_{\mathrm{melt}},\mathcal S_T,N),
\label{eq:protocol-tuple}
\end{equation}
where $T_{\mathrm{high}}$ is the high-temperature set point, $t_{\mathrm{melt}}$ is the liquid-hold time, $\mathcal S_T$ is the user-declared cooling schedule, and $N$ is the number of atoms in the cell. The present demonstrations use linear schedules parameterised by the cooling-rate magnitude $\dot T$, but the schedule driver itself is not restricted to this form. Within the user-declared candidate domain $\mathcal P(\theta^\star)$, \vitriflow{} selects
\begin{equation}
p^\star=
\arg\min_{p\in\mathcal P(\theta^\star)} J(p)
\quad\text{subject to}\quad
\mathcal G_{\mathrm{melt}},\mathcal G_{\mathrm{hold}},
\mathcal G_{\mathrm{quench}},\mathcal G_{\mathrm{size}},
\label{eq:protocol-selection}
\end{equation}
where $J(p)$ is the declared computational cost and the four $\mathcal G$ terms are the defined melt-window, hold-time, quench-equivalence, and finite-size criteria. Thus $p^\star$ is the least-cost passing candidate among the tested choices, not a universal optimum.
The selected protocol therefore defines the preparation-conditioned source population only within that tested domain.

The melt-window step identifies the first sustained liquid-like block and sets $T_{\mathrm{high}}=T_{\mathrm{liq}}+\Delta T_{\mathrm{margin}}$, where $T_{\mathrm{liq}}$ is the operational liquid onset and $\Delta T_{\mathrm{margin}}$ is a user-declared temperature margin. Independent replicas then determine the hold-time condition from $\sqrt{\mathrm{MSD}}\geq\alpha_{\mathrm{MSD}}\ell$, where $\alpha_{\mathrm{MSD}}$ is the declared displacement fraction and $\ell=(V/N)^{1/3}$ is the characteristic atomic spacing evaluated from the high-temperature cell volume $V$, together with stationarity tests comparing early and late averages of density and potential energy. Cooling schedules and cell sizes are compared with the slowest or largest tested reference using descriptor-specific equivalence tolerances. The transition score and liquid-state gates are defined in Supplementary Eqs.~\ref{eq:si-transition-indicators}--\ref{eq:si-high-temperature}; the hold-time and stationarity conditions in Eqs.~\ref{eq:si-disordering-time}--\ref{eq:si-hold-time}; and the linear-rate and finite-size comparisons used here in Eqs.~\ref{eq:si-linear-quench-steps}--\ref{eq:si-size-selection}. The corresponding a-SiO$_2$, a-Si$_3$N$_4$, and a-Sm$_2$O$_3$ calibration scans are shown in Supplementary Figs.~\ref{fig:si-cal-sio2}--\ref{fig:si-cal-sm2o3}.

\subsection{Structural screening and finite-population support}

The resulting or imported source population is denoted $\{x_b\}_{b=1}^{N_{\mathrm{gen}}}$. For prospective generation, each $x_b$ is the final relaxed cell associated with one independent cell-level sampling unit, obtained either from an independently seeded trajectory or from a decorrelated source snapshot under the declared material-specific construction; for an existing dataset, the archived cells define the source population directly. Let $\mathcal S_{\mathrm{screen}}$ denote the configured set of screens. For $s\in\mathcal S_{\mathrm{screen}}$, the package evaluates the configured descriptor vector
\begin{equation}
\mathbf d_{bs}=\Phi_s(x_b;\mathcal D_{\mathrm{screen},s})
\label{eq:screen-vector}
\end{equation}
and stores the label
\begin{equation}
A_{bs}=\mathbf 1\!\left[\Psi_s(\mathbf d_{bs})\in\mathcal I_s\right],
\label{eq:screen-label}
\end{equation}
where $\Psi_s$ is the user-declared screen rule and $\mathcal I_s$ is its allowed region. Thus $A_{bs}=1$ denotes a screen-passing cell and $A_{bs}=0$ an audit-labelled cell. The label does not itself determine exclusion. With $\epsilon_s=1$ for \code{exclude=true} and $\epsilon_s=0$ for \code{exclude=false}, the post-action mask is
\begin{equation}
R_b=\prod_{s\in\mathcal S_{\mathrm{screen}}}\left[A_{bs}+(1-A_{bs})(1-\epsilon_s)\right],
\label{eq:analysis-mask}
\end{equation}
and the analysis population is
\begin{equation}
\mathcal E_{\mathrm{ana}}=\{x_b:R_b=1\}.
\label{eq:analysis-population}
\end{equation}
The generic scalar, compound, coordination, and crystal-like screen definitions are given in Supplementary Eqs.~\ref{eq:si-screen-vector}--\ref{eq:si-generic-crystal-screen}, with the material-specific definitions in Eqs.~\ref{eq:si-sio2-defect-fraction}--\ref{eq:si-sm2o3-screen}.

For each reported descriptor, the achieved uncertainty is the confidence half-width evaluated on the declared analysis population under its stated independent sampling unit and confidence construction. For scalar descriptor $y$ and a registered population prefix of $n$ independent sampling units, let $\bar y_n$ denote the prefix mean, $h_n(y)$ the confidence half-width, and $\tau_y(\bar y_n)$ the absolute or relative resolution scale declared for that quantity. The support ratios are
\begin{equation}
R_y(n)=\frac{h_n(y)}{\tau_y(\bar y_n)},
\qquad
Q(n)=\max_{y\in\mathcal D_{\mathrm{conv}}}R_y(n),
\label{eq:support-ratio}
\end{equation}
where $\mathcal D_{\mathrm{conv}}$ is the descriptor set relevant to the materials question. The declared support condition is
\begin{equation}
Q(n)\leq q_{\mathrm{conv}},
\label{eq:support-condition}
\end{equation}
where lower $q_{\mathrm{conv}}$ requires a smaller confidence width relative to tolerance and is therefore stricter. The default production target is $q_{\mathrm{conv}}=0.2$; the descriptor-resolved uncertainty curves are shown in Supplementary Figs.~\ref{fig:si-sio2-uncertainty-bonds}--\ref{fig:si-sm2o3-uncertainty-scalars}; the statistical definitions are given in Supplementary Eqs.~\ref{eq:si-prefix-mean}--\ref{eq:si-ks-distance}.

\subsection{Stage-resolved diagnostic analysis}

Stage-resolved descriptor analysis is used diagnostically to locate when and if protocol-dependent differences enter the melt, liquid-hold, quench, or relaxation trajectory. More closely spaced quench samples are examined near the diffusion-freeze temperature $T_f$, where the diffusion coefficient $D$ falls below the configured threshold $D^\ast$ of the same units. The normalised separation and $T_f$ definitions are given in  Supplementary Eqs.~\ref{eq:si-stage-separation} and \ref{eq:si-freeze-temperature}. Differences already present in the high-temperature liquid indicate incomplete loss of initial-packing memory; differences emerging during cooling implicate the quench schedule; and differences appearing only after cooling implicate low-temperature relaxation or residual-stress effects. These diagnostics do not create an additional acceptance layer.

\section{Computational details}

The calculations were executed with the \vitriflow{} software package, version v0.4.37.0. Classical molecular dynamics was performed with LAMMPS\cite{Plimpton1995LAMMPS,Thompson2022LAMMPS}; the DFT-refinement calculations used the electronic-structure backends specified below. The package records the configured inputs, selected protocol, source and analysis populations, screen labels/actions, descriptor tables, support ratios, and provenance. The structural analysis uses the SHIK-1 a-SiO$_2$, matched MG2/PBE/HSE06 a-Si$_3$N$_4$, and Buckingham a-Sm$_2$O$_3$ datasets. Atomistic ball-and-stick representations were rendered using the Visualization Toolkit (VTK).\cite{Hanwell2015VTK}
Data plots were generated using Matplotlib.\cite{Hunter2007Matplotlib}

\subsection{Classical molecular-dynamics setup}

Classical simulations employed periodic boundary conditions and Nos\'e--Hoover dynamics with isotropic NPT coupling at zero target pressure\cite{Nose1984Thermostat,Hoover1985Canonical,Parrinello1980NPT,Martyna1994MTK}. The a-SiO$_2$ benchmark used the SHIK-1 potential\cite{Sundararaman2018SHIK}, a-Si$_3$N$_4$ used the MG2 potential as the empirical baseline\cite{Marian2000MG2}, and a-Sm$_2$O$_3$ used a Coulomb--Buckingham model based on the Sm-containing oxide parameterisation of Olsson \textit{et al.}\cite{Olsson2017SmCoO3Buckingham, Olsson2019AmorphousSm2O3, Buckingham1938ExpSix} Time steps were selected from the admissible region: 0.5~fs was used for a-SiO$_2$ and a-Sm$_2$O$_3$, while a-Si$_3$N$_4$ used 1.0~fs; for a-Sm$_2$O$_3$, the divergent Buckingham component was replaced by the tabulated $C^2$ Buckingham--ZBL splice defined in the Supplementary Information, with the Coulomb contribution unchanged.\cite{Ziegler1985ZBL} Temperature scans determined the operational melt window, after which liquid-hold times and quench rates were selected by the descriptor-equivalence criteria defined above. Source populations were constructed from the material-specific independent sampling units until the declared descriptor-support condition was evaluated on the analysis population. For a-Sm$_2$O$_3$, the 50~K~ps$^{-1}$ production rate was deliberately selected to explore the order--disorder boundary; by contrast, the earlier amorphous-only study used 100~K~ps$^{-1}$ to suppress crystal-like fragments.\cite{Olsson2019AmorphousSm2O3} The material-resolved calibration scans and selected settings are reported in Supplementary Figs.~\ref{fig:si-cal-sio2}--\ref{fig:si-cal-sm2o3} and Table~\ref{tab:autotuned-run-settings}.

\subsection[DFT-refined Si3N4 datasets]{DFT-refined Si$_3$N$_4$ datasets}

The MG2+PBE and MG2+HSE06 a-Si$_3$N$_4$ datasets were generated by cell optimisation of the MG2-derived amorphous nitride structures at the PBE and HSE06 levels, respectively. These refinements were applied to the MG2-derived structural population rather than obtained from independent DFT melt--quench trajectories.

All DFT calculations were performed spin-polarised using CP2K/\hspace{0pt}Quickstep\cite{Kuhne2020CP2K,VandeVondele2005Quickstep}. The PBE branch used the PBE exchange--correlation functional\cite{Perdew1996PBE}; the HSE06 branch used the HSE06 hybrid functional\cite{Heyd2003HSE,Heyd2006HSEErratum,Krukau2006HSE06} together with the auxiliary density matrix method (ADMM)\cite{Guidon2010ADMM} to reduce the cost of hybrid-functional exchange. DZVP-SR-MOLOPT basis sets\cite{VandeVondele2007MOLOPT} were used for valence electrons, and GTH pseudopotentials\cite{Goedecker1996GTH,Hartwigsen1998GTH,Krack2005GTH} were used for core electrons. The auxiliary plane-wave density cutoff and relative cutoff were set, after convergence testing, to 650~Ry and 70~Ry, respectively, giving an energy precision of approximately 0.1~meV~atom$^{-1}$.

Cell optimisations were performed with the BFGS algorithm\cite{Broyden1970BFGS,Fletcher1970BFGS,Goldfarb1970BFGS,Shanno1970BFGS}. The convergence criteria were $10^{-7}$~eV for energy differences and $10^{-3}$~eV~\AA$^{-1}$ for forces. After relaxation, the same screen and descriptor pipeline were applied to the MG2, MG2+PBE, and MG2+HSE06 structures.

\subsection{Descriptors and comparison metrics}

Descriptors were partitioned by role. The preflight and protocol-selection set comprised density, mean-squared-displacement-derived diffusion, simple radial-distribution peak descriptors, and energy stationarity. Final structural analysis was based on total and partial radial distribution functions, bond-angle distributions, coordination statistics evaluated on a fixed cutoff grid, chemically meaningful ring statistics\cite{Franzblau1991Rings}, and material-specific artefact metrics. For Sm$_2$O$_3$, final comparison emphasised density, pair structure, local-order metrics\cite{Steinhardt1983BondOrder,Lechner2008BondOrder}, and coordination-distribution stability rather than defect counts against a fixed allowed coordination set. Secondary diagnostics, including structure factors, void measures, and elastic or stress signatures, support the analysis, but do not determine primary acceptance decisions.

\begin{table*}[htb]
\centering
\caption{Material-specific source populations, screen labels/actions, and analysis populations. Counts are $N_{\mathrm{gen}}:N_{A=1}:N_{A=0}:N_{\mathrm{ana}}$, denoting generated, screen-passing, audit-labelled, and analysis cells. Hard screens use \code{exclude=true}; soft screens use \code{exclude=false}.}
\label{tab:materials-screening-summary}
\small
\setlength{\tabcolsep}{4pt}
\begin{tabularx}{\textwidth}{@{} ll X r @{}}
\toprule
Material & Dataset & Screen & $N_{\mathrm{gen}}:N_{A=1}:N_{A=0}:N_{\mathrm{ana}}$ \\
\midrule
a-SiO$_2$ & SHIK-1
& exact coordination; hard
& 2533:1872:661:1872 \\
a-Si$_3$N$_4$ & MG2
& coordination-outlier; soft
& 455:420:35:455 \\
a-Si$_3$N$_4$ & MG2+PBE
& coordination-outlier; soft
& 455:451:4:455 \\
a-Si$_3$N$_4$ & MG2+HSE06
& coordination-outlier; soft
& 455:452:3:455 \\
a-Sm$_2$O$_3$ & Buckingham
& crystal-like-order; hard
& 510:430:80:430 \\
\bottomrule
\end{tabularx}
\end{table*}
\FloatBarrier

\section{Results and discussion}

Each benchmark defines a different analysis population because each asks a different structural question. In a-SiO$_2$, exact Si:O~=~4 and O:Si~=~2 first-shell coordination defines the defect-free analysis population; failed cells form an oxygen-bridge-defective comparison population. In a-Si$_3$N$_4$, the coordination-outlier screen uses \code{exclude=false}: the label is retained as a refinement-response diagnostic, while the full matched MG2-derived population is included in the analysis. In a-Sm$_2$O$_3$, the screen targets crystal-like order rather than fixed coordination and excludes labelled cells from the mixed-coordination amorphous analysis population. The common operation is therefore not automatic rejection, but explicit construction of the source population, label, action, analysis population, and claim-support test.

Unless otherwise stated, uncertainties on population means are reported as standard errors over the relevant analysis or audit-labelled population.

\subsection{Material-specific screening and ensemble yield}

Table~\ref{tab:materials-screening-summary} summarises the generated populations, screen labels, and analysis populations used in the main-text results. For a-SiO$_2$, 1872 of 2533 generated SHIK-1 cells satisfy the exact tetrahedral-network criterion, while 661 form an audited oxygen-bridge-defective comparison population. For a-Si$_3$N$_4$, the coordination-outlier screen is advisory: all 455 cells in each refinement level are used for descriptor statistics, while 35 MG2, 4 MG2+PBE, and 3 MG2+HSE06 cells carry the high-defect label. This decreasing labelled subset is part of the refinement response. The MG2 ensemble satisfies the short-, medium-, and long-range convergence groups; the PBE and HSE06 descriptors are evaluated on the MG2-derived structures relaxed at each level rather than from independently converged ensembles. For a-Sm$_2$O$_3$, the screen retains 430 of 510 generated cells as amorphous and identifies 80 cells as containing significant crystal-like character.

\subsection[Tetrahedral-network fidelity in a-SiO2]
{Tetrahedral-network fidelity in a-SiO$_2$}

\begin{figure*}[htb!]
  \centering
  \includegraphics[width=\textwidth]{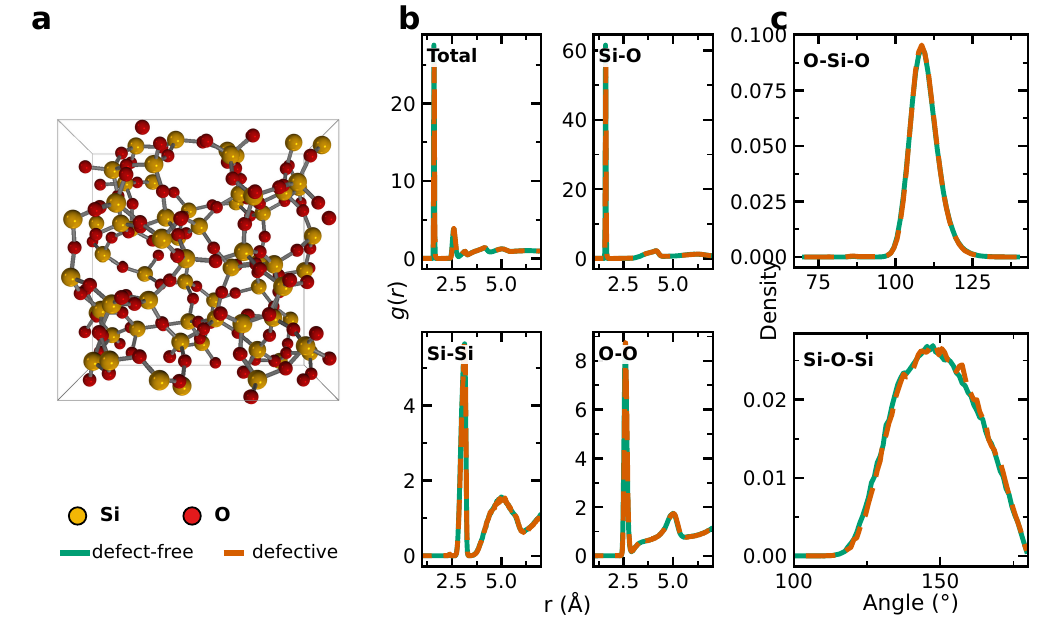}
  \caption{Structural characterisation of the sampled a-SiO$_2$ populations. a) Representative defect-free structure, selected as describing the median density, first-peak, angular-tail and ring-count of the ensemble. Oxygen is shown as red spheres, silicon as yellow spheres and Si--O bonds use a 2.07~\AA{} cutoff. b) Equal-cell total and partial radial distribution functions for the defect-free and defective populations. c) Equal-cell O--Si--O and Si--O--Si bond angle distributions.}
  \label{fig:sio2-overview}
\end{figure*}

Silica provides a stringent test because its average structure is well constrained, while many physically important environments occupy the tails of that structure. An ideal continuous random network consists of fourfold-coordinated Si tetrahedra connected by twofold bridging O atoms.\cite{Zachariasen1932Glass,Keen1999SilicaTotalScattering} These dominant environments determine the radial and angular signatures against which atomistic silica models are commonly assessed. Small rings, compressed Si--O--Si bridges and coordination defects are much less prevalent, but have been associated with strained and potentially active environments in amorphous silica.\cite{Uchino2000SilicaRings, Awazu2003Strained} The relevant modelling problem is therefore not only whether a sampled population reproduces silica on average, but whether it resolves the minority structures sufficiently to develop a meaningful structure-property relationship.

The defect-free and defective populations both retain the characteristic tetrahedral-network structure. Their total and partial radial distribution functions are closely matched, and their O--Si--O and Si--O--Si distributions preserve the expected intra- and inter-tetrahedral geometries (Fig.~\ref{fig:sio2-overview}; Supplementary Fig.~\ref{fig:si-sio2-structural}). These conventional descriptors establish that both populations represent credible amorphous silica. Their close agreement also exposes the limitation of a mean-structure comparison: minority environments can be distributed differently through the sampled population while remaining weakly expressed in ensemble-averaged pair and angle distributions.

Using the Si--O graph, the exact Si:O${}=4$ and O:Si${}=2$ coordination criterion partitions the 2,533 generated structures into 1,872 defect-free and 661 containing at least one coordination defect. Only 0.408\% of O sites are non-ideal, yet such sites occur in 26.10\% of the source cells. Local prevalence and population incidence are therefore distinct structural quantities. The deviations are predominantly oxygen-centred, while the tetrahedral Si framework remains largely intact, identifying local rearrangements embedded within an otherwise credible glass network rather than globally collapsed structures. Finally, these coordination defects are understood to originate with the artificially high cooling rate and across the tested cooling rates show a near linear dependence with lower cooling rates producing fewer defective cells albeit at a significantly higher computational cost.

This distinction has a direct modelling consequence. Pooling all atomic sites, or increasing the size of a single model, may improve the estimate of local motif prevalence, but does not determine how that motif is distributed between sampled structures or quantify the uncertainty of its population incidence. A coordination-defect claim must therefore state both its site-level denominator and the population of structures over which it is observed. Full selector attribution and screening sensitivity are reported in Supplementary Figs.~\ref{fig:si-sio2-screening} and \ref{fig:si-sio2-screen-sensitivity}; the independent-cell uncertainty analysis is reported in Fig.~\ref{fig:si-sio2-uncertainty-bonds}. The 26.10\% value is reported as cell incidence across the independent source cells.

\begin{figure*}[htb!]
  \centering
  \includegraphics[width=\textwidth]{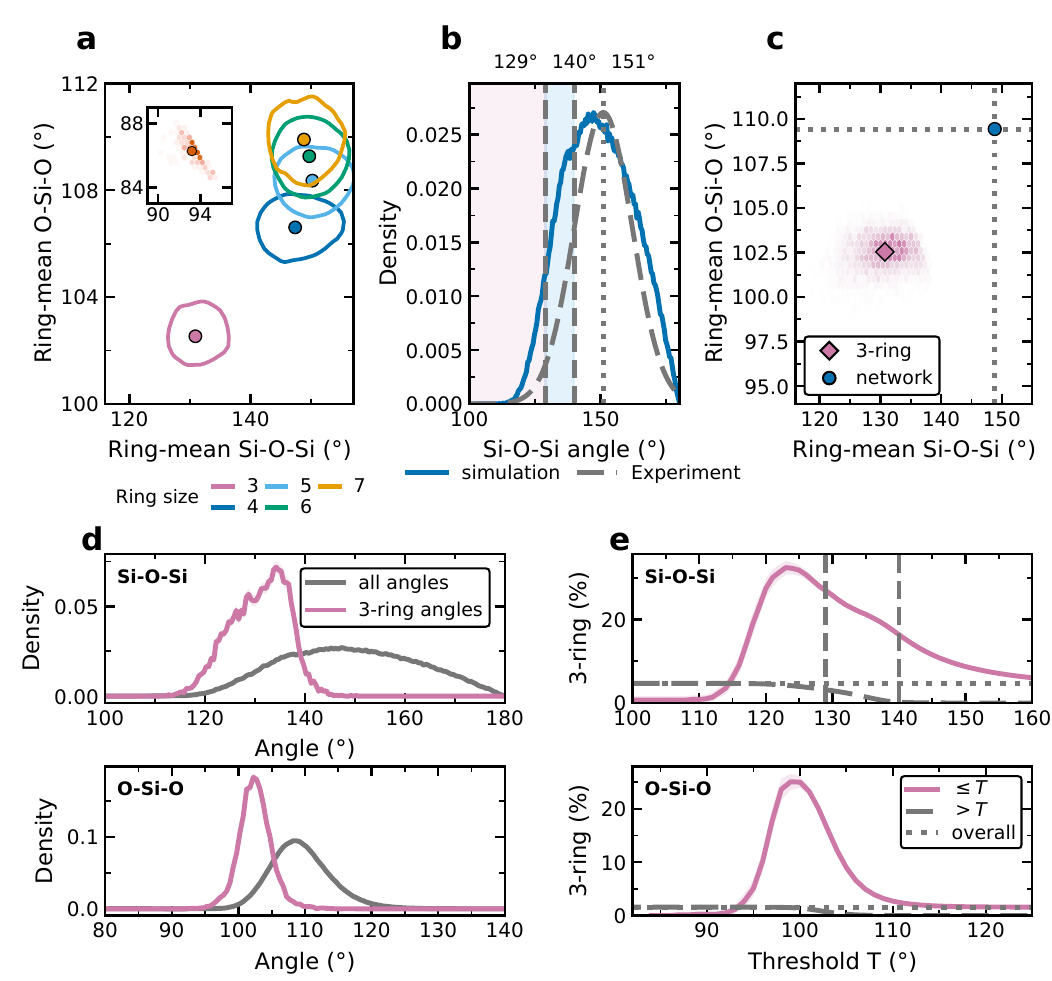}
  \caption{Ring-resolved Si--O--Si and O--Si--O geometry in amorphous silica. a) Joint ring-mean angles after assigning one shortest zero-winding ring to each exactly twofold O bridge. Contours show classes 3--7 and the inset resolves the 2-ring basin; this bridge assignment is distinct from the complete primitive-cycle census. b) Exactly twofold Si--O--Si distribution compared with the NMR-derived normal guide $\mu=151^{\circ}$ and $\sigma=11^{\circ}$ from Mauri \textit{et al.} \cite{Mauri2000SiOSi}. c) Joint mean angles for 5,169 validated primitive 3-rings. d) Equal-cell angle distributions for the full network (grey) and 3-rings (pink). e) Probability of 3-ring membership below (pink) and above (grey dashed) angular threshold $\theta_c$, with the overall probability shown by grey dotted lines.}
  \label{fig:sio2-focused}
\end{figure*}

Coordination defects are not the only minority environments obscured by conventional average-structure descriptors. Within the coordination-screened population, ring topology organises the tails of the angular distributions. Assigning one shortest zero-winding ring to each exactly twofold O bridge maps the coupled Si--O--Si and O--Si--O geometry across the observed ring classes (Fig.~\ref{fig:sio2-focused}a). The sparse 2-ring basin corresponds to edge-sharing tetrahedra, whereas the larger 3-ring population provides a statistically resolved case in which local angular deformation can be related directly to medium-range network topology. The experimental NMR mean and width place the simulated Si--O--Si distribution in a physical reference frame (Fig.~\ref{fig:sio2-focused}b). The values $151^{\circ}$, $140^{\circ}=\mu-\sigma$ and $129^{\circ}=\mu-2\sigma$ are consequently reference marks, rather than fitted boundaries or discrete structural classes. At $129^{\circ}$, 6.392\% of exactly twofold bridges occupy the lower Si--O--Si tail. The complete threshold dependence is important: three-ring association evolves continuously through the distribution and is not created by selecting a single angular cutoff.

Within the 5,169 validated primitive 3-rings, reduced inter-tetrahedral Si--O--Si angles are accompanied by a systematic shift in the intra-tetrahedral O--Si--O geometry of the same ring (Fig.~\ref{fig:sio2-focused}c). The complete marginal distributions confirm that both angle families are displaced relative to the wider network (Fig.~\ref{fig:sio2-focused}d). At the $129^{\circ}$ reference, lower-tail Si--O--Si bridges are enriched in primitive 3-rings by a factor of 5.746. The association remains substantial between $129^{\circ}$ and $140^{\circ}$ and becomes negligible only above approximately $140^{\circ}$ (Fig.~\ref{fig:sio2-focused}e). The materials consequence is that the lower angular tail is not unstructured statistical variation about the network mean. It is preferentially associated with a defined medium-range topology and coupled to deformation of the O--Si--O geometry within the same ring. Small-ring and compressed-bridge populations may therefore provide a structural basis for interpretations involving strained or active silica environments, although the present analysis establishes structural association rather than a chemical, mechanical or electronic mechanism.

The modelling consequence is equally direct. A force field or preparation protocol can reproduce density, radial distribution functions and mean bond angles while misrepresenting the population or network context of the tail states relevant to a materials claim. Validation against bulk averages alone is therefore insufficient where the intended inference concerns coordination defects, compressed bridges or small rings. The required test is whether the sampled population resolves the prevalence, incidence and topology of the specific structural quantity being invoked.

These results show that reproducing the mean tetrahedral network does not by itself resolve the minority structures on which tail-based materials interpretations depend. Site prevalence, cell incidence, angular-tail occupancy and ring context are distinct structural quantities, and each carries its own effective uncertainty
under the declared population and sampling construction. That uncertainty generally decreases as the sampled population grows, but not at a common rate and not towards a unique population size at which the silica ensemble becomes globally converged. A preparation protocol or force field may therefore support a precise description of the mean network while leaving the prevalence or topology of a relevant tail comparatively uncertain. The appropriate question is not whether the ensemble is ``large enough'', but what uncertainty has been achieved for the specific structural quantity used and how it relates to the materials question being addressed. The descriptor-resolved uncertainty is reported in Fig.~\ref{fig:si-sio2-uncertainty-bonds}, the family- and descriptor-level persistence audits in Tables~\ref{tab:si-family-uncertainty-thresholds} and \ref{tab:si-descriptor-uncertainty-thresholds}, and the screening sensitivity in Fig.~\ref{fig:si-sio2-screen-sensitivity}.

\subsection[DFT-refinement hierarchy in a-Si3N4]{DFT-refinement hierarchy in a-Si$_3$N$_4$}

\begin{figure*}[htb!]
  \centering
  \includegraphics[width=\textwidth]{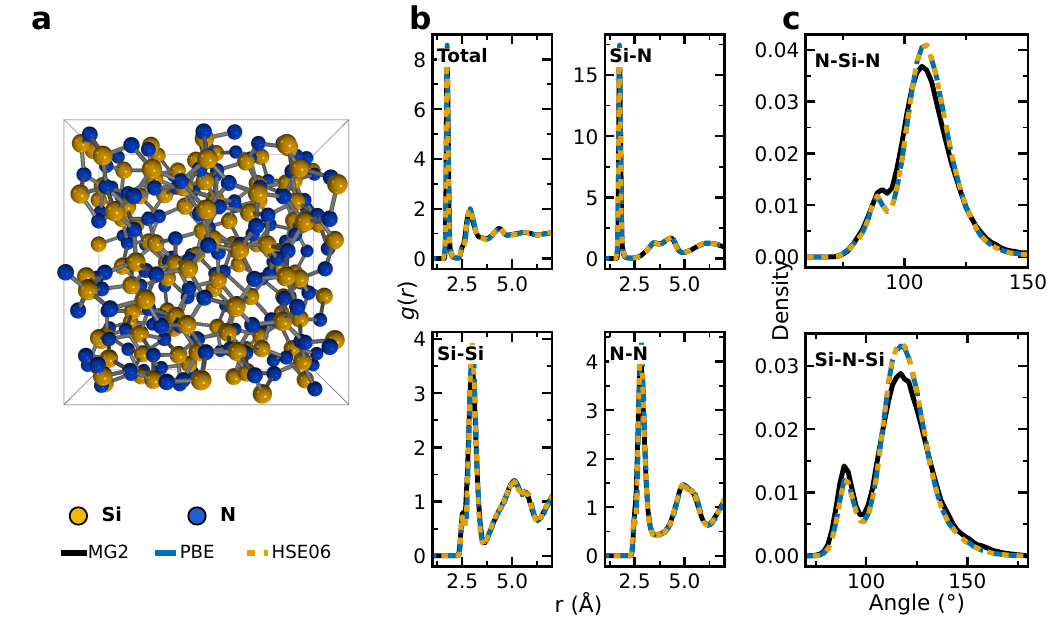}
  \caption{General structural characterisation of the matched a-Si$_3$N$_4$ population. a) Representative MG2 structure from the matched MG2, PBE and HSE06 parent set. Nitrogen is shown in blue, silicon in yellow and Si--N bonds use a 2.33~\AA{} cutoff. b) Equal-parent total and partial radial distribution functions for MG2, PBE and HSE06. c) Equal-parent N--Si--N and Si--N--Si bond angle distributions.}
  \label{fig:si3n4-overview}
\end{figure*}

The a-Si$_3$N$_4$ benchmark addresses a common hierarchy in amorphous materials modelling: structures are generated using a computationally tractable force field and subsequently refined with electronic-structure methods before their properties are evaluated. The unresolved question is what this refinement actually changes. A local DFT optimisation can correct packing, bond lengths and marginal coordination environments, but it cannot be assumed to resample the network topology selected during melt-quench generation. This distinction is important because the electronic, chemical and mechanical response of amorphous nitrides depends on the distribution of local coordination and connectivity, not only on their ensemble-averaged pair structure.\cite{Marian2000MG2,Ippolito2011ASi3N4,Dasmahapatra2018ASi3N4,Huckmann2024ChargeTrapping,Cottom2024HydrogenSiN,Cottom2025ForgedCharge}

We test this distinction using 455 matched MG2 parent structures, each refined independently along PBE and HSE06 branches. The PBE and HSE06 datasets are therefore paired descendants of the same generated
population rather than independent DFT melt-quench ensembles. Atom identity and parent lineage are retained throughout, so structural changes can be evaluated within the same underlying network. Because the branches also differ in their optimisation paths, they are treated as method branches rather than as an unconfounded comparison of exchange-correlation functionals.

The equal-parent radial and angular distributions show that all three branches retain the characteristic heteropolar a-Si$_3$N$_4$ network (Fig.~\ref{fig:si3n4-overview}). Refinement produces visible changes in the first Si--N shell and smaller shifts in the angular distributions, but does not replace the broad network structure inherited from MG2. These descriptors establish the general structural response. They do not, however, distinguish a change in radial scale from a change in distribution shape, or either of these from reconstruction of the first-neighbour graph. Are the changes driven by a sharpening of the neighbour environments at a broadly fixed network topology or is the network being remade?

\begin{figure*}[htb!]
  \centering
  \includegraphics[width=\textwidth]{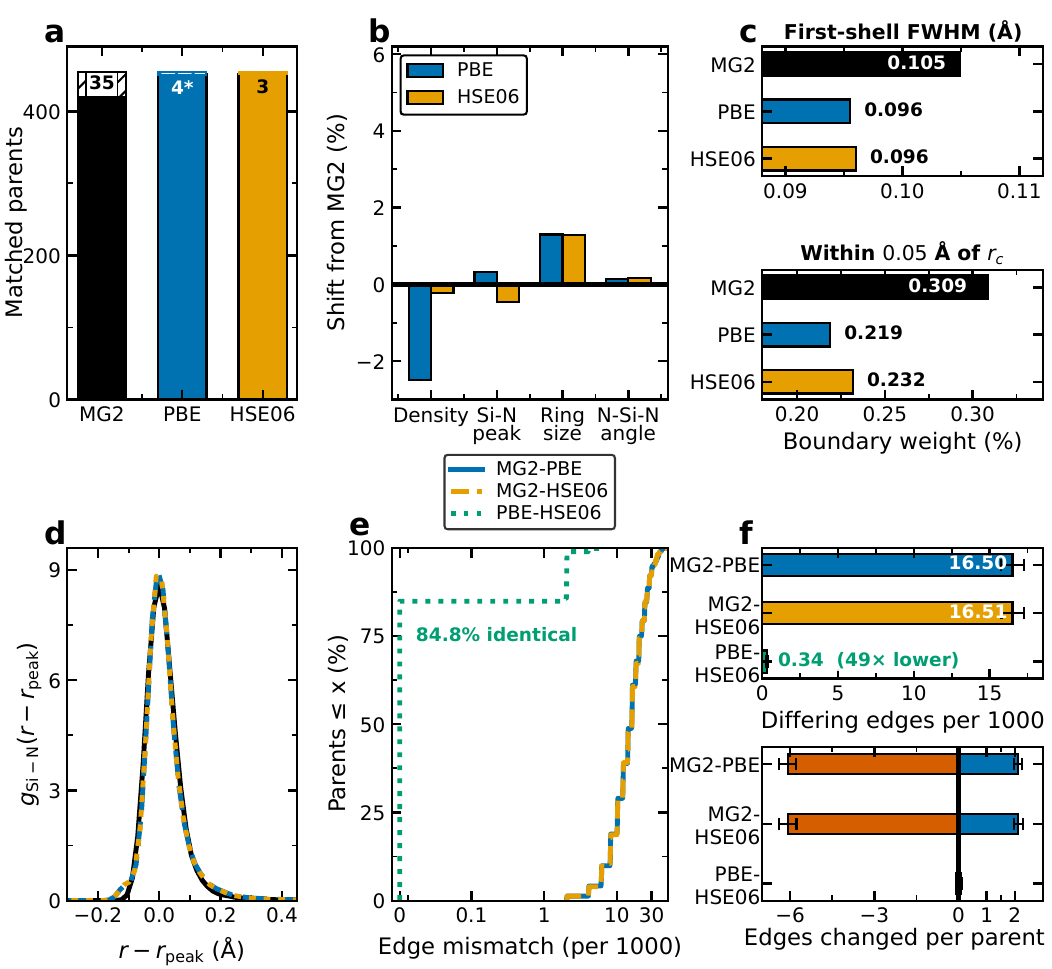}
  \caption{First-shell sharpening and topology across matched a-Si$_3$N$_4$ refinement branches. a) Matched-parent population and diagnostic audit labels; all 455 parents are retained. The PBE count is $4^{\ast}$ because one parent lies on the audit boundary. b) Relative changes from MG2 in density, Si--N first-peak position, mean ring size and mean N--Si--N angle. c) Si--N first-shell width and graph-boundary pair weight. d) Peak-referenced Si--N distributions. In panel c, $r_c=r_c^{\mathrm{RDF},M}$ is the branch-specific registered RDF boundary, distinct from the audit cutoff and the common-topology cutoff; in panel d, $r_{\mathrm{peak}}=r_{\mathrm{pk}}^{(M)}$ is the branch-specific Si--N first-peak position. e) Common-graph edge mismatch across matched parents; 84.8\% of PBE-HSE06 pairs have identical first-neighbour graphs. f) Aggregate edge mismatch and removed and created edges per parent. PBE and HSE06 are treated as method branches.}
  \label{fig:si3n4-mg2-pbe-hse}
\end{figure*}

With \code{exclude=false}, all 455 matched parents remain in each branch. The audit-labelled highly defective subset decreases from 35 parents for MG2 to $4^{\ast}$ for PBE and 3 for HSE06 (Fig.~\ref{fig:si3n4-mg2-pbe-hse}a). The asterisk records one PBE parent lying on the registered audit boundary. The reduction in audit labels indicates that DFT refinement removes many of the marginal coordination environments present in the force-field structures.

The largest refinement response is associated with packing and absolute radial scale. Relative to MG2, the PBE density decreases by 2.49\%, whereas the HSE06 density differs by only $-0.23$\% (Fig.~\ref{fig:si3n4-mg2-pbe-hse}b). The registered Si--N first-peak position shifts by $+0.32$\% for PBE and $-0.46$\% for HSE06, while the mean ring-size and angular changes are smaller. The branches therefore do not respond through one common scalar displacement: cell packing, first-neighbour scale and medium-range descriptors are impacted to differing degrees.

Both DFT branches nevertheless sharpen the force-field first shell in a closely similar manner. The Si--N peak full width at half maximum decreases from 0.105~\AA{} for MG2 to 0.096~\AA{} for both PBE and HSE06. At the same time, the weight of Si--N pairs lying within 0.05~\AA{} of the registered graph boundary falls from 0.309\% for MG2 to 0.219\% for PBE and 0.232\% for HSE06 (Fig.~\ref{fig:si3n4-mg2-pbe-hse}c). The reduction in audit-labelled parents is therefore accompanied by a narrower first shell and fewer pairs whose first-neighbour assignment is sensitive to the graph boundary. DFT refinement primarily repairs marginal force-field environments rather than broadly reordering the network. Referencing each branch to its own first-shell peak separates this metric response from the residual distribution shape (Fig.~\ref{fig:si3n4-mg2-pbe-hse}d). After removal of the branch-specific radial displacement, the PBE and HSE06 Si--N distributions are nearly indistinguishable and both remain sharper than MG2. The apparent difference between the two DFT branches in absolute coordinates is dominated by packing and bond-length scale, rather than by substantial differences in first-shell shape.

The graph analysis provides a more stringent topological test. Under the registered common 2.33~\AA{} graph, 84.8\% of matched PBE-HSE06 parent pairs have identical first-neighbour edge sets (Fig.~\ref{fig:si3n4-mg2-pbe-hse}e). Across the complete population, the PBE-HSE06 edge-union mismatch is only 0.34 edges per 1,000 union edges, compared with 16.50 and 16.51 edges per 1,000 for MG2-PBE and MG2-HSE06, respectively (Fig.~\ref{fig:si3n4-mg2-pbe-hse}f). The mismatch between the two DFT descendants is consequently approximately 49 times smaller than that between either descendant and its MG2 parent. The PBE-HSE06 comparison also contains fewer than 0.2 removed and created edges per parent, whereas each MG2-DFT comparison contains approximately eight.

These results show that DFT refinement changes different levels of the amorphous structure by markedly different amounts. Packing and first-neighbour length scale respond measurably, the force-field first shell sharpens and a small population of marginal contacts is repaired, but the PBE and HSE06 descendants retain nearly identical first-neighbour topology. The two DFT branches therefore converge on essentially the same local network after correcting closely related features of the MG2 population.

The consequence is that structural differences between the PBE and HSE06 branches can be interpreted without conflating them with wholesale reconstruction of the amorphous network. Conversely, the larger MG2-DFT differences include both metric relaxation and local topological repair. This separation is necessary wherever electronic states, trapping sites, chemical response or mechanical transformation are attributed to a specific coordination environment: a change in the electronic-structure description need not imply that a different network population has been sampled.

Each level of this comparison carries a distinct effective uncertainty under the matched-parent sampling construction. Density, first-shell shape, audit-label prevalence and graph mismatch are not resolved to a common precision by one nominal population size. Complete structural distributions and scalar descriptors are reported in Figs.~\ref{fig:si-si3n4-structural} and \ref{fig:si-si3n4-scalars}; descriptor-resolved uncertainty is shown in Fig.~\ref{fig:si-si3n4-uncertainty-pairs}; and the separate coordination-audit and graph-boundary sensitivities are reported in Figs.~\ref{fig:si-si3n4-audit-sensitivity} and \ref{fig:si-si3n4-boundary}.

\subsection[Crystal-like order and mixed coordination in a-Sm2O3]
{Crystal-like order and mixed coordination in a-Sm$_2$O$_3$}

\begin{figure*}[htb!]
  \centering
  \includegraphics[width=\textwidth]{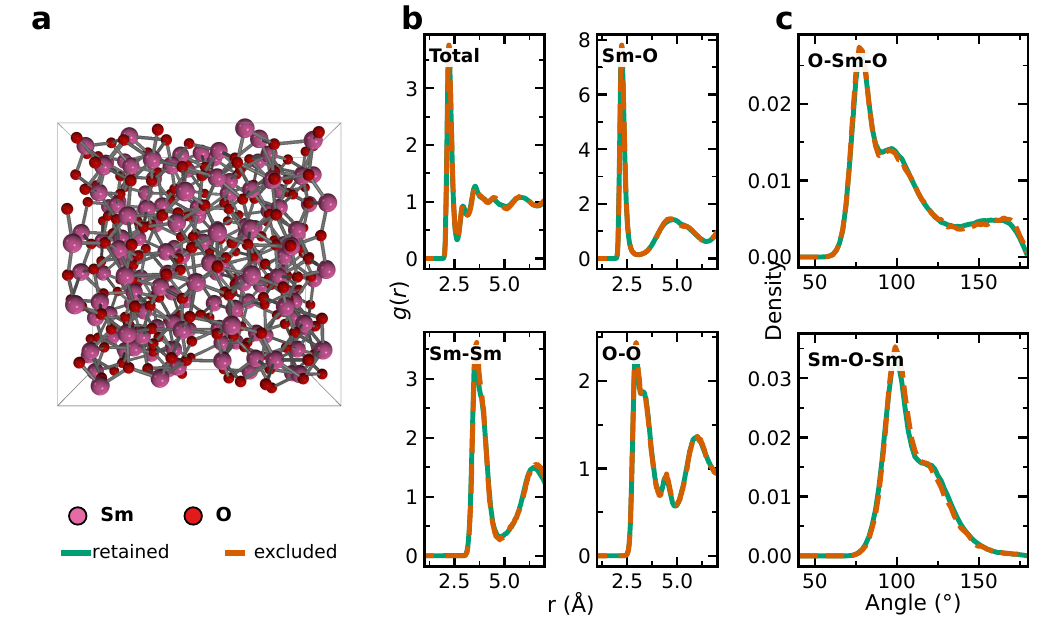}
  \caption{Structural characterisation of the sampled a-Sm$_2$O$_3$ populations. a) Representative amorphous structure. Oxygen is shown as red spheres, samarium as pink spheres and Sm--O bonds use a 2.95~\AA{} cutoff. b) Equal-cell total and partial radial distribution functions for the amorphous and partially ordered populations. c) Equal-cell O--Sm--O and Sm--O--Sm bond angle distributions.}
  \label{fig:sm2o3-overview}
\end{figure*}

Sm$_2$O$_3$ poses a different population problem from the Si-based network materials. Broad Sm--O and O--Sm coordination environments are intrinsic to the amorphous rare-earth oxide, so a fixed-coordination screen would remove legitimate local environments and bias the population towards an
artificially idealised network.\cite{Olsson2019AmorphousSm2O3} The competing preparation outcome is the formation of extended regions of crystal-like order, which become more prevalent as the cooling schedule allows greater structural ordering. Treating every generated cell as amorphous would therefore mix predominantly disordered structures with partially ordered products, whereas assigning crystallographic phases from a local-order scalar would exceed what the calculation establishes.

This distinction is equally important when melt-quench structures are used to represent partially crystallised or polycrystalline films. Such a description should be supported by a measured prevalence and connectivity of ordered environments, rather than by visual inspection, density or a single order parameter. Here, the screen is consequently used not merely to discard undesirable cells, but to quantify the ordered material generated by the preparation protocol and to construct
separate amorphous and partially ordered populations.

The two populations retain the broad pair and angular structure of the disordered Sm--O network, but differ consistently in their short- and medium-range correlations (Fig.~\ref{fig:sm2o3-overview}). The comparison therefore concerns different degrees of order within one chemically related rare-earth-oxide population rather than unrelated structures. It also shows why density or one mean local-order value is insufficient: the relevant distinction is whether ordered sites remain isolated within a disordered network or form an extended connected region.

We identify these environments using the local bond-order construction of Steinhardt and the neighbour-averaged extension of Lechner and Dellago.\cite{Steinhardt1983BondOrder,Lechner2008BondOrder} Predominantly amorphous cells have mean crystalline and largest-connected-region fractions $f_{\mathrm{cryst}}=0.0257\pm0.0015$ and $f_{\mathrm{cluster}}=0.0232\pm0.0014$, respectively, whereas the partially ordered population has $f_{\mathrm{cryst}}=f_{\mathrm{cluster}}=0.348\pm0.024$. The near equality of the two values in the partially ordered population demonstrates that most ordered sites belong to one connected region rather than appearing as isolated high-order fluctuations.

\begin{figure*}[htb!]
  \centering
  \includegraphics[width=\textwidth]{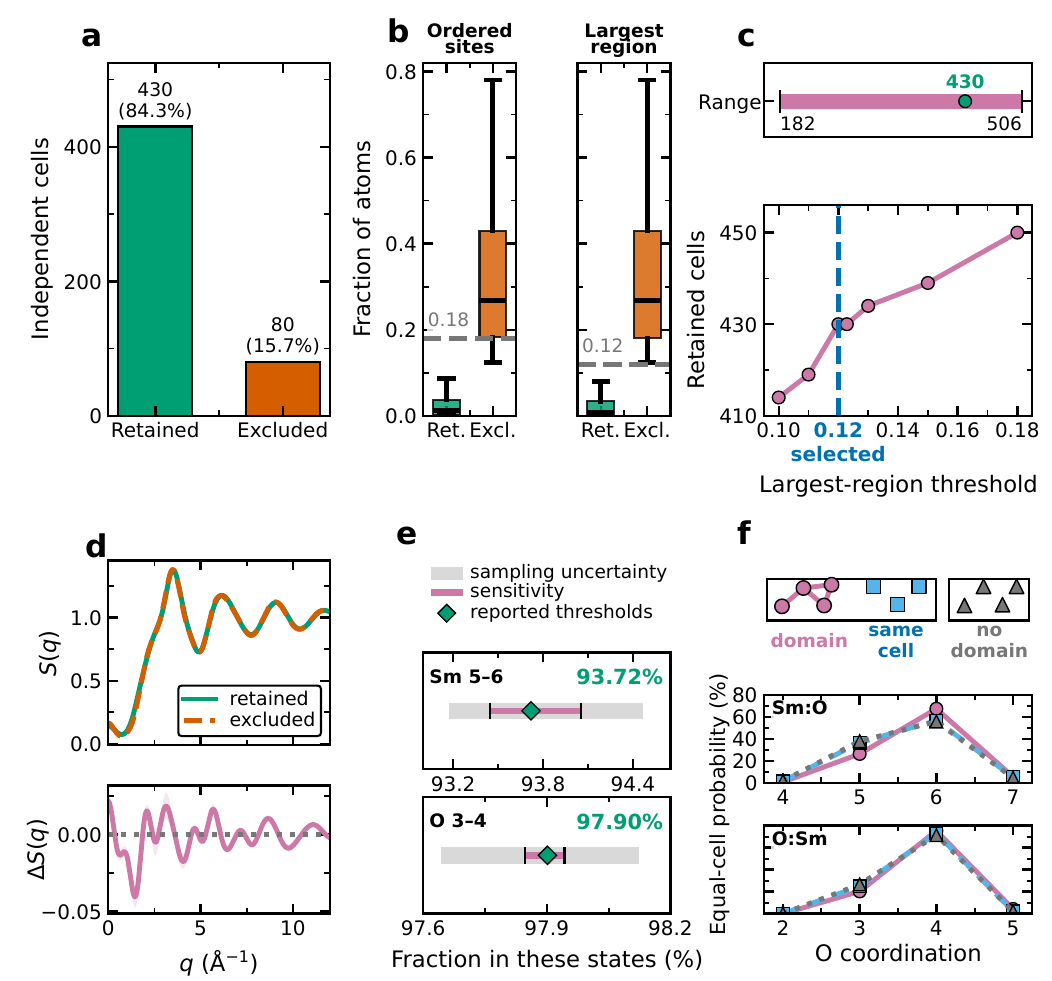}
  \caption{Crystal-like order and structural robustness in a-Sm$_2$O$_3$. a) Operational split into predominantly amorphous and partially ordered populations. b) Ordered-site and largest-connected-region fractions with the selected thresholds. c) Amorphous-population size across the tested threshold family and the largest-region-threshold slice; the selected setting retains 430 cells. d) Equal-cell total structure factors for the two populations and their difference with 95\% uncertainty. e) Dominant five- and sixfold Sm and three- and fourfold O fractions; grey shows sampling uncertainty, magenta sensitivity and the diamond the selected thresholds. f) Coordination distributions within connected ordered regions, outside those regions in the same cells and in cells containing no connected ordered region.}
  \label{fig:sm2o3-crystal-screening}
\end{figure*}

At the selected thresholds, 430 of the 510 independently generated cells form the predominantly amorphous population and 80 form the partially ordered comparison population (Fig.~\ref{fig:sm2o3-crystal-screening}a,b). The largest-region criterion identifies all 80 partially ordered cells, while the ordered-site criterion identifies a nested subset of 60. Bragg sharpness is non-binding and the archived reference-overlap diagnostic is non-discriminative; both are retained in the Supporting Information as audit quantities rather than treated as coequal determinants of the population. The classification is therefore attributed to connected local order. It measures the extent of crystal-like structure without making the link to a crystallographic phase.

A single threshold choice would conceal how strongly the reported amount of partially ordered material depends on the operational definition. Across the declared 108-setting classifier family, the predominantly amorphous population ranges from 182 to 506 cells, or 35.7--99.2\% of the source population (Fig.~\ref{fig:sm2o3-crystal-screening}c). This variation is not evidence for one uniquely correct boundary. It quantifies the range of population assignments supported by the declared family of local-order definitions and makes the resulting uncertainty in the amorphous or partially ordered fraction explicit.

The two populations also differ outside the quantities used to construct the classifier. Their total structure factors retain the broad response of a disordered oxide, but the partially ordered-minus-amorphous difference resolves systematic changes in reciprocal space (Fig.~\ref{fig:sm2o3-crystal-screening}d). Together with the real-space pair and angle distributions in Fig.~\ref{fig:sm2o3-overview}, this shows that the population split corresponds to a broader structural change rather than merely to displacement along the classifier coordinates. These correlations are consistent with increased medium-range order, but do not identify a particular crystal structure.

The central result is that in this case the operational order--disorder boundary does not destabilise the dominant local chemistry. At the selected setting, five- and sixfold environments account for 93.72\% of Sm sites, while three- and fourfold environments account for 97.90\% of O sites (Fig.~\ref{fig:sm2o3-crystal-screening}e). Across the complete classifier family, these equal-cell shares vary only from 93.447--94.055\% for Sm and 97.846--97.943\% for O. The number of cells assigned to the amorphous population therefore changes strongly, while the conclusion that a-Sm$_2$O$_3$ is dominated by mixed five-/sixfold Sm and three-/fourfold O coordination remains stable.

Connected ordered regions nevertheless select a distinct part of these broad coordination distributions. The Sm population within the ordered region shifts towards sixfold coordination relative to the remainder of the same cell and to cells containing no connected ordered region; the O population shifts correspondingly towards fourfold coordination (Fig.~\ref{fig:sm2o3-crystal-screening}f). Equal-cell mean coordinations are 5.776 for Sm and 3.843 for O inside the connected region, compared with 5.658 and 3.771 outside it in the same cells and 5.633 and 3.755 in the no-region population. The supported interpretation is therefore a higher-coordination, crystal-like domain embedded within a disordered mixed-coordinate network, rather than a coordination-idealised phase.

For calculations intended to describe amorphous Sm$_2$O$_3$, the partition prevents cells containing extended order from silently biasing ensemble averages. For calculations intended to represent partially crystallised or polycrystalline material, the same ordered-site and connectivity measures provide quantitative descriptors of how much ordered material has formed and whether it is distributed as isolated motifs or connected domains. Crystallographic phase identity would require additional phase-resolved evidence, but the degree and connectivity of crystal-like order are already measurable from the generated population.

The modelling consequence is that population membership and materials inference must be tested separately. The exact number of cells labelled amorphous or partially ordered depends strongly on the selected classifier thresholds, whereas the dominant mixed coordination is robust throughout the declared family. Reporting only the selected 430/80 split would hide the first fact; rejecting the classifier because that split moves would ignore the second. A defensible model should therefore report the operational population definition, its sensitivity and the stability of every downstream structural conclusion. 

Reciprocal-space order, pair and angle distributions, classifier prevalence and coordination populations also carry different effective uncertainties under the same set of independent cells. Adding cells generally reduces these uncertainties, but does not define one population size at which all descriptions of Sm$_2$O$_3$ become globally converged. Structural distributions and screening descriptors are reported in Figs.~\ref{fig:si-sm2o3-structural} and \ref{fig:si-sm2o3-screening}; threshold-family sensitivity and classifier-component attribution in Figs.~\ref{fig:si-sm2o3-classifier-sensitivity} and \ref{fig:si-sm2o3-component-response}; and descriptor-resolved uncertainty in Fig.~\ref{fig:si-sm2o3-uncertainty-scalars}.

\subsection{Systematic modelling of amorphous populations}

Together, the three benchmark systems demonstrate a systematic route from ensemble generation to materials-level structural interpretation. In a-SiO$_2$, conventional pair and angle distributions validate the mean tetrahedral network, while the ring-resolved analysis shows that the lower angular tail is organised by a specific medium-range topology that is not apparent from the average structure. In a-Si$_3$N$_4$, the matched refinement hierarchy separates geometric relaxation from network reconstruction: DFT refinement redistributes density and first-neighbour length scale, sharpens the force-field first shell and repairs marginal contacts, while the PBE and HSE06 descendants retain essentially the same first-neighbour topology. In a-Sm$_2$O$_3$, local-order screening quantifies the emergence and connectivity of partially ordered regions from a disordered parent population, while preserving the mixed coordination that characterises the amorphous rare-earth oxide. The three systems therefore resolve complementary aspects of an amorphous ensemble: mean structure, distribution tails, inherited topology and the development of connected order.

Vitriflow makes these distinctions controllable rather than implicit. Calibration defines the preparation-conditioned source population; explicit labels and actions construct the population relevant to the materials question; and cell- or parent-level uncertainty quantifies the precision of each resulting structural statement. Preparation protocols, force fields and refinement methods can consequently be compared without conflating changes in packing, local geometry, network topology or population membership. The result is an auditable connection between what was generated, which structures were analysed and what the sampled population establishes. Quantity-specific uncertainty is reported directly rather than being hidden behind a single assertion of ensemble convergence, while Fig.~\ref{fig:si-declaration-consequences} shows how changes in an operational population definition propagate into the materials result.

\section{Conclusions}

Structure-property relations in amorphous materials are inherently challenging: they are encoded not only in the mean structure, but in the prevalence, correlations and spatial organisation of local environments across length scales. A model can reproduce density, pair correlations and mean coordination while misrepresenting the strained motifs, distribution tails or connected domains required to capture a material response. The relevant computational object is therefore not a nominal structure, or an ensemble declared globally converged, but a finite population that resolves the property-relevant structural quantity and states the uncertainty with which its prevalence and context are known.

The three materials demonstrate this connection directly. In a-SiO$_2$, strained bonds and small rings have been implicated in electronic, optical, radiation and chemical responses.\cite{Uchino2000SilicaRings,Awazu2003Strained} Here, strain is followed from intra-tetrahedral O--Si--O deformation, through compressed inter-tetrahedral Si--O--Si bridges, to primitive 3-ring topology. The lower angular tail is thereby identified as an organised medium-range population rather than undifferentiated variation about the network mean. In a-Si$_3$N$_4$, higher-level refinement is widely used to improve empirically generated structures, but its action on the amorphous network is rarely quantified. The matched hierarchy shows that DFT redistributes density and first-neighbour length scale, sharpens the force-field first shell and repairs marginal contacts, while the PBE and HSE06 descendants retain essentially the same first-neighbour topology. This distinction is central when coordination and connectivity are used to interpret charge trapping, hydrogen response, oxidation and structural transformation.\cite{Huckmann2024ChargeTrapping,Cottom2024HydrogenSiN,Huckmann2026DryOxidation,Cottom2025ForgedCharge,Wilhelmer2026ElectronTrapping} In Sm$_2$O$_3$, where amorphous-to-partially ordered or polycrystalline evolution is relevant to the dielectric and optical behaviour of rare-earth-oxide films,\cite{Leskela2003RareEarthOxides,Dakhel2004Sm2O3ThinFilms, Olsson2019AmorphousSm2O3} the analysis quantifies both the amount and connectivity of crystal-like order while showing that the mixed coordination characteristic of the disordered oxide remains robust. Structural heterogeneity is thus converted from an uncontrolled feature of the generated cells into a measurable materials variable.

\vitriflow{} makes this analysis executable and auditable within a user-defined parameter space. For newly generated ensembles, it connects a calibrated preparation protocol to the resulting source population; for existing datasets, it begins from the supplied structures and their declared provenance. In either case, it applies the same population construction, screening, structural analysis and uncertainty quantification, records screen labels separately from their actions, and reports the achieved uncertainty of each quantity over its stated sampling unit. Scientific choices remain explicit, but their consequences for population membership, structural resolution and interpretation are measured rather than assumed. The resulting record links how the structures were obtained, which population was analysed and what that finite population establishes, providing a reproducible route from amorphous structure datasets to quantitatively supported structure-property hypotheses.

\section*{CRediT authorship contribution statement}
\noindent
\textbf{Jonathon Cottom:} Conceptualization, Methodology, Software, Validation, Formal analysis, Investigation, Data curation, Visualization, Writing -- original draft, Writing -- review \& editing. 
\textbf{Robin Delhomme:} Methodology, Software, Validation, Formal analysis, Investigation, Data curation, Visualization, Writing -- review \& editing. 
\textbf{Emilia Olsson:} Conceptualization, Methodology, Software, Resources, Supervision, Project administration, Funding acquisition, Writing -- review \& editing.

\section*{Acknowledgements}
The authors would like to thank Marsal Colitt Marull, Lina-Henriette Schwering, Xander de Haan, and all members of the Olsson Materials Theory and Modelling group whose assistance has been invaluable in testing and refining \vitriflow{}. 
This work was conducted at the Advanced Research Center for Nanolithography, a public-private partnership between the University of Amsterdam (UvA), Vrije Universiteit Amsterdam (VU), Rijksuniversiteit Groningen (RUG), the Netherlands Organization for Scientific Research (NWO), and the semiconductor equipment manufacturer ASML. This work made use of the Dutch national e-infrastructure with the support of the SURF Cooperative using grant no. EINF-16595 and EINF-8981. The authors thank SURF (www.surf.nl) for the support in using the National Supercomputer Snellius. This work made use of the ARCNL Minerva cluster hosted at HOPSTER at UvA. E.O. is grateful for a WISE Fellowship from the NWO and acknowledges support via Holland High Tech through a public-private partnership in research and development within the Dutch top sector of High-Tech Systems and Materials (HTSM).

\section*{Data availability}
The \vitriflow{} package is hosted on GitHub (\url{https://github.com/Olsson-Materials-Modelling/Vitriflow}); the version specific to this paper is archived in the Zenodo repository [10.5281/zenodo.21111823]. The structures and post-analysis JSON files are freely available in the same Zenodo repository [10.5281/zenodo.21111823].

\section*{Declaration of competing interest}
The authors declare that they have no known competing financial interests or personal relationships that could have appeared to influence the work reported in this paper.

\bibliographystyle{elsarticle-num}
\bibliography{refs}

\end{document}


\begin{frontmatter}
\title{Supplementary Information for ``\texttt{Vitriflow}: calibrated amorphous structure ensembles from melt--quench simulations''}

\author[inst1,inst2]{Jonathon Cottom}
\author[inst1,inst2]{Robin Delhomme}
\author[inst1,inst2]{Emilia Olsson\corref{cor1}}
\ead{k.i.e.olsson@uva.nl}
\affiliation[inst1]{organization={Institute for Theoretical Physics, University of Amsterdam},
            addressline={Science Park 904},
            city={Amsterdam},
            postcode={1098 XH},
            country={the Netherlands}}
\affiliation[inst2]{organization={Advanced Research Center for Nanolithography},
            addressline={Science Park 106},
            city={Amsterdam},
            postcode={1098 XG},
            country={the Netherlands}}
\cortext[cor1]{Corresponding author.}

\end{frontmatter}

\section{Supplementary methodological details}
\label{sec:si-methods}

\subsection{Scope, ordering, and data lineage}
\label{sec:si-scope}

The supplementary methodology follows the order of the main Results section. The scientific framework defines the decision chain; the \vitriflow{} software package executes and records it. First, the common operations are specified: descriptor extraction, numerical preflight, protocol calibration, user-defined artefact screening, analysis-population construction, and statistical support. The material-specific implementations are then described in the order used in the main text: a-SiO$_2$ as a tetrahedral-network fidelity case, a-Si$_3$N$_4$ as a matched DFT-refinement case, and a-Sm$_2$O$_3$ as a mixed-coordination rare-earth oxide requiring crystal-like screening.

The source-data deposit contains the workflow configuration files, final JSON summaries, run inputs, final relaxed structures, and descriptor tables. A generated cell denotes a completed material-specific source construction followed by the configured quench and final relaxation sequence; its independent cell-level sampling unit may originate from an independently seeded trajectory or a decorrelated source snapshot. Numerical failures during preflight or calibration are treated as rejected numerical settings, not as generated cells. Cells excluded by the material-specific structural screen are retained in the audit trail and, where useful, analysed as comparison populations.

For a generated final cell $x_b$, with cell index $b=1,\ldots,N_{\mathrm{gen}}$, let $\mathcal S_{\mathrm{screen}}$ denote the configured screens. Each screen $s$ stores a label

\begin{equation}
A_{bs}\in\{0,1\},
\end{equation}

where $A_{bs}=1$ denotes a screen-passing cell and $A_{bs}=0$ denotes an audit-labelled cell. Each configured screen also stores an exclude setting,
\begin{equation}
\epsilon_s=\begin{cases} 1, & \code{exclude=true},\\
0, & \code{exclude=false}.
\end{cases}
\end{equation}
The post-action analysis mask is
\begin{equation}
R_b=\prod_{s\in\mathcal S_{\mathrm{screen}}}\left[A_{bs}+(1-A_{bs})(1-\epsilon_s)\right],
\end{equation}
with $R_b=1$ when no screen is configured. The analysis population is
\begin{equation}
\mathcal E_{\mathrm{ana}}=\{x_b:R_b=1\},
\qquad
N_{\mathrm{ana}}=|\mathcal E_{\mathrm{ana}}|.
\end{equation}
For each screen, the screen-passing and audit-labelled populations are
\begin{equation}
\mathcal E_{A=1}^{(s)}=\{x_b:A_{bs}=1\},
\qquad
\mathcal E_{A=0}^{(s)}=\{x_b:A_{bs}=0\}.
\end{equation}
Their screen-specific counts are $N_{A=a}^{(s)}=|\mathcal E_{A=a}^{(s)}|$ for $a\in\{0,1\}$; the superscript is suppressed for the single material-level screen reported in the main table. For \code{exclude=true}, audit-labelled cells are excluded from $\mathcal E_{\mathrm{ana}}$. For \code{exclude=false}, they are included in $\mathcal E_{\mathrm{ana}}$ with the label preserved. Counts are reported as $N_{\mathrm{gen}}:N_{A=1}:N_{A=0}:N_{\mathrm{ana}}$, matching the notation used in the main text.

\subsection{Descriptor map and descriptor roles}
\label{sec:si-descriptor-map}

The workflow is descriptor-driven. For a structure or trajectory segment $x$, \vitriflow{} evaluates a material-specific descriptor map
\begin{equation}
\Phi(x;\mathcal D)
=
\left(d_1(x),d_2(x),\ldots,d_{n_{\mathcal D}}(x)\right),
\label{eq:si-descriptor-map}
\end{equation}
where $\mathcal D$ is the descriptor set selected for the materials question and $n_{\mathcal D}=|\mathcal D|$. Descriptor sets are partitioned by role:
\begin{equation}
\mathcal D
=
\mathcal D_{\mathrm{pf}}
\cup
\mathcal D_{\mathrm{cal}}
\cup
\mathcal D_{\mathrm{screen}}
\cup
\mathcal D_{\mathrm{conv}}
\cup
\mathcal D_{\mathrm{report}}.
\end{equation}
These subsets need not be disjoint. Here, $\mathcal D_{\mathrm{pf}}$ contains quantities used for numerical stability preflight, $\mathcal D_{\mathrm{cal}}$ for protocol calibration, $\mathcal D_{\mathrm{screen}}$ for material-specific artefact screening, $\mathcal D_{\mathrm{conv}}$ for statistical convergence, and $\mathcal D_{\mathrm{report}}$ for the final structural comparisons. The configured rate- and size-equivalence descriptor subsets satisfy $\mathcal D_{\mathrm{quench}},\mathcal D_{\mathrm{size}}\subseteq\mathcal D_{\mathrm{cal}}$.

The core descriptor family comprises density, total and partial radial distribution functions, first-shell coordination numbers, chemically relevant bond-angle distributions, and medium-range descriptors.\cite{Allen2017ComputerSimulation,Fischer2006Diffraction} The medium-range descriptors are chosen to avoid graph-construction artefacts: projected Si-ring statistics are used for SiO$_2$, primitive bond-graph rings are used for Si$_3$N$_4$, and local-order metrics are used for Sm$_2$O$_3$.\cite{Franzblau1991Rings,Steinhardt1983BondOrder,Lechner2008BondOrder} Structure factors, void-clearance distributions, elastic fingerprints, and residual-stress summaries are retained as secondary diagnostics and are not used as primary acceptance criteria unless explicitly configured. This descriptor family is modular and material-specific, not complete or universally transferable. Users may supply additional invariant local representations, including atom-centred symmetry functions, SOAP, ACE-based representations, or other invariant descriptor bases; none is treated as a universal default.

\subsection{Structural descriptor extraction}
\label{sec:si-descriptor-extraction}

For a final relaxed cell $x_b$ with volume $V_b$, the mass density is
\begin{equation}
\rho_b=\frac{M_b}{V_b},
\end{equation}
where $M_b$ is the total cell mass. Partial radial distribution functions are evaluated on a fixed radial grid $r_k=k\Delta r$ with $k\geq1$ and bin width $\Delta r>0$.\cite{Allen2017ComputerSimulation} For species $\alpha$ and $\beta$,
\begin{equation}
g_{\alpha\beta}^{(b)}(r_k)
=
\frac{V_b}
{4\pi r_k^2\Delta r\,N_\alpha^{(b)}
\left(N_\beta^{(b)}-\delta_{\alpha\beta}\right)}
\sum_{i\in\alpha}
\sum_{\substack{j\in\beta\\ j\neq i}}
\mathbf 1
\left[
\left|r_{ij}-r_k\right|
<
\frac{\Delta r}{2}
\right],
\end{equation}
where $N_\alpha^{(b)}$ is the number of atoms of species $\alpha$ in cell $b$, $r_{ij}$ is the minimum-image pair distance, and $\delta_{\alpha\beta}$ is the Kronecker delta. Total radial distribution functions are obtained by composition-weighted sums of the corresponding partial functions. Where a total structure factor is reported, $q$ is the scattering-vector magnitude in \AA$^{-1}$, $S_{\alpha\beta}(q)$ is the dimensionless partial structure factor returned by the fixed archived evaluator, and $w_{\alpha\beta}(q)$ is its dimensionless registered composition/scattering weight. Thus $S(q)=\sum_{\alpha\beta}w_{\alpha\beta}(q)S_{\alpha\beta}(q)$ is dimensionless; the same evaluator convention, weights, and $q$ grid are used for every compared population.

First-shell coordination numbers are computed from fixed cutoff radii. For an atom $i$ of species $\alpha$, the number of neighbours of species $\beta$ is
\begin{equation}
n_{i:\beta}^{(b)}
=
\sum_{\substack{j\in\beta\\j\neq i}}
\mathbf 1
\left[
r_{ij}\leq r_{\alpha\beta}^{c}
\right].
\end{equation}
The cutoff $r_{\alpha\beta}^{c}$ is obtained from the first minimum following the primary peak in the smoothed pair-distance histogram and is then held fixed for all cells compared within a material-specific analysis. For cross-model comparisons, the same cutoff-selection rule and fixed descriptor grid are applied to each dataset before ensemble statistics are computed.

Bond-angle probability densities are evaluated on the bond graph implied by the same first-shell cutoffs and on the fixed angular grid $\theta_k=k\Delta\theta$, where $\Delta\theta>0$ is the angular bin width. For a triplet $\alpha$--$\beta$--$\gamma$, where $\beta$ is the central species,
\begin{equation}
P_{\alpha\beta\gamma}^{(b)}(\theta_k)
=
\frac{1}{M_{\alpha\beta\gamma}^{(b)}\Delta\theta}
\sum_{j\in\beta}
\sum_{i\in\mathcal N_\alpha(j)}
\sum_{\substack{\ell\in\mathcal N_\gamma(j)\\ \ell\neq i}}
\mathbf 1
\left[
\left|\theta_{ij\ell}-\theta_k\right|
<
\frac{\Delta\theta}{2}
\right],
\end{equation}
where $\mathcal N_\alpha(j)$ is the set of bonded $\alpha$ neighbours of central atom $j$, $\theta_{ij\ell}$ is the angle subtended by atoms $i$--$j$--$\ell$, and $M_{\alpha\beta\gamma}^{(b)}>0$ is the number of counted triplets after excluding self-pairs with $i=\ell$. No normalised angular density is assigned to a cell with no triplet of the requested type.

Ring statistics are graph-based.\cite{Franzblau1991Rings} For SiO$_2$, the projected Si graph is an O-indexed multigraph with one edge for each exactly twofold bridging O atom:
\begin{equation}
E_{\mathrm{SiO_2}}^{(b)}
=
\left\{
e_o=(\{i,j\},o):
i,j\in\mathrm{Si},\ i\neq j,\ o\in\mathrm O,\
r_{io}\leq r_{\mathrm{SiO}}^c,\
r_{jo}\leq r_{\mathrm{SiO}}^c,\
n_{o:\mathrm{Si}}^{(b)}=2
\right\}.
\end{equation}
Distinct bridging O atoms therefore produce distinct parallel edges between the same Si pair; two such edges form a projected 2-cycle. For the bridge-resolved analysis, each edge $e_o$ is assigned one shortest zero-winding cycle returned by the registered deterministic graph search. This assignment is distinct from the complete validated primitive-cycle census used for the ring fractions. For Si$_3$N$_4$, rings are computed directly on the bipartite Si--N bond graph,
\begin{equation}
(i,j)\in E_{\mathrm{Si_3N_4}}^{(b)}
\quad\Longleftrightarrow\quad
i\in\mathrm{Si},\;
j\in\mathrm N,\;
r_{ij}\leq r_{\mathrm{SiN}}^c.
\end{equation}
Projected rings are not used for Si$_3$N$_4$ because projection through threefold N centres can generate artificial short cycles. If $R_s^{(b)}$ is the number of primitive rings of size $s$, the reported ring fraction is
\begin{equation}
p_s^{(b)}
=
\frac{R_s^{(b)}}{\sum_{s'}R_{s'}^{(b)}},
\qquad
\sum_{s'}R_{s'}^{(b)}>0.
\label{eq:si-ring-fraction}
\end{equation}
Ring-free cells are recorded separately and are not assigned a normalised ring fraction.

\subsection{Numerical stability preflight}
\label{sec:si-preflight}

Before physical protocol calibration, \vitriflow{} screens the declared numerical settings of the molecular-dynamics calculation. Validation of the force provider, integrator, timestep domain, backend, and relevant state space is an upstream responsibility; the preflight does not certify those elements generally. Let
\begin{equation}
\theta=(\Delta t,\tau_T,\tau_P,s_{\mathrm{skin}},\mathcal C_{\mathrm{core}})
\label{eq:si-preflight-settings}
\end{equation}
denote the time step, thermostat damping time, barostat damping time, neighbour skin, and optional tabulated $C^2$ short-range core-splice specification.\cite{Nose1984Thermostat,Hoover1985Canonical,Parrinello1980NPT,Martyna1994MTK} For each candidate $\theta$ and each probe temperature
\begin{equation}
T^\ast\in\mathcal T_{\mathrm{probe}},
\end{equation}
short trajectories are run and tail averages are computed over the final fraction of the thermodynamic record.

The dimensionless temperature-control error is
\begin{equation}
\epsilon_T(\theta,T^\ast)
=
\frac{
\left|
\langle T\rangle_{\mathrm{tail}}-T^\ast
\right|
}
{T_{\mathrm{scale}}},
\qquad
T_{\mathrm{scale}}
=
\max(T^\ast,T_{\mathrm{floor}}),
\end{equation}
with $T_{\mathrm{floor}}=1~\mathrm{K}$ used only to avoid singular scaling near zero temperature. The pressure-control error and volume sanity factor are
\begin{equation}
\epsilon_P(\theta,T^\ast)
=
\frac{
\left|
\langle P\rangle_{\mathrm{tail}}-P_{\mathrm{target}}
\right|
}
{P_{\mathrm{tol}}},
\end{equation}
and
\begin{equation}
R_V(\theta,T^\ast)
=
\max
\left(
\frac{\langle V\rangle_{\mathrm{tail}}}{V_{\mathrm{ref}}},
\frac{V_{\mathrm{ref}}}{\langle V\rangle_{\mathrm{tail}}}
\right).
\end{equation}
Here, $P_{\mathrm{target}}$ is the applied pressure target, $P_{\mathrm{tol}}$ is the pressure tolerance, and $V_{\mathrm{ref}}$ is the reference volume used to detect large box changes. The admissible numerical region is
\begin{equation}
\begin{aligned}
\Theta_{\mathrm{adm}} = \bigg\{ \theta : \;& \epsilon_T(\theta,T^\ast)\leq\epsilon_{T,\max}, \\
& \epsilon_P(\theta,T^\ast)\leq\epsilon_{P,\max}, \\
& R_V(\theta,T^\ast)\leq R_{V,\max}, \\
& \text{for all }T^\ast\in\mathcal T_{\mathrm{probe}}, \\
& \text{and no numerical failure occurs} \bigg\}.
\end{aligned}
\end{equation}
Here, $\epsilon_{T,\max}$ and $\epsilon_{P,\max}$ are the configured dimensionless control-error limits and $R_{V,\max}>1$ is the configured volume-ratio limit.
Numerical failures include force-evaluation failures, neighbour-list failures, unstable volume excursions, non-finite thermodynamic quantities, and backend-specific molecular-dynamics errors.\cite{Kim2023NeighborList}

Surviving candidates are ranked by the preflight score
\begin{equation}
S_{\mathrm{pf}}(\theta)
=
\max_{T^\ast\in\mathcal T_{\mathrm{probe}}}
\left[
\epsilon_T(\theta,T^\ast)
+
\epsilon_P(\theta,T^\ast)
+
\frac{R_V(\theta,T^\ast)-1}
{R_{V,\max}-1}
\right].
\end{equation}
The selected numerical setting $\theta^\star$ is the lowest-scoring admissible candidate within the tested domain. When scores are comparable, conservative barostat damping is preferred for NPT production. A short NVE segment is used only as a fail-safe for gross instability or execution failure; it is not a user-facing metric, an Autotune objective, or a general certification of timestep stability. This automated fail-safe is distinct from the separate production-timestep NVE and vibrational-density-of-states diagnostics shown in Figs.~\ref{fig:si-nve-timestep} and \ref{fig:si-vdos-timestep}. For spectroscopic wavenumber $\tilde\nu$, the associated vibrational period is $P=(c\tilde\nu)^{-1}$, where $c$ is the speed of light in units consistent with $\tilde\nu$ and $P$. The timestep diagnostic is therefore the dimensionless ratio $P/\Delta t=(c\tilde\nu\Delta t)^{-1}$. Longer high-temperature verification runs are then performed at $\theta^\star$ before physical protocol calibration.

\begingroup
\makeatletter
\setlength{\@fptop}{0pt}
\makeatother

\begin{figure}[p]
  \centering
  \includegraphics[width=\textwidth]{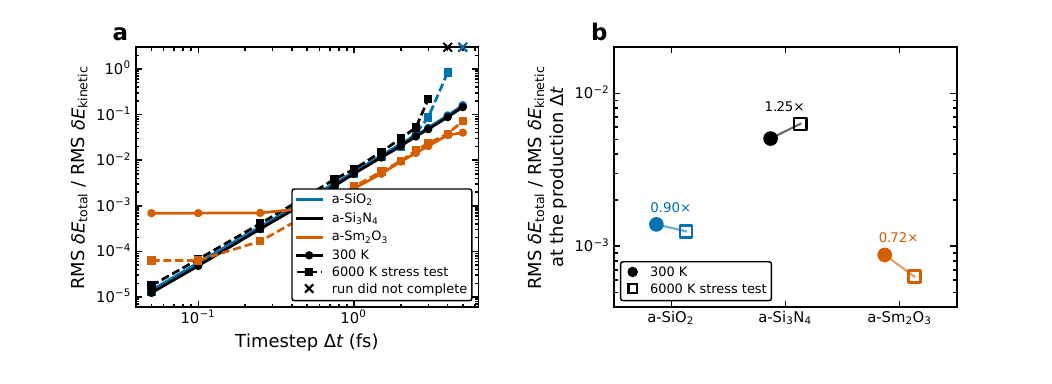}
  \caption{Production-timestep validation by microcanonical energy conservation. (a) Ratio of the root-mean-square total-energy fluctuation to the root-mean-square kinetic-energy fluctuation as a function of timestep for a-SiO$_2$, a-Si$_3$N$_4$, and a-Sm$_2$O$_3$ at 300~K and at the common 6000~K high-temperature stress-test condition, which exceeds every selected $T_{\mathrm{high}}$. Crosses mark runs that did not complete. (b) The same ratio at the production timestep for each material and temperature. The annotated factors are the ratios of the 6000~K values to the corresponding 300~K values.}
  \label{fig:si-nve-timestep}
\end{figure}

\begin{figure}[p]
  \centering
  \includegraphics[width=\textwidth]{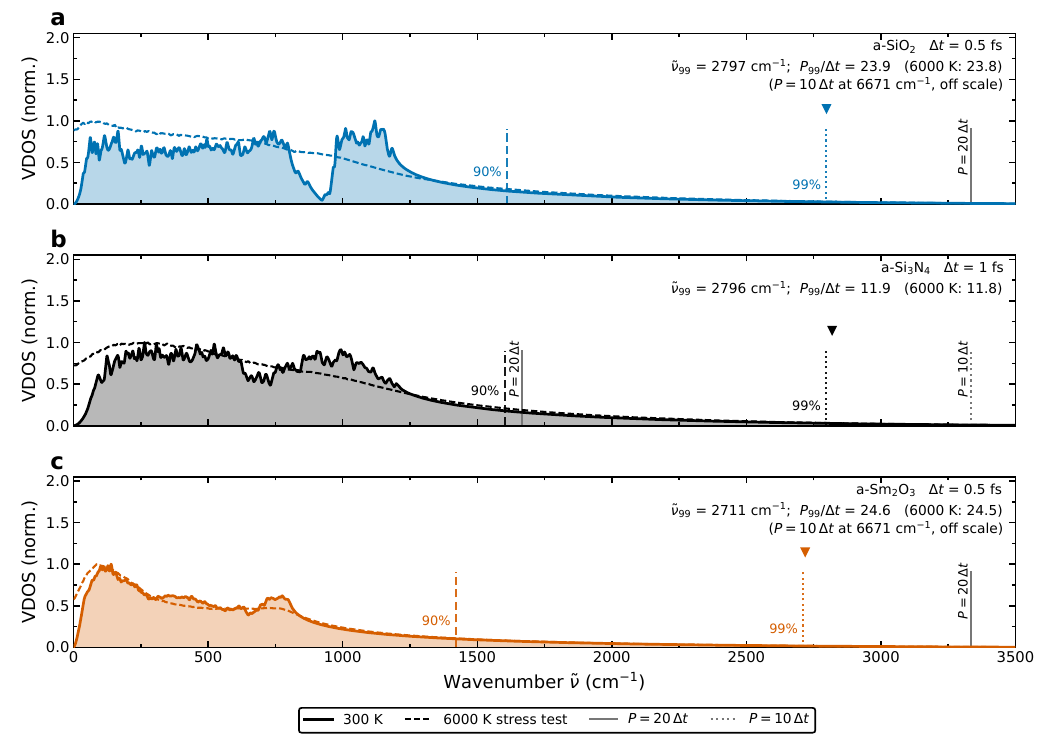}
  \caption{Vibrational-density-of-states timestep check for (a) a-SiO$_2$, (b) a-Si$_3$N$_4$, and (c) a-Sm$_2$O$_3$. Normalised spectra are shown for 300~K and the common 6000~K high-temperature stress-test condition. The 300~K 90th- and 99th-percentile wavenumbers are marked. The wavenumbers corresponding to vibrational periods of 20 and 10 integration steps are marked when they fall within the plotted range and are otherwise reported as off scale. Each panel reports the production timestep and $P_{99}/\Delta t$, where $P_{99}=(c\tilde\nu_{99})^{-1}$; the corresponding 6000~K value is given in parentheses.}
  \label{fig:si-vdos-timestep}
\end{figure}

\clearpage
\endgroup

\subsection[C2 Buckingham--ZBL core splice]{$C^2$ Buckingham--ZBL core splice}
\label{sec:si-core}

For a Buckingham component $U_{{\rm B},ij}(r)$, \vitriflow{} replaces the divergent short-range branch by a pair-specific, tabulated $C^2$ splice to the ZBL component $U_{{\rm Z},ij}(r)$; Coulomb and any additional Morse components are retained unchanged.\cite{Buckingham1938ExpSix,Ziegler1985ZBL} With $F_{{\rm B},ij}=-U_{{\rm B},ij}'$, $F_{{\rm Z},ij}=-U_{{\rm Z},ij}'$, and resolved radii $r_{{\rm in},ij}<r_{{\rm out},ij}$, define
\begin{equation}
x_{ij}(r)=\frac{r-r_{{\rm in},ij}}{r_{{\rm out},ij}-r_{{\rm in},ij}},
\qquad
w(x)=(1-x)^3(1+3x+6x^2).
\end{equation}
The spliced force is
\begin{equation}
F_{{\rm sp},ij}(r)=
\begin{cases}
F_{{\rm Z},ij}(r), & r\le r_{{\rm in},ij},\\
w(x_{ij})F_{{\rm Z},ij}(r)+[1-w(x_{ij})]F_{{\rm B},ij}(r),
& r_{{\rm in},ij}<r<r_{{\rm out},ij},\\
F_{{\rm B},ij}(r), & r\ge r_{{\rm out},ij}.
\end{cases}
\end{equation}
The corresponding energy is anchored to the Buckingham energy at the outer join:
\begin{equation}
\begin{aligned}
U_{{\rm sp},ij}(r)&=
\begin{cases}
U_{{\rm Z},ij}(r)+\Delta U_{ij}, & r\le r_{{\rm in},ij},\\
U_{{\rm B},ij}(r_{{\rm out},ij})
+\displaystyle\int_r^{r_{{\rm out},ij}}F_{{\rm sp},ij}(s)\,\mathrm ds,
& r_{{\rm in},ij}<r<r_{{\rm out},ij},\\
U_{{\rm B},ij}(r), & r\ge r_{{\rm out},ij},
\end{cases}\\[-2pt]
\Delta U_{ij}&=U_{{\rm B},ij}(r_{{\rm out},ij})
+\int_{r_{{\rm in},ij}}^{r_{{\rm out},ij}}F_{{\rm sp},ij}(s)\,\mathrm ds
-U_{{\rm Z},ij}(r_{{\rm in},ij}).
\end{aligned}
\label{eq:si-core-splice}
\end{equation}
The quintic satisfies $w(0)=1$, $w(1)=0$, and $w'=w''=0$ at both endpoints; hence $U_{{\rm sp},ij}$, $U_{{\rm sp},ij}'=-F_{{\rm sp},ij}$, and $U_{{\rm sp},ij}''$ are continuous at both joins. The transition integral is evaluated by fixed 32-point Gauss--Legendre quadrature. Preflight initialises the requested global outer radius as $r_{\rm out}^{\rm req}=\operatorname{clip}(f_{\rm out}\tilde r_{\rm nn},r_{\rm out,min},r_{\rm out,max})$ and $r_{\rm in}^{\rm req}=f_{\rm in}r_{\rm out}^{\rm req}$; failed trials multiply $r_{\rm out}^{\rm req}$ by $g>1$ up to $r_{\rm out,max}$. Here $\tilde r_{\rm nn}$ is the median nearest-neighbour distance, $f_{\rm out}$ is the requested outer-radius factor, $f_{\rm in}$ is the inner-to-outer radius ratio, $r_{\rm out,min}$ and $r_{\rm out,max}$ are the configured outer-radius bounds, and $g$ is the multiplicative growth factor after a failed trial. The realised join interval is then resolved separately for each species pair on a connected repulsive branch of the complete original pair interaction. The resulting table is verified before dynamics and fixed for production.

\subsection{Systematic protocol calibration}
\label{sec:si-protocol-calibration}

For an admissible numerical setting $\theta^\star$, the thermal protocol is written
\begin{equation}
p=(T_{\mathrm{high}},t_{\mathrm{melt}},\mathcal S_T,N),
\end{equation}
where $T_{\mathrm{high}}$ is the high-temperature state point, $t_{\mathrm{melt}}$ is the liquid-hold time, $\mathcal S_T$ is the cooling schedule, and $N$ is the number of atoms. For the linear schedules used here, $\mathcal S_T$ is parameterised by the positive cooling-rate magnitude $\dot T$. Calibration proceeds sequentially through melt-window identification, liquid-hold selection, quench-rate selection, and finite-size selection. At stage $k$, the workflow selects
\begin{equation}
p_k^\star
=
\operatorname*{arg\,min}_{p_k\in\mathcal P_k(p_1^\star,\ldots,p_{k-1}^\star)}
J_k(p_k)
\quad\text{subject to}\quad
\mathcal G_k(p_k)=1.
\end{equation}
Here $\mathcal P_k$ is the finite candidate set conditional on the selections retained at preceding stages, $J_k$ is the stage cost, and $\mathcal G_k$ is the corresponding Boolean pass predicate. The final protocol $p^\star=(T_{\mathrm{high}}^\star,t_{\mathrm{melt}}^\star,\mathcal S_T^\star,N^\star)$ is the result of this stepwise least-cost selection within the declared and tested domain. If a stage has no passing candidate, calibration is marked unresolved and no final $p^\star$ is assigned.

\subsubsection{Melt-window identification}
\label{sec:si-melt-window}

A temperature scan is performed on a ladder
\begin{equation}
T_1<T_2<\cdots<T_M.
\end{equation}
At each $T_i$, replicas provide estimates of diffusion $D_i$, first-peak height $H_i$, first-peak width $W_i$, volume $V_i$, and mean-squared displacement $\mathrm{MSD}_i$.\cite{Allen2017ComputerSimulation} Representative values are taken as medians across replicas unless otherwise specified.

The transformed indicators are
\begin{equation}
Y_D(T_i)=\log\!\left(\frac{D_i+\varepsilon_D}{D_{\mathrm{scale}}}\right),
\qquad
Y_H(T_i)=-\log\!\left(\frac{H_i+\varepsilon_H}{H_{\mathrm{scale}}}\right),
\qquad
Y_W(T_i)=\log\!\left(\frac{W_i+\varepsilon_W}{W_{\mathrm{scale}}}\right).
\label{eq:si-transition-indicators}
\end{equation}
The registered three-point smoothing operator is applied to each transformed indicator before its finite-difference temperature derivative per kelvin, denoted by $\Delta_TY(T_i)$; the derivative is centred at interior scan points and one-sided at the endpoints. The transition score is
\begin{equation}
\Sigma(T_i)
=
w_D Z[\Delta_TY_D]_i
+
w_H Z[\Delta_TY_H]_i
+
w_W Z[\Delta_TY_W]_i,
\end{equation}
where $w_D,w_H,w_W\geq0$ are configured dimensionless weights. For a derivative sequence $u$, the registered standardisation is
\begin{equation}
Z[u]_i
=
\frac{u_i-\bar u}{s_u}.
\end{equation}
Here $\bar u$ and $s_u$ are its arithmetic mean and sample standard deviation over the scanned temperatures; $Z[u]_i=0$ when $s_u=0$. The positive floors $\varepsilon_D$, $\varepsilon_H$, and $\varepsilon_W$ have the units of $D_i$, $H_i$, and $W_i$, respectively, and the positive fixed scales $D_{\mathrm{scale}}$, $H_{\mathrm{scale}}$, and $W_{\mathrm{scale}}$ have the matching units. These scales add constants only and do not change the finite-difference score.
The score $\Sigma(T_i)$ identifies the transition region. Melt confirmation additionally uses displacement and structural gates. Define
\begin{equation}
\ell_i
=
\left(\frac{V_i}{N}\right)^{1/3},
\qquad
\chi_i
=
\frac{\sqrt{\mathrm{MSD}_i}}{\ell_i}.
\end{equation}
The displacement gate is
\begin{equation}
I_\chi(i)
=
\mathbf 1[\chi_i\geq\chi_\star].
\end{equation}
When structural gates are active,
\begin{equation}
I_H(i)=\mathbf 1[H_i\leq \eta_HH_0],
\qquad
I_W(i)=\mathbf 1[W_i\geq \eta_WW_0],
\end{equation}
where $H_0$ and $W_0$ are the corresponding low-temperature or initial-state first-peak descriptors; $\chi_\star$, $\eta_H$, and $\eta_W$ are configured dimensionless gate thresholds. Inactive structural gates are set to one. The confirmed-melt indicator is
\begin{equation}
M_i=I_\chi(i)I_H(i)I_W(i).
\end{equation}
If $i_m$ is the first index beginning a block of $m\geq1$ consecutive values $M_i=1$, the nominal operational melt-reference temperature is the preceding sampled temperature; otherwise it is the temperature of the maximum transition score:
\begin{equation}
T_m=
\begin{cases}
T_{\max(i_m-1,1)}, & i_m\ \text{exists},\\
T_{\operatorname*{arg\,max}_i\Sigma(T_i)}, & \text{otherwise}.
\end{cases}
\end{equation}
For liquid-onset selection, let $D_{\mathrm{top}}$ be the median of the positive diffusion estimates at the highest $k_{\mathrm{liq}}$ sampled temperatures. With configured fraction $f_{\mathrm{liq}}$, mobility threshold $\chi_{\mathrm{liq}}$, and sustained-block length $m_{\mathrm{liq}}$,
\begin{equation}
D_{\mathrm{liq}}^\ast=f_{\mathrm{liq}}D_{\mathrm{top}},
\qquad
L_i=\mathbf 1[D_i\geq D_{\mathrm{liq}}^\ast]\,
\mathbf 1[\chi_i\geq\chi_{\mathrm{liq}}].
\end{equation}
The operational liquid onset is
\begin{equation}
T_{\mathrm{liq}}
=
\min\left\{
T_i:
L_i=L_{i+1}=\cdots=L_{i+m_{\mathrm{liq}}-1}=1
\right\}.
\end{equation}
If this set is empty, no $T_{\mathrm{liq}}$ is assigned. The marker $T_m$ is an operational transition reference, not a thermodynamic melting-point estimate; it is distinct from both $T_{\mathrm{liq}}$ and the liquid-hold time $t_{\mathrm{melt}}$.
The high-temperature production set point is then
\begin{equation}
T_{\mathrm{high}}
=
T_{\mathrm{liq}}
+
\Delta T_{\mathrm{margin}},
\label{eq:si-high-temperature}
\end{equation}
unless a stricter user-supplied material bound is imposed.

\subsubsection{Liquid-hold time}
\label{sec:si-liquid-hold}

At $T_{\mathrm{high}}$, independent replicas determine the minimum liquid-hold time. With $\ell=(V/N)^{1/3}$ evaluated in the high-temperature state, $t_{\min}$ the configured minimum observation time, and $\alpha_{\mathrm{MSD}}$ the dimensionless displacement fraction, the disordering time for replica $k$ is
\begin{equation}
t_{\mathrm{dis},k}
=
\inf
\left\{
t\geq t_{\min}:
\sqrt{\mathrm{MSD}_k(t)}\geq \alpha_{\mathrm{MSD}}\ell
\right\},
\label{eq:si-disordering-time}
\end{equation}
subject to stationarity of density and potential energy. For a scalar thermodynamic quantity $x$, stationarity between early and late windows is measured by
\begin{equation}
\delta_x
=
\frac{
|\bar x_{\mathrm{late}}-\bar x_{\mathrm{early}}|
}
{\max(|\bar x_{\mathrm{late}}|,\varepsilon_x)}.
\end{equation}
The stationarity condition is
\begin{equation}
\delta_\rho\leq \eta_\rho,
\qquad
\delta_U\leq \eta_U,
\end{equation}
where $U$ is the potential energy, $\varepsilon_x>0$ has the same units as $x$, and $\eta_\rho$ and $\eta_U$ are configured dimensionless stationarity limits. The assigned liquid-hold time is the worst-case disordering time across replicas:
\begin{equation}
t_{\mathrm{melt}}^\star
=
\max_k t_{\mathrm{dis},k}.
\label{eq:si-hold-time}
\end{equation}
If no finite $t_{\mathrm{dis},k}$ is found within the configured maximum hold time, the calibration is marked unresolved and the workflow either extends the hold or returns to the melt-window selection step, depending on the configured policy.

\subsubsection{Quench-rate selection}
\label{sec:si-quench-rate}

Candidate cooling rates are denoted by $\dot T_i>0$, with the slowest tested rate $\dot T_{\mathrm{ref}}$ used as the reference. For candidate rate $\dot T_i$, the number of quench integration steps is
\begin{equation}
n_q(\dot T_i)
=
\left\lceil
\frac{T_{\mathrm{high}}-T_{\mathrm{final}}}
{\dot T_i\,\Delta t}
\right\rceil.
\label{eq:si-linear-quench-steps}
\end{equation}
Here $T_{\mathrm{final}}$ is the configured final quench temperature. In this expression, $\dot T_i$ and $\Delta t$ use a common time unit; for the tabulated K~ps$^{-1}$ rates, $\Delta t$ is converted from fs to ps. For scalar descriptor $j$, let $\hat\mu_{ij}$ and $\mathrm{SE}_{ij}$ be the candidate replica-population mean and standard error, and let $\hat\mu_{\mathrm{ref},j}$ and $\mathrm{SE}_{\mathrm{ref},j}$ denote the independently sampled slow-rate reference. The equivalence error is
\begin{equation}
E_{ij}
=
|\hat\mu_{ij}-\hat\mu_{\mathrm{ref},j}|
+
c_{\alpha_{\mathrm{CI}}}
\sqrt{
\mathrm{SE}_{ij}^{2}
+
\mathrm{SE}_{\mathrm{ref},j}^{2}
}.
\end{equation}
Here $c_{\alpha_{\mathrm{CI}}}$ is the declared two-sided confidence multiplier at significance $0<\alpha_{\mathrm{CI}}<1$. The absolute tolerance $\tau_{j,\mathrm{abs}}>0$ has the units of descriptor $j$, whereas $\tau_{j,\mathrm{rel}}\geq0$ is dimensionless.
The descriptor-specific tolerance is
\begin{equation}
\tau_j(\mu)
=
\max
\left(
\tau_{j,\mathrm{abs}},
\tau_{j,\mathrm{rel}}|\mu|
\right).
\end{equation}
The candidate rate is descriptor-equivalent to the reference if
\begin{equation}
E_{ij}
\leq
\tau_j(\hat\mu_{\mathrm{ref},j})
\qquad
\forall j\in\mathcal D_{\mathrm{quench}}.
\end{equation}
The selected cooling rate is the fastest equivalent candidate:
\begin{equation}
\dot T^\star
=
\max
\left(
\{\dot T_{\mathrm{ref}}\}\cup
\left\{\dot T_i>\dot T_{\mathrm{ref}}:
E_{ij}
\leq
\tau_j(\hat\mu_{\mathrm{ref},j})
\;\forall j\in\mathcal D_{\mathrm{quench}}
\right\}
\right).
\end{equation}
If no faster candidate satisfies the equivalence condition, the slowest tested rate is retained.

\subsubsection{Finite-size selection}
\label{sec:si-size-selection}

When size scans are performed, near-cubic cells of increasing size $N_s$ are compared against the largest tested cell $N_{\mathrm{ref}}$. The size-specific and reference replica populations in this construction are independently generated; for a paired construction, the standard error of the paired differences replaces the quadrature term below. For descriptor $j$,
\begin{equation}
E_{sj}
=
|\hat\mu_{sj}-\hat\mu_{\mathrm{ref},j}|
+
c_{\alpha_{\mathrm{CI}}}
\sqrt{
\mathrm{SE}_{sj}^{2}
+
\mathrm{SE}_{\mathrm{ref},j}^{2}
}.
\end{equation}
The selected size is the smallest size satisfying
\begin{equation}
N^\star
=
\min
\left(
\{N_{\mathrm{ref}}\}\cup
\left\{N_s<N_{\mathrm{ref}}:
E_{sj}
\leq
\tau_j(\hat\mu_{\mathrm{ref},j})
\;\forall j\in\mathcal D_{\mathrm{size}}
\right\}
\right).
\label{eq:si-size-selection}
\end{equation}
If no smaller cell is equivalent to the largest tested cell, the largest cell is retained.

\subsubsection{Calibration and structural targets}
\label{sec:si-calibration-results}

\begin{landscape}
\begin{figure}[p]
  \centering
  \includegraphics[width=\linewidth,height=0.80\textheight,keepaspectratio]{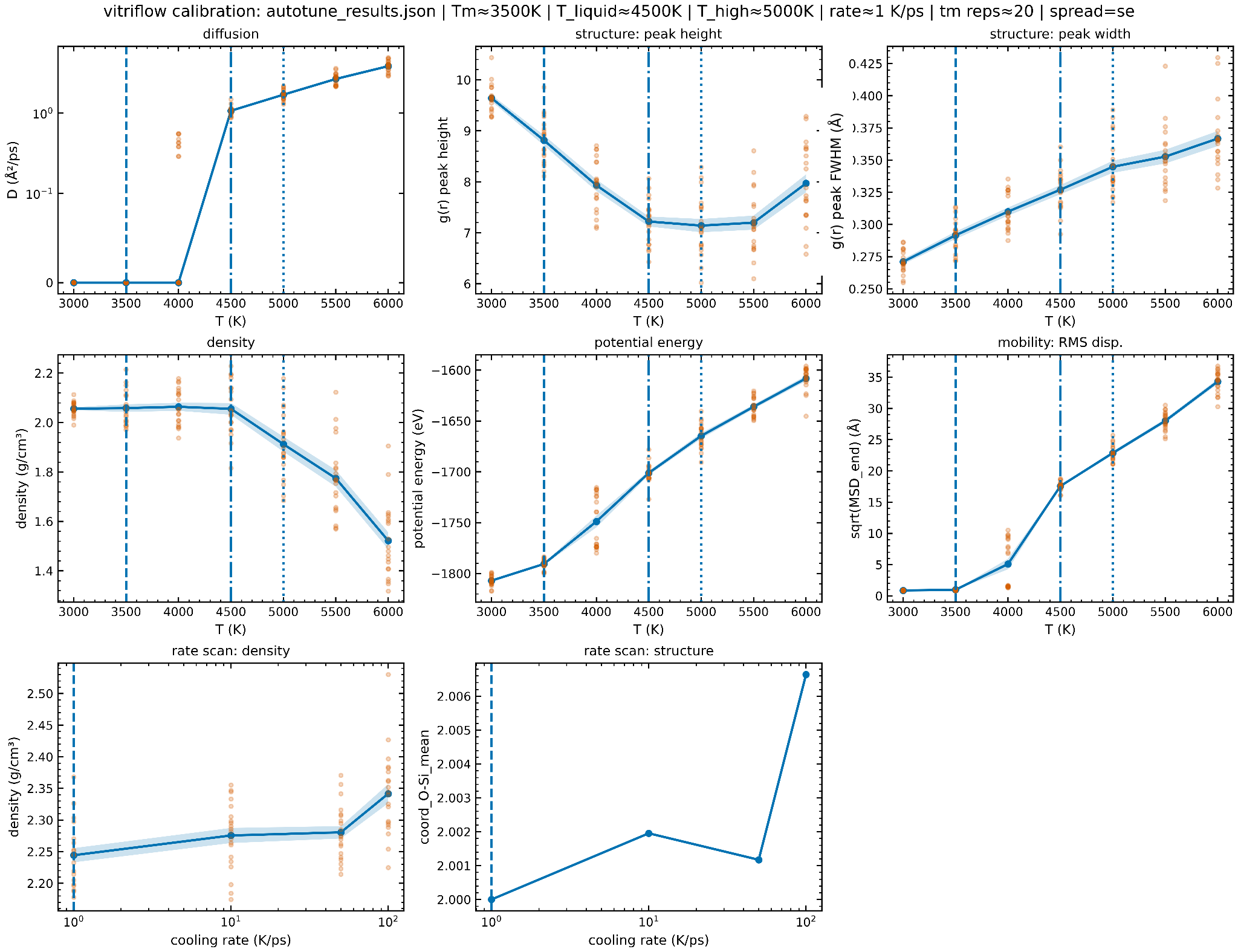}
  \caption{a-SiO$_2$ \vitriflow{} calibration scan. The temperature ladder identifies the melt and liquid regimes from diffusion, radial-distribution first-peak height and FWHM $W$ (\AA), density, potential energy, and root-mean-square displacement. The cooling-rate panels compare density and mean O--Si coordination. Each sampled temperature has $n_{\rm rep}=20$ replicas and the shaded bands show the standard error.}
  \label{fig:si-cal-sio2}
\end{figure}
\end{landscape}

\begin{landscape}
\begin{figure}[p]
  \centering
  \includegraphics[width=\linewidth,height=0.80\textheight,keepaspectratio]{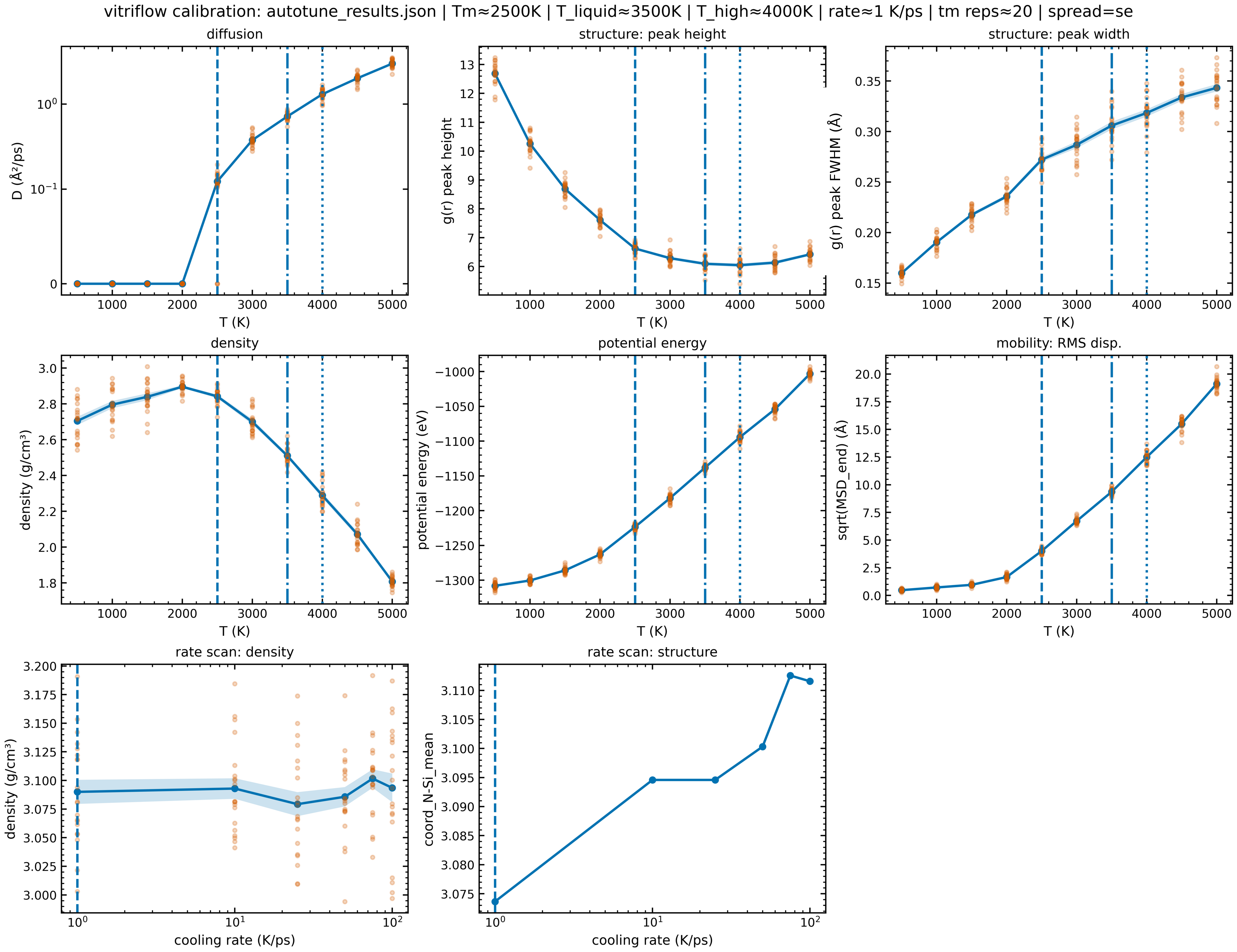}
  \caption{a-Si$_3$N$_4$ \vitriflow{} calibration scan. The temperature ladder identifies the melt and liquid regimes from diffusion, radial-distribution first-peak height and FWHM $W$ (\AA), density, potential energy, and root-mean-square displacement. The cooling-rate panels compare density and mean N--Si coordination. Each sampled temperature has $n_{\rm rep}=20$ replicas and the shaded bands show the standard error.}
  \label{fig:si-cal-si3n4}
\end{figure}
\end{landscape}

\begin{landscape}
\begin{figure}[p]
  \centering
  \includegraphics[width=\linewidth,height=0.80\textheight,keepaspectratio]{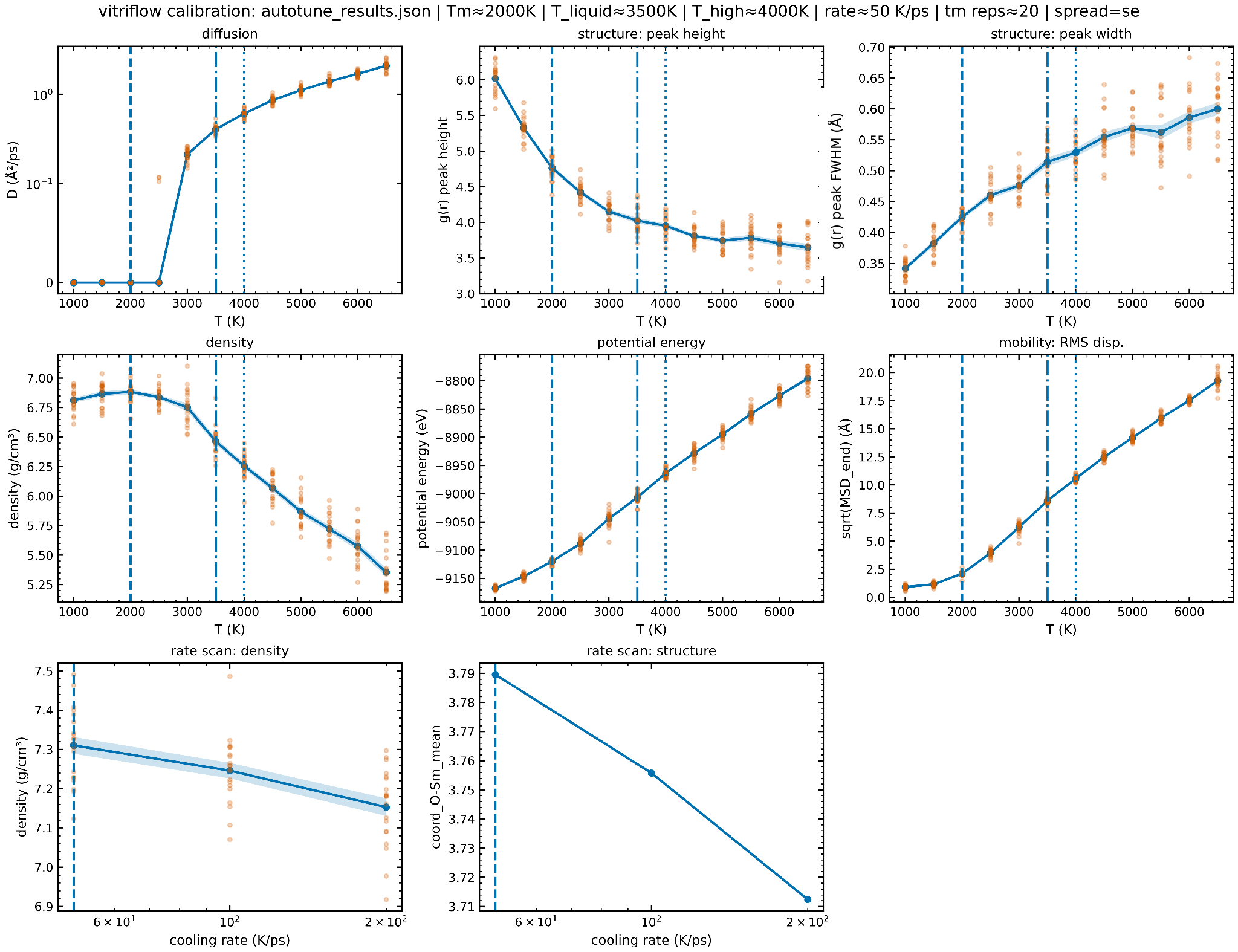}
  \caption{a-Sm$_2$O$_3$ \vitriflow{} calibration scan. The temperature ladder identifies the melt and liquid regimes from diffusion, radial-distribution first-peak height and FWHM $W$ (\AA), density, potential energy, and root-mean-square displacement. The cooling-rate panels compare density and mean O--Sm coordination. Each sampled temperature has $n_{\rm rep}=20$ replicas and the shaded bands show the standard error.}
  \label{fig:si-cal-sm2o3}
\end{figure}
\end{landscape}

The material-resolved calibration scans are shown in Figs.~\ref{fig:si-cal-sio2}--\ref{fig:si-cal-sm2o3}, and the selected numerical and production settings are collected in Table~\ref{tab:autotuned-run-settings}. The upper six panels of each figure show the temperature-ladder diffusion, first-peak, density, potential-energy, and displacement diagnostics used to identify the melt and liquid regimes. The lower panels show the cooling-rate comparisons. Orange points are the $n_{\rm rep}=20$ replica values at each sampled temperature, blue curves are the corresponding summaries, and shaded bands show the standard error across replicas. The operational marker $T_m$ is returned by the transition score and melt-confirmation gates; it is not a thermodynamic melting-point estimate and is distinct from $T_{\rm liq}$ and the liquid-hold time $t_{\rm melt}$. Vertical markers identify $T_m$, $T_{\rm liq}$, and $T_{\rm high}$. In the plot headers, the plain-text fields \code{Tm}, \code{T\_liquid}, \code{rate}, and \code{tm reps} denote $T_m$, $T_{\rm liq}$, $\dot T$, and $n_{\rm rep}$, respectively. First-peak FWHM values $W$ are in \AA; the retained production parameters are those in Table~\ref{tab:autotuned-run-settings}.

\begin{table}[htbp]
\centering
\caption{Selected numerical and production settings for the melt--quench source populations. Classical runs used periodic NPT dynamics at a target pressure of 0~bar. Here $n_{\rm rep}$ is the dimensionless calibration-replica count, $t_{\rm snap}$ is the source-snapshot spacing after melt equilibration, and the core values are the requested global $C^2$ splice radii $r_{\rm in}^{\rm req};r_{\rm out}^{\rm req}$.}
\label{tab:autotuned-run-settings}
\footnotesize
\setlength{\tabcolsep}{3.0pt}
\renewcommand{\arraystretch}{1.03}
\begin{tabular}{@{}lccc@{}}
\toprule
Parameter & a-SiO$_2$ & a-Si$_3$N$_4$ & a-Sm$_2$O$_3$ \\
\midrule
model & SHIK-1 & MG2 & Buck. \\
atoms & 192 & 280 & 350 \\
$\Delta t$ (fs) & 0.5 & 1.0 & 0.5 \\
$\tau_T\,;\,\tau_P$ (ps) & 0.05; 0.5 & 0.50; 5.0 & 0.10; 1.0 \\
$s_{\mathrm{skin}}$ (\AA) & 0.25 & 0.50 & 1.00 \\
$T_{\rm liq}\,;\,T_{\rm high}$ (K) & 4500; 5000 & 3500; 4000 & 3500; 4000 \\
$n_{\rm rep}$ (count); $t_{\rm melt}^{\star}$ (ps) & 20; 50.0 & 20; 107.8 & 20; 44.0 \\
$t_{\rm snap}$ (ps) & 10 & -- & -- \\
$\dot T^{\star}$ (K~ps$^{-1}$) & 1.0 & 1.0 & 50.0 \\
$T_{\rm final}$ (K); equilibration (ps) & 300; 100 & 300; 100 & 300; 100 \\
$C^2$ splice $r_{\rm in}^{\rm req};r_{\rm out}^{\rm req}$ & none & none & 1.100; 1.375~\AA \\
\bottomrule
\end{tabular}
\end{table}

\clearpage

\clearpage

\subsection{User-defined artefact screening}
\label{sec:si-screening}

Artefact screening is optional and user-defined. The code does not distinguish between simple and compound descriptors: a screen evaluates a configured descriptor vector and records whether the result falls inside the allowed region. For screen $s$,
\begin{equation}
\mathbf d_{bs}=\Phi_s(x_b;\mathcal D_{\mathrm{screen},s}),
\label{eq:si-screen-vector}
\end{equation}
\begin{equation}
\mathbf z_{bs}=\Psi_s(\mathbf d_{bs}),
\end{equation}
and
\begin{equation}
A_{bs}=\mathbf 1
\left[\mathbf z_{bs}\in\mathcal I_s\right],
\end{equation}
where $\Psi_s$ is the configured screen rule and $\mathcal I_s$ is the allowed region. The label $A_{bs}$ is then interpreted through the screen's \code{exclude} action, as defined in Sec.~\ref{sec:si-scope}.

A scalar screen is recovered when $\mathbf z_{bs}$ has one component,
\begin{equation}
A_{bs}^{\mathrm{scalar}}
=
\mathbf 1
\left[z_{bs}\in[z_s^{\min},z_s^{\max}]\right].
\end{equation}
One-sided thresholds are obtained by setting one bound to $\pm\infty$. A reference-shifted scalar screen is the equivalent form
\begin{equation}
A_{bs}^{\mathrm{shift}}
=
\mathbf 1
\left[-\eta_s^{-}\leq z_{bs}-z_s^{\mathrm{ref}}\leq\eta_s^{+}\right],
\end{equation}
which is useful for density windows, peak-position tolerances, or other descriptors evaluated relative to a reference value.

A compound screen is a joint rule on several scalar components,
\begin{equation}
A_{bs}^{\mathrm{comp}}
=
\prod_{u=1}^{Q_s}
\mathbf 1
\left[
z_{bsu}\in I_{su}
\right],
\end{equation}
where $Q_s$ is the number of scalar components and $I_{su}$ is the allowed interval for component $u$; equivalently $\mathbf z_{bs}\in\prod_{u=1}^{Q_s} I_{su}$. The SiO$_2$ coordination screen and Sm$_2$O$_3$ crystal-like-order screen are compound examples.

For coordination-based screening, the defect fraction for central species $\alpha$ and neighbour species $\beta$ is
\begin{equation}
f_{\alpha:\beta}^{(b)}
=
\frac{1}{N_\alpha^{(b)}}
\sum_{i\in\alpha}
\mathbf 1
\left[
n_{i:\beta}^{(b)}
\notin
\mathcal A_{\alpha:\beta}
\right],
\end{equation}
where $\mathcal A_{\alpha:\beta}$ is the allowed coordination set. A
coordination screen has the form
\begin{equation}
A_b^{\mathrm{coord}}
=
\prod_{\alpha:\beta}
\mathbf 1
\left[
f_{\alpha:\beta}^{(b)}
\leq
\eta_{\alpha:\beta}
\right].
\end{equation}
Exact coordination screening is recovered by setting the relevant
$\eta_{\alpha:\beta}=0$.

For crystal-like screening, a local-order classifier assigns
\begin{equation}
c_i^{(b)}\in\{0,1\},
\end{equation}
where $c_i^{(b)}=1$ denotes a crystal-like local environment. Here $N_b$ is the number of sites to which the classifier is applied, and $\mathcal C_{\max}^{(b)}$ is the largest connected component of $c_i^{(b)}=1$ sites under the registered neighbour graph. The crystalline
fraction and largest crystal-like cluster fraction are
\begin{equation}
f_{\mathrm{cryst}}^{(b)}
=
\frac{1}{N_b}
\sum_{i=1}^{N_b}c_i^{(b)},
\qquad
f_{\mathrm{cluster}}^{(b)}
=
\frac{|\mathcal C_{\max}^{(b)}|}{N_b}.
\end{equation}
The corresponding compound screen is
\begin{equation}
A_b^{\mathrm{cryst}}
=
\mathbf 1
\left[
f_{\mathrm{cryst}}^{(b)}
\leq
\eta_{\mathrm{cryst}}
\right]
\mathbf 1
\left[
f_{\mathrm{cluster}}^{(b)}
\leq
\eta_{\mathrm{cluster}}
\right].
\label{eq:si-generic-crystal-screen}
\end{equation}

\subsection{Statistical summaries and convergence criteria}
\label{sec:si-statistics}

Statistical convergence is assessed on the analysis population $\mathcal E_{\mathrm{ana}}$ unless explicitly stated otherwise. Let $b_1,\ldots,b_{N_{\mathrm{ana}}}$ be the fixed archived ordering of its independent sampling units. For scalar descriptor $y$, the analysis-population prefix mean for $2\leq n\leq N_{\mathrm{ana}}$ is
\begin{equation}
\bar y_n
=
\frac{1}{n}
\sum_{k=1}^{n}y_{b_k},
\label{eq:si-prefix-mean}
\end{equation}
and the sample variance is
\begin{equation}
s_n^2(y)
=
\frac{1}{n-1}
\sum_{k=1}^{n}
(y_{b_k}-\bar y_n)^2.
\end{equation}
The standard error is
\begin{equation}
\mathrm{SE}_n(y)
=
\frac{s_n(y)}{\sqrt n}.
\end{equation}
The confidence half-width is
\begin{equation}
h_n(y)
=
c_{\alpha_{\mathrm{CI}},n}
\frac{s_n(y)}{\sqrt n},
\end{equation}
where $0<\alpha_{\mathrm{CI}}<1$ is the two-sided significance level and $c_{\alpha_{\mathrm{CI}},n}$ is the Student-$t$ multiplier $t_{1-\alpha_{\mathrm{CI}}/2,n-1}$ or its large-sample normal approximation.\cite{Student1908ProbableError}

For prospective production checks, let $\alpha_{\mathrm{family}}$ be the configured familywise error probability and $m_{\mathrm{test}}$ the number of simultaneous scalar and fixed-grid component checks. The per-test level is
\begin{equation}
\alpha_{\mathrm{test}}
=
\begin{cases}
\alpha_{\mathrm{family}}/m_{\mathrm{test}}, & \text{Bonferroni mode},\\
\alpha_{\mathrm{family}}, & \text{otherwise}.
\end{cases}
\end{equation}
Thus $\alpha_{\mathrm{eff}}=\alpha_{\mathrm{test}}$ for a production check and $\alpha_{\mathrm{eff}}=\alpha_{\mathrm{CI}}$ for the retrospective pointwise intervals reported here.

The descriptor-specific convergence tolerance is
\begin{equation}
\tau_y(\bar y_n)
=
\max
\left(
\tau_{y,\mathrm{abs}},
\tau_{y,\mathrm{rel}}|\bar y_n|
\right).
\end{equation}
The confidence-width ratio is
\begin{equation}
R_y(n)
=
\frac{h_n(y)}
{\tau_y(\bar y_n)}.
\end{equation}
Descriptor $y$ is converged at prefix size $n$ when
\begin{equation}
R_y(n)\leq q_{\mathrm{conv}}.
\end{equation}
For the convergence descriptor set $\mathcal D_{\mathrm{conv}}$, the population-level convergence ratio is
\begin{equation}
Q(n)
=
\max_{y\in\mathcal D_{\mathrm{conv}}}
R_y(n).
\end{equation}
The analysis population satisfies the defined convergence criterion when
\begin{equation}
Q(n)\leq q_{\mathrm{conv}}.
\end{equation}
The declared target used in this work is $q_{\mathrm{conv}}=0.2$; smaller values are stricter. This notation is the same as in the main text. This statement is limited to the declared descriptor set and tolerances; it is not a global label for every observable or every materials question. The confidence-width/tolerance curves shown in the main figures are plots of $R_y(n)$ or $Q(n)$.

For bounded descriptors scaled to $[0,1]$, empirical Bernstein and Hoeffding bounds are also evaluated\cite{Hoeffding1963Bounded,MaurerPontil2009EmpiricalBernstein}:
\begin{equation}
h_n^{\mathrm{EB}}
=
\sqrt{
\frac{2\widetilde v_n\ln(3/\alpha_{\mathrm{eff}})}{n}
}
+
\frac{3\ln(3/\alpha_{\mathrm{eff}})}{n},
\end{equation}
\begin{equation}
h_n^{\mathrm{H}}
=
\sqrt{
\frac{\ln(2/\alpha_{\mathrm{eff}})}{2n}
},
\end{equation}
where $\widetilde v_n=\min[s_n^2(y),1/4]$ after scaling the descriptor to $[0,1]$. For descriptors originally defined on $[a,b]$, the bounds are evaluated after affine scaling to $[0,1]$ and then rescaled to the original units.

Distributional stability is monitored for fixed-grid descriptors such as radial distribution functions, angular distributions, and ring histograms. For empirical cumulative distribution functions $F_A$ and $F_B$, the Wasserstein-1 distance is\cite{Villani2009OptimalTransport}
\begin{equation}
W_1(F_A,F_B)
=
\int_{-\infty}^{\infty}
|F_A(t)-F_B(t)|\,dt,
\end{equation}
and the Kolmogorov--Smirnov distance is\cite{Massey1951KolmogorovSmirnov}
\begin{equation}
D_{\mathrm{KS}}(F_A,F_B)
=
\sup_t |F_A(t)-F_B(t)|.
\label{eq:si-ks-distance}
\end{equation}
For fixed-grid histograms, the same definitions are evaluated on the common grid. Distributional tests are used to identify drift between analysis-cell prefixes, between analysis and audit-labelled populations, or between interaction-model datasets. When prospective production uses both confidence-width and stability modes, both must pass. The stability check compares the registered population blocks using the configured Wasserstein or Kolmogorov--Smirnov distance; when bootstrapping is enabled, its configured upper quantile must not exceed the corresponding tolerance.

\paragraph{Quantity-specific effective uncertainty}\label{par:si-effective-uncertainty}
For every reported structural quantity, the finite population has an achieved effective uncertainty defined for the stated population, sampling unit, and confidence construction. As independent sampling units are added this uncertainty generally decreases, but no uniquely correct population size follows. Conditional threshold statements are made only against the declared resolution rule already defined in this Supporting Information; the achieved uncertainty for the available population remains the primary reported quantity. The descriptor-resolved curves are shown in Figs.~\ref{fig:si-sio2-uncertainty-bonds}--\ref{fig:si-sm2o3-uncertainty-scalars}; their family- and descriptor-level crossing audits are reported in Tables~\ref{tab:si-family-uncertainty-thresholds} and \ref{tab:si-descriptor-uncertainty-thresholds}, respectively.

\begingroup
\makeatletter
\setlength{\@fptop}{0pt}
\makeatother

\begin{figure}[!htbp]
\centering
\includegraphics[page=1,width=\textwidth]{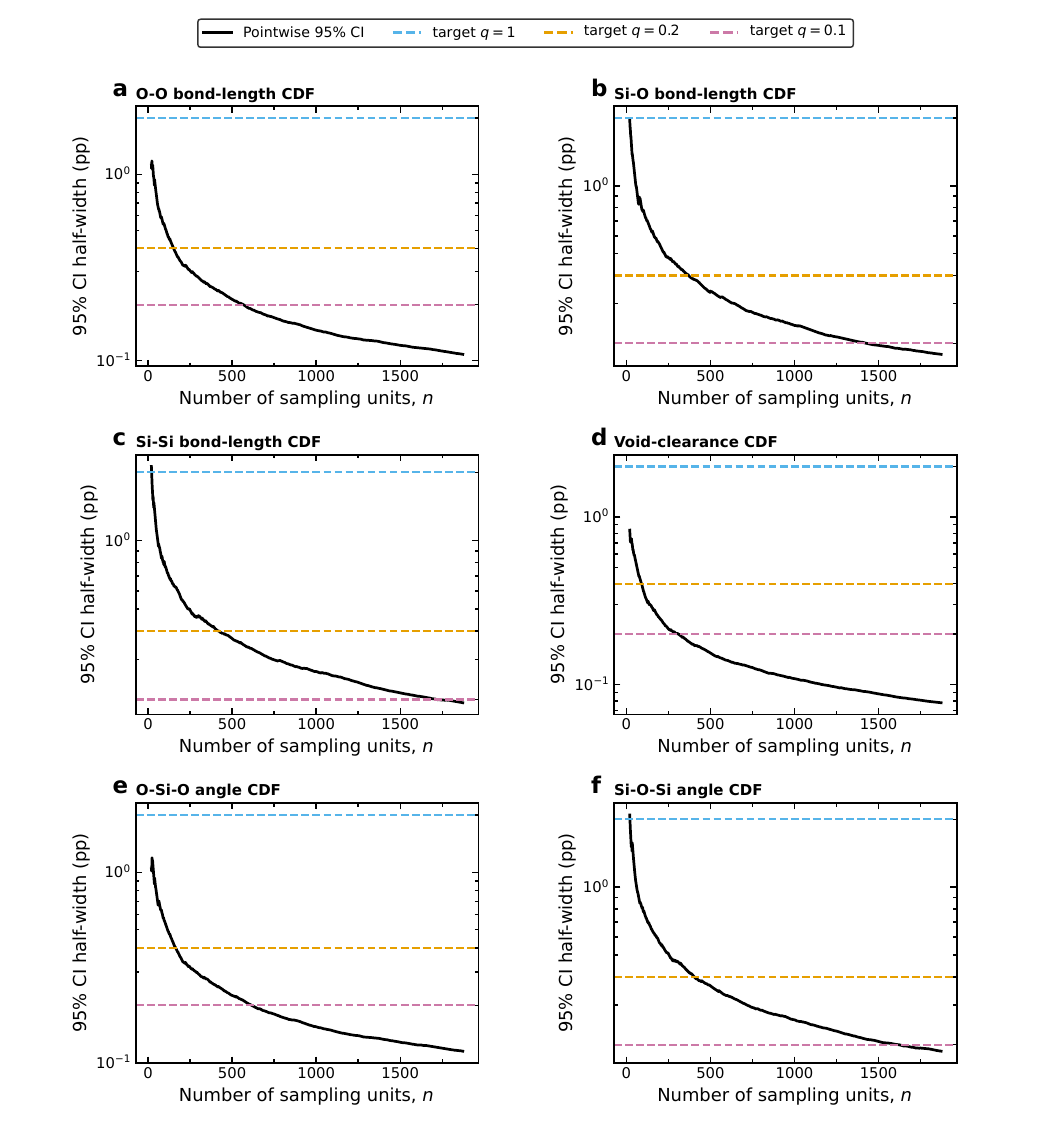}
\caption{Retrospective pointwise 95\% Student-$t$ confidence half-widths for a-SiO$_2$ pair-distance, void-clearance and angular CDFs over prefixes of the 1872 exact-coordination analysis cells. The horizontal guides are declared fractions of the registered descriptor scale and are conditional resolution references, not material-quality thresholds. The archived horizontal axis uses the term ``cells''; the intervals are independent-cell confidence intervals. The exact-coordination CDFs are not shown: the plotted population satisfies the coordination condition by construction, so their half-width is zero at every prefix.}
\label{fig:si-sio2-uncertainty-bonds}
\end{figure}

\begin{figure}[p]
\ContinuedFloat
\centering
\includegraphics[page=2,width=\textwidth]{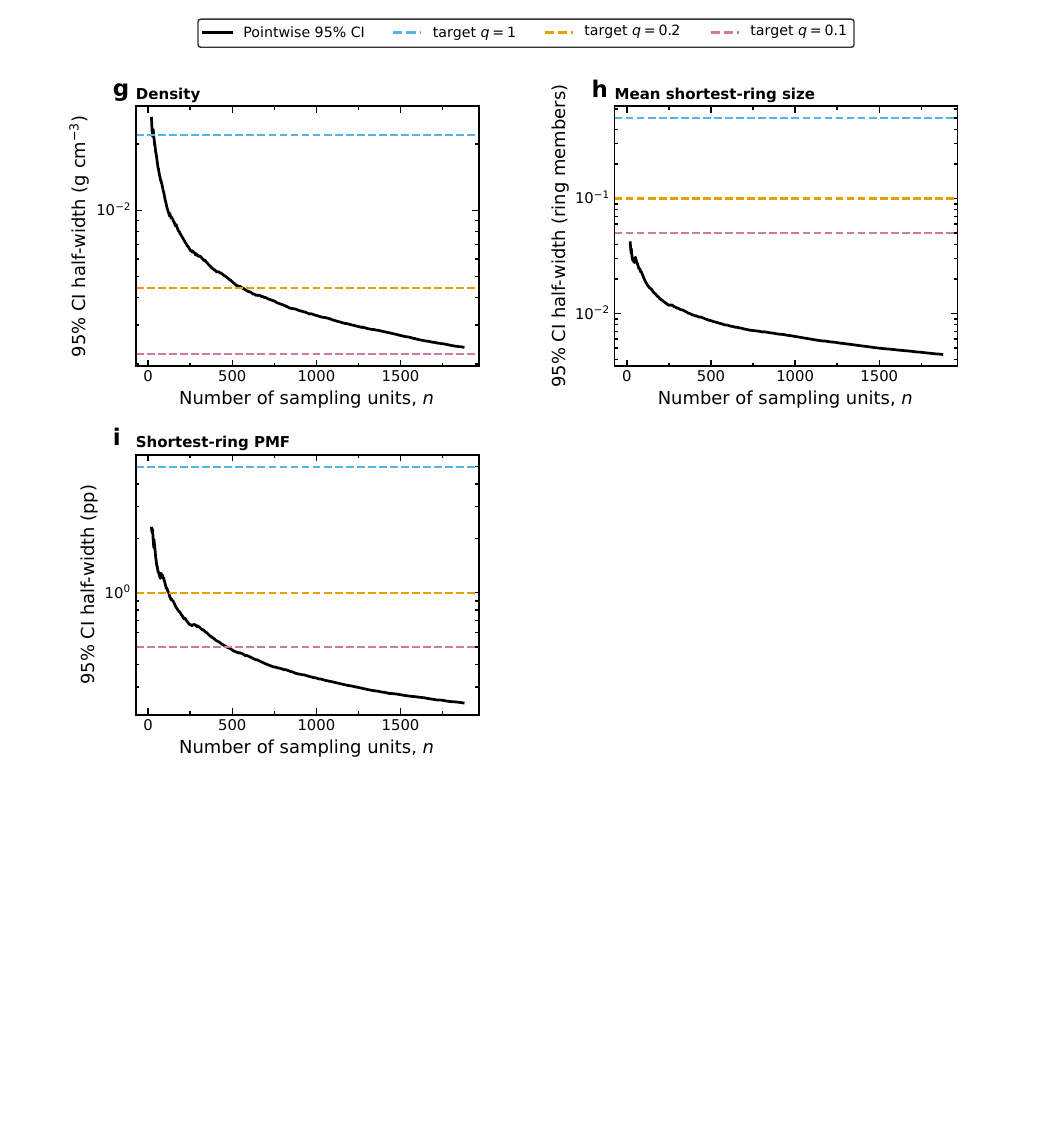}
\caption{Continued: retrospective pointwise 95\% confidence half-widths for the a-SiO$_2$ ring-size PMF, mean ring size, and density. The population, uncertainty convention, declared-resolution guides, and independent-cell confidence-interval construction are those in Fig.~\ref{fig:si-sio2-uncertainty-bonds}.}
\label{fig:si-sio2-uncertainty-rings-density}
\end{figure}

\begin{figure}[!htbp]
\centering
\includegraphics[page=1,width=\textwidth]{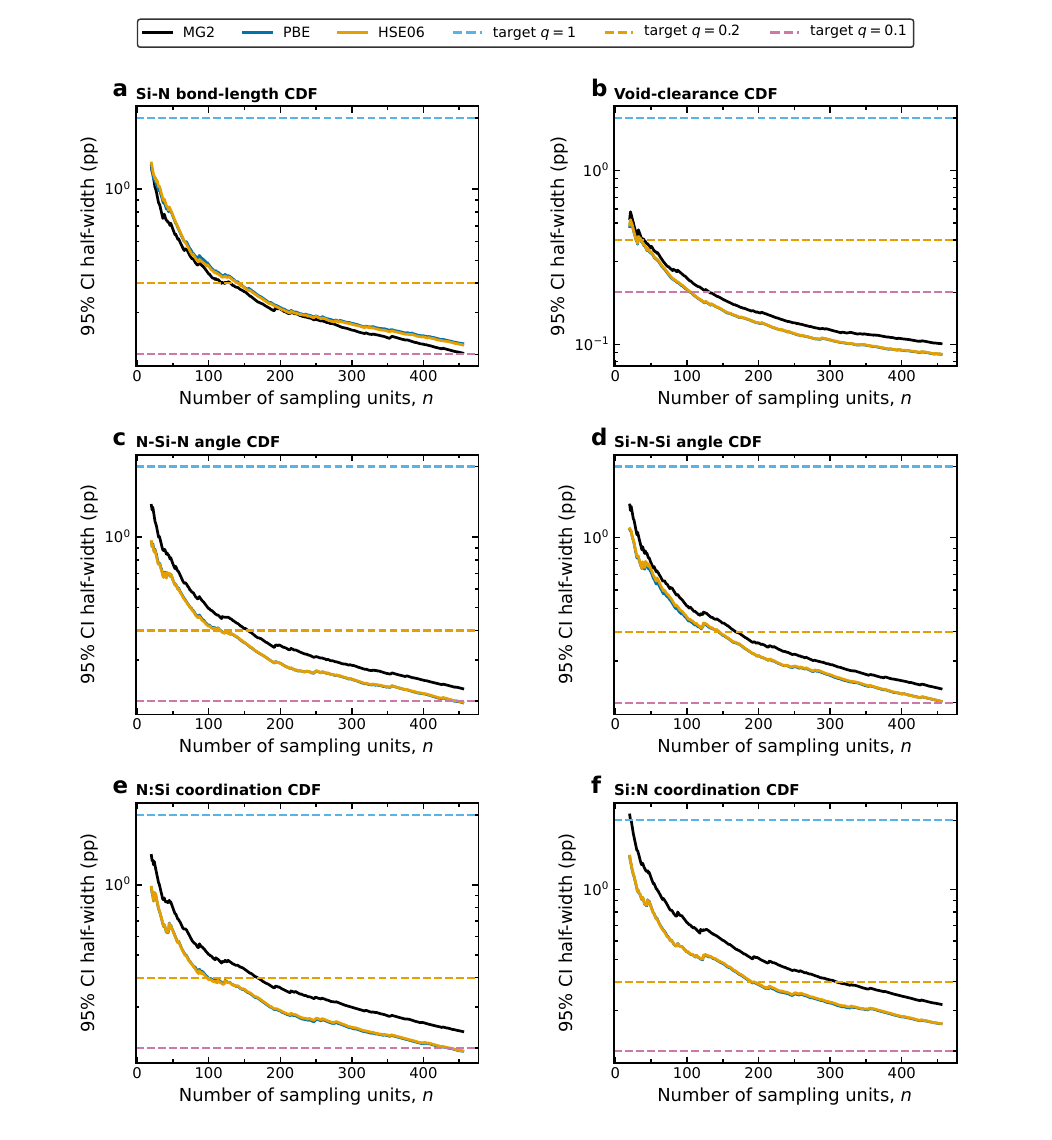}
\caption{Retrospective achieved uncertainty for Si--N pair-distance, void-clearance, angular and coordination CDFs across the 455 matched MG2, PBE, and HSE06 parent branches. The limiting fixed-grid component is selected separately by branch and prefix; the target line is the median branch-specific declared level and the band spans branch levels where they differ. These are method branches, not an unconfounded exchange--correlation-functional comparison.}
\label{fig:si-si3n4-uncertainty-pairs}
\end{figure}

\begin{figure}[p]
\ContinuedFloat
\centering
\includegraphics[page=2,width=\textwidth]{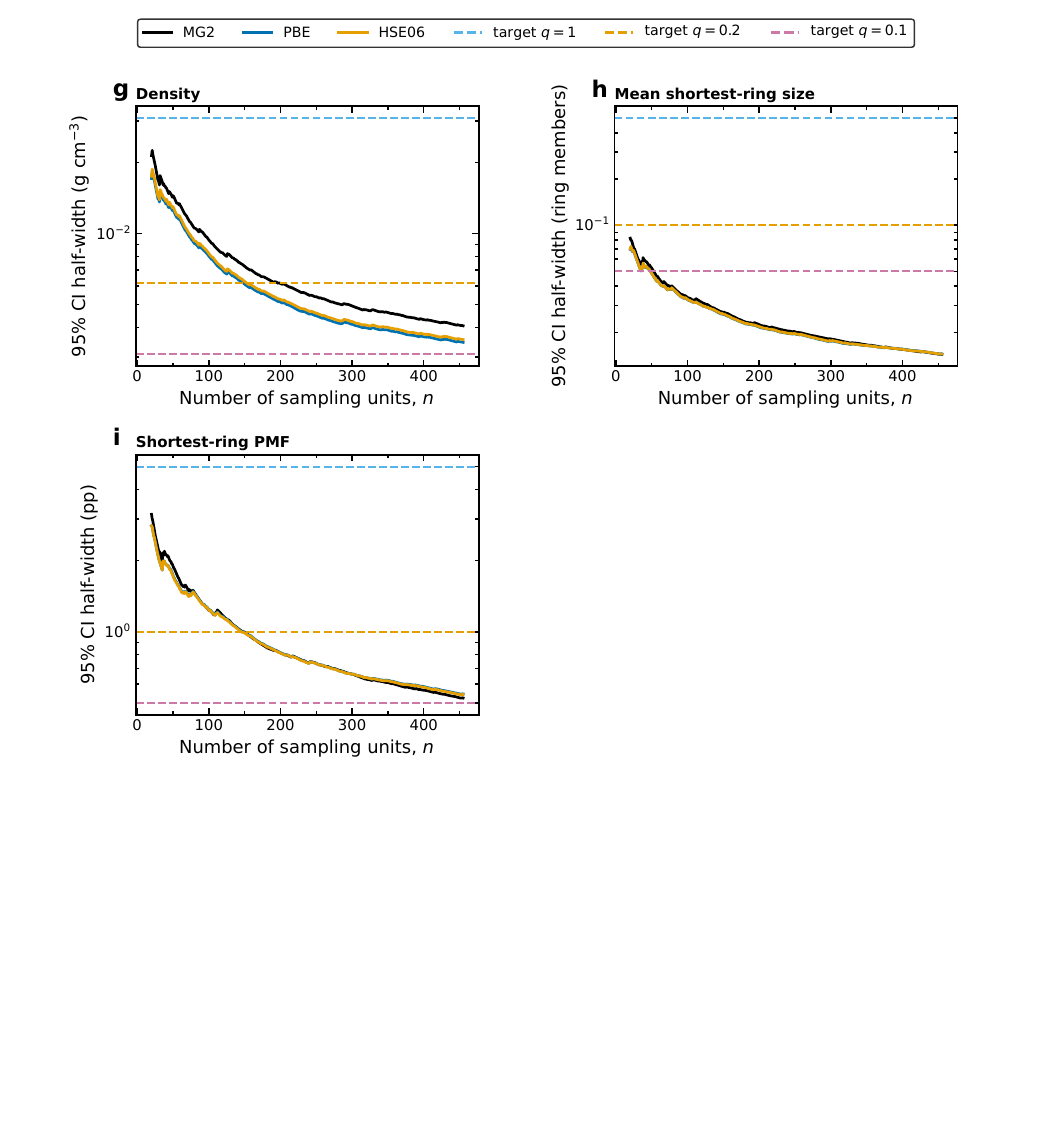}
\caption{Continued: retrospective achieved uncertainty for the a-Si$_3$N$_4$ ring-size PMF, mean ring size, and density over the matched parent branches. The method-branch and conditional-resolution qualifications are those in Fig.~\ref{fig:si-si3n4-uncertainty-pairs}.}
\label{fig:si-si3n4-uncertainty-rings-density}
\end{figure}

\begin{figure}[!htbp]
\centering
\includegraphics[page=1,width=\textwidth]{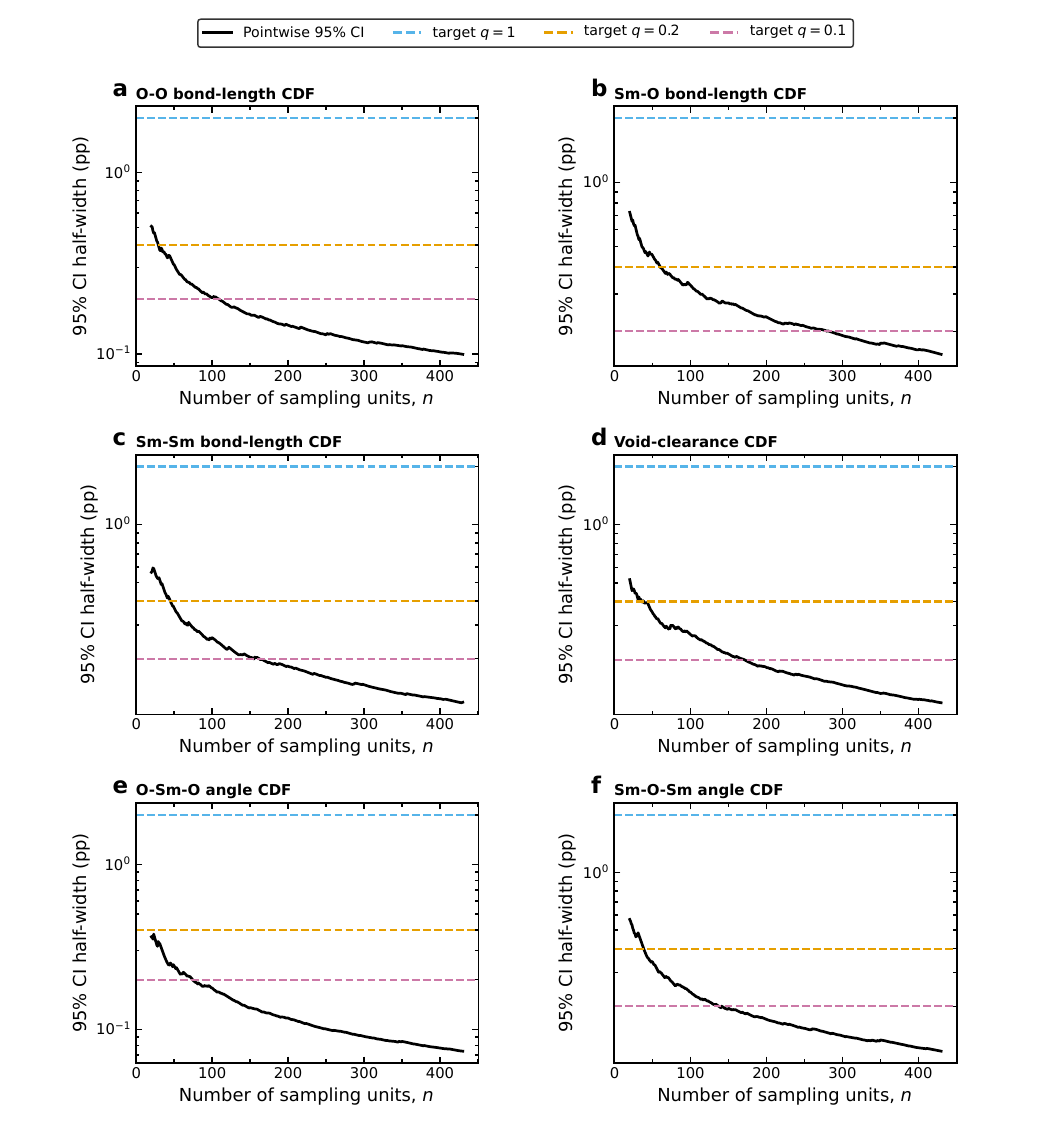}
\caption{Retrospective pointwise 95\% confidence half-widths for a-Sm$_2$O$_3$ pair-distance, void-clearance and angular CDFs over prefixes of the 430 retained-amorphous cells. The horizontal guides are conditional declared-resolution references, not a global population-size criterion.}
\label{fig:si-sm2o3-uncertainty-pairs}
\end{figure}

\begin{figure}[p]
\ContinuedFloat
\centering
\includegraphics[page=2,width=\textwidth]{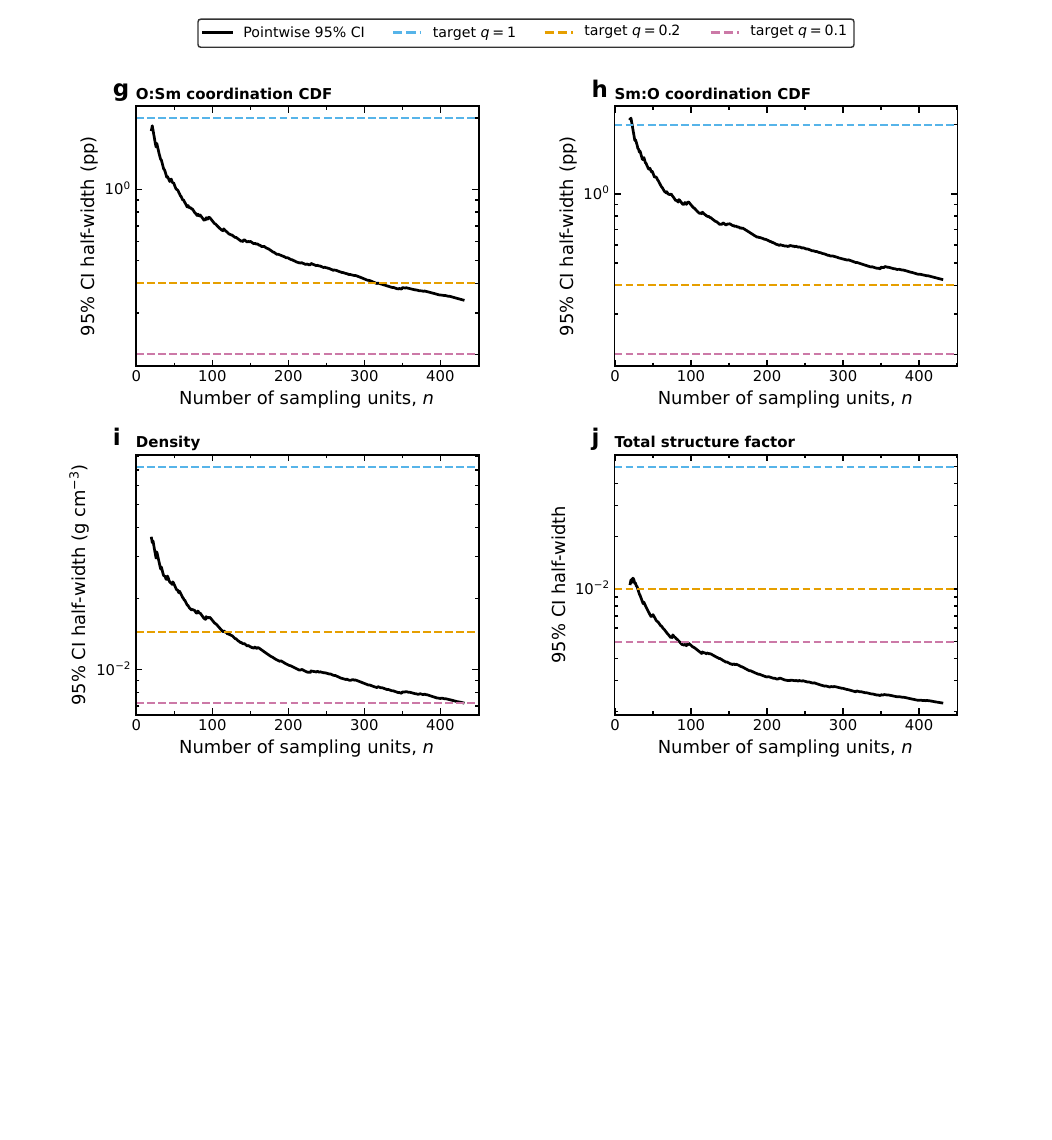}
\caption{Continued: retrospective pointwise 95\% confidence half-widths for the a-Sm$_2$O$_3$ coordination CDFs, density and total $S(q)$. The population and conditional-resolution conventions are those in Fig.~\ref{fig:si-sm2o3-uncertainty-pairs}.}
\label{fig:si-sm2o3-uncertainty-scalars}
\end{figure}

\clearpage
\endgroup

\begin{landscape}
\centering
\captionof{table}{Retrospective crossings of declared uncertainty levels for the descriptor families displayed in Figs.~\ref{fig:si-sio2-uncertainty-bonds}--\ref{fig:si-sm2o3-uncertainty-scalars}. Each entry is $n^{\mathrm{pers}}_{0.2}/n^{\mathrm{pers}}_{0.1}$, where $n^{\mathrm{pers}}_{q_{\mathrm{conv}},j}$ is the first observed population prefix after which the pointwise 95\% Student-$t$ confidence half-width remains no larger than $q_{\mathrm{conv}}\tau_j$ through the final recorded population. The family value is the maximum over its displayed descriptors. $>N_{\mathrm{ana}}$ denotes a level not reached; ``screen-defined'' denotes the exact-coordination a-SiO$_2$ population, for which coordination variance is imposed by the population definition; and -- denotes a family not displayed. These retrospective values are quantity-, population-, confidence-, and resolution-specific and are not universal population-size requirements.}
\label{tab:si-family-uncertainty-thresholds}
\footnotesize
\setlength{\tabcolsep}{3.0pt}
\renewcommand{\arraystretch}{1.08}
\begin{tabularx}{\linewidth}{@{}l X c c c c c@{}}
\toprule
& & \shortstack{a-SiO$_2$\\SHIK-1\\$N_{\rm ana}=1872$} & \multicolumn{3}{c}{a-Si$_3$N$_4$; $N_{\rm ana}=455$} & \shortstack{a-Sm$_2$O$_3$\\Buckingham\\$N_{\rm ana}=430$} \\
\cmidrule(lr){4-6}
Metric family & Declared claim scale $\tau$ & $n_{0.2}/n_{0.1}$ & \shortstack{MG2\\$n_{0.2}/n_{0.1}$} & \shortstack{MG2+PBE\\$n_{0.2}/n_{0.1}$} & \shortstack{MG2+HSE06\\$n_{0.2}/n_{0.1}$} & $n_{0.2}/n_{0.1}$ \\
\midrule
Density & $\max(0.001~\mathrm{g\,cm^{-3}},0.01|\hat\mu|)$ & 560 / $>1872$ & 199 / $>455$ & 153 / $>455$ & 154 / $>455$ & 116 / 428 \\
Pair-distance CDFs & 2 percentage points & 420 / 1718 & 131 / $>455$ & 280 / $>455$ & 272 / $>455$ & 61 / 283 \\
Angular CDFs & 2 percentage points & 404 / 1616 & 168 / $>455$ & 106 / 394 & 109 / 391 & 39 / 142 \\
Coordination CDFs / defect content & 2 percentage points & \shortstack{screen-\\defined} & 308 / $>455$ & 70 / 318 & 70 / 312 & $>430$ / $>430$ \\
Ring statistics & 5 percentage points (PMF); 0.5 members (mean) & 122 / 466 & 147 / $>455$ & 146 / $>455$ & 146 / $>455$ & -- \\
Void-clearance CDF & 2 percentage points & 89 / 313 & 40 / 132 & 34 / 108 & 34 / 107 & 39 / 171 \\
Total $S(q)$ & $\max(0.05,0.05|\hat\mu|)$ & -- & -- & -- & -- & 30 / 85 \\
\bottomrule
\end{tabularx}
\end{landscape}

\clearpage

\begingroup
\scriptsize
\setlength{\tabcolsep}{3.0pt}
\renewcommand{\arraystretch}{1.04}
\begin{longtable}{@{}>{\raggedright\arraybackslash}p{2.15cm}>{\raggedright\arraybackslash}p{4.5cm}>{\raggedright\arraybackslash}p{2.35cm}rr>{\raggedright\arraybackslash}p{2.7cm}@{}}
\caption{Descriptor-level retrospective crossing audit underlying Table~\ref{tab:si-family-uncertainty-thresholds}. $n_{q_{\mathrm{conv}},j}^{\mathrm{pers}}$ records the first observed population prefix after which the reported confidence half-width remains below $q_{\mathrm{conv}}\tau_j$ through $N_{\mathrm{ana}}$. The persistence column identifies a raw first crossing followed by a recrossing. Provisional and right-censored entries retain their archived status.}\label{tab:si-descriptor-uncertainty-thresholds}\\
\toprule
Dataset & Descriptor & $\tau_j$ at $N_{\mathrm{ana}}$ & $n_{0.2}^{\mathrm{pers}}$ & $n_{0.1}^{\mathrm{pers}}$ & First $\to$ persistent when different \\
\midrule
\endfirsthead
\multicolumn{6}{@{}l}{\textbf{Table~\thetable{} (continued)}}\\[0.5ex]
\toprule
Dataset & Descriptor & $\tau_j$ at $N_{\mathrm{ana}}$ & $n_{0.2}^{\mathrm{pers}}$ & $n_{0.1}^{\mathrm{pers}}$ & First $\to$ persistent when different \\
\midrule
\endhead
\midrule
\multicolumn{6}{r@{}}{Continued on next page}\\
\endfoot
\bottomrule
\endlastfoot
a-SiO$_2$ / SHIK-1 & Density & 0.02217 g cm$^{-3}$ & 560 & $>1872$ & -- \\
 & O--O pair-distance CDF & 2 percentage points & 152 & 567 & -- \\
 & Si--O pair-distance CDF & 2 percentage points & 371 & 1429 & -- \\
 & Si--Si pair-distance CDF & 2 percentage points & 420 & 1718 & -- \\
 & O--Si--O angle CDF & 2 percentage points & 165 & 617 & -- \\
 & Si--O--Si angle CDF & 2 percentage points & 404 & 1616 & $q_{\rm conv}=0.1$: 1612$\to$1616 \\
 & O:Si coordination CDF & 2 percentage points & \textit{screen-defined} & \textit{screen-defined} & -- \\
 & Si:O coordination CDF & 2 percentage points & \textit{screen-defined} & \textit{screen-defined} & -- \\
 & Projected Si-ring-size PMF & 5 percentage points & 122 & 466 & $q_{\rm conv}=0.1$: 464$\to$466 \\
 & Mean projected Si-ring size & 0.5 members & 20 & 20 & -- \\
 & Void-clearance CDF & 2 percentage points & 89 & 313 & $q_{\rm conv}=0.1$: 311$\to$313 \\
\addlinespace[2pt]
a-Si$_3$N$_4$ / MG2 & Density & 0.03087 g cm$^{-3}$ & 199 & $>455$ & $q_{\rm conv}=0.2$: 197$\to$199 \\
 & Si--N pair-distance CDF & 2 percentage points & 131 & $>455$ & $q_{\rm conv}=0.2$: 118$\to$131 \\
 & N--Si--N angle CDF & 2 percentage points & 155 & $>455$ & -- \\
 & Si--N--Si angle CDF & 2 percentage points & 168 & $>455$ & -- \\
 & N:Si coordination CDF & 2 percentage points & 169 & $>455$ & -- \\
 & Si:N coordination CDF & 2 percentage points & 308 & $>455$ & -- \\
 & Bond-graph ring-size PMF & 5 percentage points & 147 & $>455$ & -- \\
 & Mean bond-graph ring size & 0.5 members & 20 & 53 & -- \\
 & Void-clearance CDF & 2 percentage points & 40 & 132 & -- \\
\addlinespace[2pt]
a-Si$_3$N$_4$ / MG2+PBE & Density & 0.03011 g cm$^{-3}$ & 153 & $>455$ & -- \\
 & Si--N pair-distance CDF & 2 percentage points & 280 & $>455$ & -- \\
 & N--Si--N angle CDF & 2 percentage points & 75 & 274 & -- \\
 & Si--N--Si angle CDF & 2 percentage points & 106 & 394 & $q_{\rm conv}=0.2$: 103$\to$106 \\
 & N:Si coordination CDF & 2 percentage points & 42 & 184 & -- \\
 & Si:N coordination CDF & 2 percentage points & 70 & 318 & -- \\
 & Bond-graph ring-size PMF & 5 percentage points & 146 & $>455$ & -- \\
 & Mean bond-graph ring size & 0.5 members & 20 & 42 & $q_{\rm conv}=0.1$: 22$\to$42 \\
 & Void-clearance CDF & 2 percentage points & 34 & 108 & $q_{\rm conv}=0.2$: 29$\to$34 \\
\addlinespace[2pt]
\pagebreak[4]
a-Si$_3$N$_4$ / MG2+HSE06 & Density & 0.03081 g cm$^{-3}$ & 154 & $>455$ & -- \\
 & Si--N pair-distance CDF & 2 percentage points & 272 & $>455$ & -- \\
 & N--Si--N angle CDF & 2 percentage points & 73 & 264 & $q_{\rm conv}=0.1$: 256$\to$264 \\
 & Si--N--Si angle CDF & 2 percentage points & 109 & 391 & $q_{\rm conv}=0.1$: 387$\to$391 \\
 & N:Si coordination CDF & 2 percentage points & 43 & 175 & -- \\
 & Si:N coordination CDF & 2 percentage points & 70 & 312 & -- \\
 & Bond-graph ring-size PMF & 5 percentage points & 146 & $>455$ & -- \\
 & Mean bond-graph ring size & 0.5 members & 20 & 44 & $q_{\rm conv}=0.1$: 20$\to$44 \\
 & Void-clearance CDF & 2 percentage points & 34 & 107 & $q_{\rm conv}=0.2$: 29$\to$34 \\
\addlinespace[2pt]
a-Sm$_2$O$_3$ / Buck. & Density & 0.07258 g cm$^{-3}$ & 116 & 428 & $q_{\rm conv}=0.2$: 114$\to$116 \\
 & O--O pair-distance CDF & 2 percentage points & 30 & 110 & -- \\
 & Sm--O pair-distance CDF & 2 percentage points & 61 & 283 & -- \\
 & Sm--Sm pair-distance CDF & 2 percentage points & 45 & 162 & $q_{\rm conv}=0.1$: 155$\to$162 \\
 & O--Sm--O angle CDF & 2 percentage points & 20 & 75 & -- \\
 & Sm--O--Sm angle CDF & 2 percentage points & 39 & 142 & $q_{\rm conv}=0.1$: 138$\to$142 \\
 & O:Sm coordination CDF & 2 percentage points & 319 & $>430$ & $q_{\rm conv}=0.2$: 316$\to$319 \\
 & Sm:O coordination CDF & 2 percentage points & $>430$ & $>430$ & -- \\
 & Void-clearance CDF & 2 percentage points & 39 & 171 & $q_{\rm conv}=0.2$: 36$\to$39 \\
 & Total $S(q)$ & 0.05 & 30 & 85 & -- \\
\end{longtable}
\endgroup

\subsection{Stage-resolved diagnostic analysis}
\label{sec:si-stage-resolved}

Stage-resolved descriptor analysis is diagnostic rather than a separate acceptance layer. For each trajectory, descriptors are evaluated along the melt, liquid-hold, quench, and relaxation segments. Let $d_j(s,t)$ denote descriptor $j$ evaluated at time $t$ in stage $s$, and let $\hat\mu_{A,j}(s,t)$ be its sampling-unit mean over population $A$ at that stage and time. For two populations $A$ and $B$, the normalised descriptor separation is
\begin{equation}
\Delta_j^{A,B}(s,t)
=
\frac{
|\hat\mu_{A,j}(s,t)-\hat\mu_{B,j}(s,t)|
}
{\tau_j(\hat\mu_{B,j}(s,t))}.
\label{eq:si-stage-separation}
\end{equation}
A divergence is considered structurally relevant when $\Delta_j^{A,B}(s,t)>1$ and persists over the configured time or temperature window.

During cooling, let $D(T_j)$ be the diffusion estimate at sampled cooling temperature $T_j$ and let $D^\ast$ be the configured diffusion threshold in the same units. The diffusion-freeze temperature is defined as
\begin{equation}
T_f
=
\max
\left\{
T_i:
D(T_j)\leq D^\ast
\;\text{for all}\;
T_j\leq T_i
\right\}.
\label{eq:si-freeze-temperature}
\end{equation}
If this set is empty, no diffusion-freeze temperature is assigned for that trajectory.
Divergence already present in the high-temperature liquid indicates incomplete loss of structural memory. Divergence emerging between $T_{\mathrm{high}}$ and $T_f$ identifies the quench schedule as the dominant control parameter. Divergence appearing only after cooling indicates low-temperature relaxation, residual-stress effects, or final minimisation effects. Secondary diagnostics such as $S(q)$, void-clearance distributions, elastic fingerprints, and residual-stress summaries are used only when the primary descriptor set does not resolve the origin of the divergence.

\subsection[a-SiO2: tetrahedral-network fidelity]{a-SiO$_2$: tetrahedral-network fidelity}
\label{sec:si-sio2-methods}

The a-SiO$_2$ benchmark tests whether the generated amorphous silica population satisfies the chemically ideal tetrahedral network definition. The SHIK-1 source melt was initialised by random placement of 64 Si and 128 O atoms in a cubic cell of side length 21~\AA{}.\cite{Sundararaman2018SHIK} After heating for 5~ps and equilibrating at 5000~K for 50~ps, the first source snapshot and 2532 further snapshots at 10-ps intervals produced 2533 source configurations over a common heat/melt trajectory of $5+50+(2533-1)\times10=25{,}375$~ps. No residual correlation relevant to the reported estimands was detected at the 10-ps lag, and each resulting quenched and relaxed box is treated as an independent cell-level sampling unit.

The primary structural screen is exact first-shell coordination. For each cell,
\begin{equation}
f_{\mathrm{Si:O}}^{(b)}
=
\frac{1}{N_{\mathrm{Si}}^{(b)}}
\sum_{i\in\mathrm{Si}}
\mathbf 1
\left[
n_{i:\mathrm O}^{(b)}\neq 4
\right],
\label{eq:si-sio2-defect-fraction}
\end{equation}
and
\begin{equation}
f_{\mathrm{O:Si}}^{(b)}
=
\frac{1}{N_{\mathrm O}^{(b)}}
\sum_{i\in\mathrm O}
\mathbf 1
\left[
n_{i:\mathrm{Si}}^{(b)}\neq 2
\right].
\end{equation}
The SiO$_2$ coordination-screen label is
\begin{equation}
A_b^{\mathrm{SiO_2}}
=
\mathbf 1
\left[
f_{\mathrm{Si:O}}^{(b)}=0
\right]
\mathbf 1
\left[
f_{\mathrm{O:Si}}^{(b)}=0
\right].
\end{equation}
This screen is configured with \code{exclude=true}. Thus, $A_b^{\mathrm{SiO_2}}=1$ cells form the defect-free silica analysis population: all Si centres are fourfold coordinated by O and all O centres are twofold coordinated by Si. Cells with $A_b^{\mathrm{SiO_2}}=0$ are excluded from the silica analysis population and retained as the oxygen-bridge-defective audit population analysed in the main text.

The SiO$_2$ reporting descriptor set comprises density, total and partial radial distribution functions, the Si--O first-peak position and height, O--Si--O and Si--O--Si angular distributions, and projected Si-ring statistics. The convergence descriptor groups include representative scalar descriptors from the local, angular, and medium-range sets. The projected-ring multigraph is built only through bridging O atoms satisfying $n_{o:\mathrm{Si}}^{(b)}=2$, ensuring that the ring topology describes the chemically admissible tetrahedral network rather than defect-mediated graph artefacts.

Threshold sensitivity is evaluated by recomputing the analysis fraction and key descriptor means under controlled perturbations of the first-shell cutoff grid. The final reported SiO$_2$ analysis population uses the fixed cutoff table archived with the source data.

\paragraph{Prevalence, incidence, and network context.}\label{par:si-sio2-population-context}
At the frozen 2.07~\AA{} graph, non-ideal O sites account for 0.408\% of O sites and occur in 26.10\% of the source cells; the latter is cell incidence. For an angular threshold $\theta_c$, the three-member-ring enrichment is $\mathcal E_3(\theta_c)=\Pr(\text{bridge belongs to a 3-ring}\mid\theta_{\mathrm{SiOSi}}<\theta_c)/\Pr(\text{bridge belongs to a 3-ring})$. Among exactly twofold O bridges, the $\theta_c=129^\circ$ lower-angle population is 6.392\%, with $\mathcal E_3(129^\circ)=5.746$ under the registered ring definition. These quantities distinguish site prevalence, cell incidence across the source population, and network association; they do not assign a mechanism. The complete structural comparison, selector attribution, and screen sensitivity are shown in Figs.~\ref{fig:si-sio2-structural}--\ref{fig:si-sio2-screen-sensitivity}.

\begingroup
\makeatletter
\setlength{\@fptop}{0pt}
\makeatother

\begin{figure}[!htbp]
  \centering
  \includegraphics[width=\textwidth]{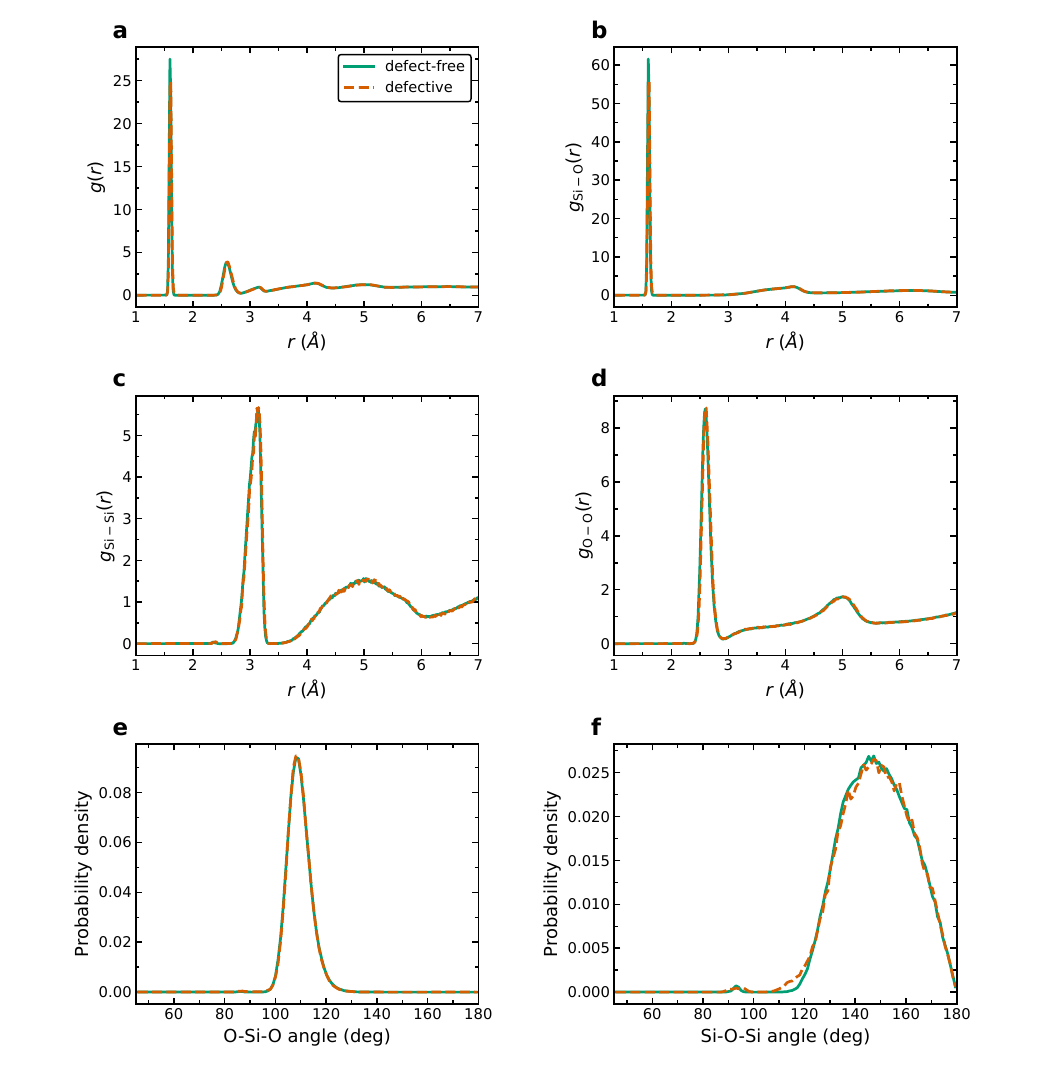}
  \caption{a-SiO$_2$ structural descriptors for the defect-free and defective populations. (a) Total radial distribution function. (b) Si--O partial radial distribution function. (c) Si--Si partial radial distribution function. (d) O--O partial radial distribution function. (e) O--Si--O angular distribution. (f) Si--O--Si angular distribution. Radial distribution functions are restricted to the structurally informative 1--7~\AA{} range.}
  \label{fig:si-sio2-structural}
\end{figure}

\begin{figure}[!htbp]
  \centering
  \includegraphics[width=\textwidth]{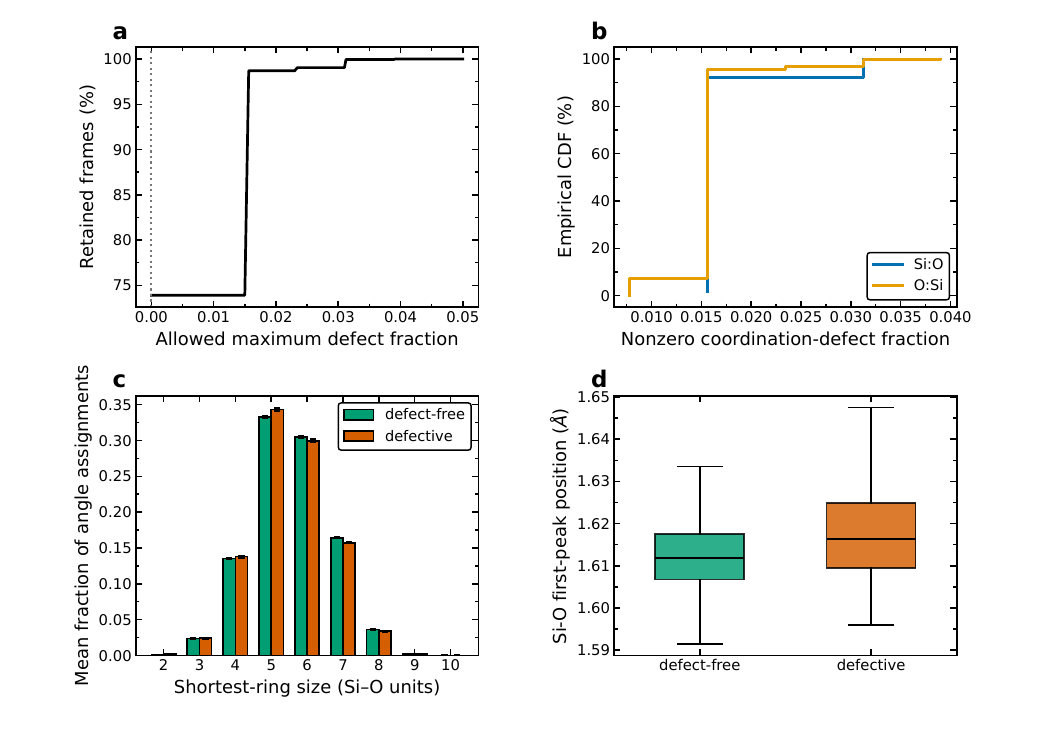}
  \caption{a-SiO$_2$ screening details. (a) Retained fraction as a function of the allowed maximum coordination-defect fraction. (b) Distribution of nonzero defect fractions for the Si:O and O:Si selectors. (c) Shortest projected Si-ring-size statistics for the defect-free and defective populations, resolved over ring sizes 2--10; ring size counts the Si of the projected cycle, each consecutive pair joined by one exactly twofold bridging O. (d) Si--O first-peak-position distributions.}
  \label{fig:si-sio2-screening}
\end{figure}

\begin{figure}[!htbp]
\centering
\includegraphics[page=1,width=\textwidth]{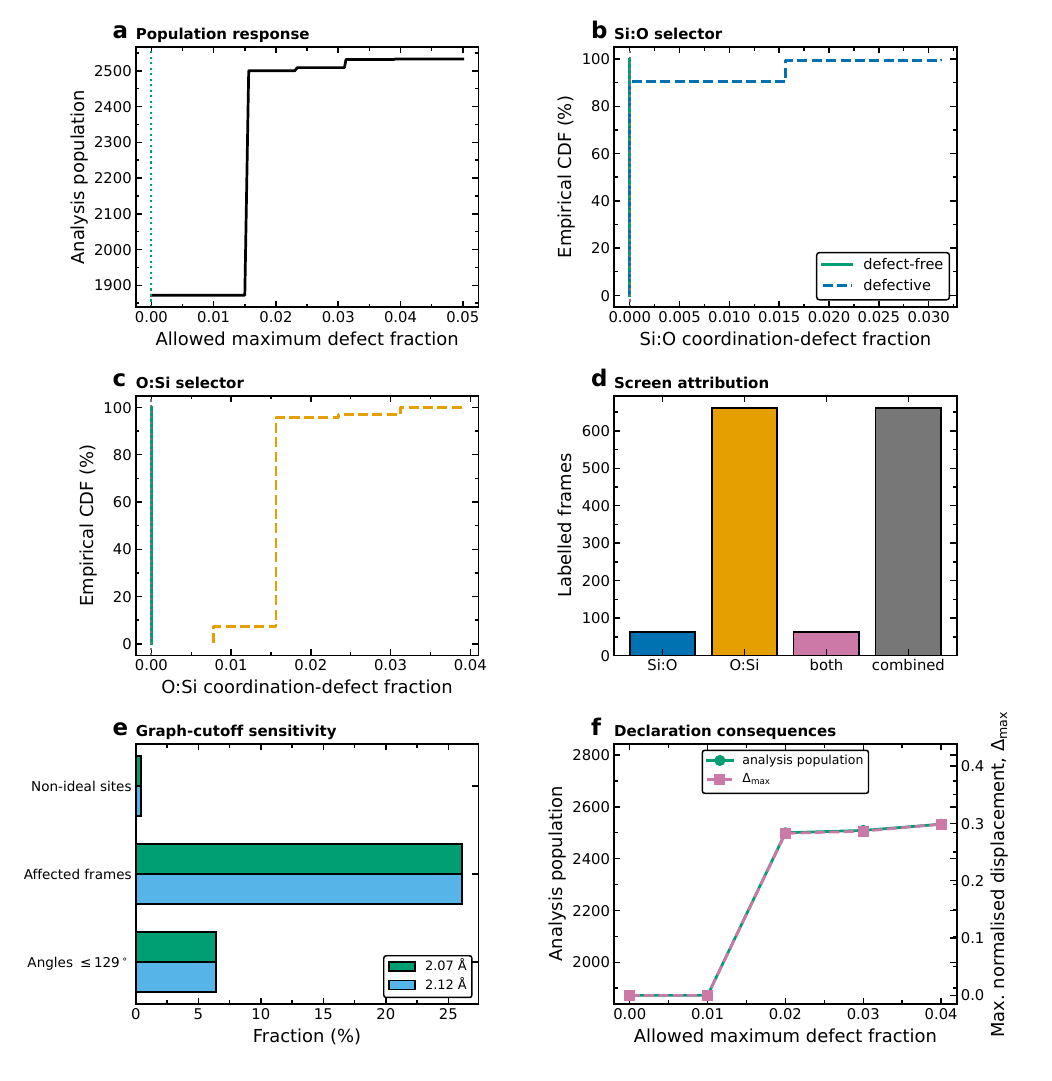}
\caption{Sensitivity of the a-SiO$_2$ screen and declared populations for the 2533 source cells. The 1872 exact-coordination analysis cells and 661 coordination-defect cells are defined by the primary 2.07~\AA{} graph; 2.12~\AA{} remains sensitivity-only. All 63 Si:O failures also fail O:Si. The reported cell incidence uses the independent cells as sampling units.}
\label{fig:si-sio2-screen-sensitivity}
\end{figure}

\clearpage
\endgroup

\subsection[a-Si3N4: controlled DFT refinement]{a-Si$_3$N$_4$: controlled DFT refinement}
\label{sec:si-si3n4-methods}

The a-Si$_3$N$_4$ benchmark tests controlled multi-fidelity refinement in a heteropolar amorphous network. The MG2 dataset is the empirical melt--quench population.\cite{Marian2000MG2} The MG2+PBE and MG2+HSE06 datasets are PBE- and HSE06-relaxed descendants of the MG2-derived amorphous population, analysed with the same screen definitions, descriptor pipeline, and statistical workflow.\cite{Perdew1996PBE,Heyd2003HSE,Heyd2006HSEErratum,Krukau2006HSE06} For $M\in\{\mathrm{MG2},\mathrm{PBE},\mathrm{HSE06}\}$, $x_b^{(M)}$ denotes the branch-$M$ structure descended from shared parent $b$. The comparison is therefore a controlled matched-parent refinement analysis under common descriptors, not an independent DFT melt--quench comparison, an unconfounded exchange--correlation-functional benchmark, or a single-score ranking of interaction models.

Nominal coordination defects are defined using
\begin{equation}
\mathcal A_{\mathrm{Si:N}}=\{4\},
\qquad
\mathcal A_{\mathrm{N:Si}}=\{3\}.
\end{equation}
The corresponding defect fractions are
\begin{equation}
f_{\mathrm{Si:N}}^{(b,M)}
=
\frac{1}{N_{\mathrm{Si}}^{(b,M)}}
\sum_{i\in\mathrm{Si}}
\mathbf 1
\left[
n_{i:\mathrm N}^{(b,M)}\notin\mathcal A_{\mathrm{Si:N}}
\right],
\end{equation}
and
\begin{equation}
f_{\mathrm{N:Si}}^{(b,M)}
=
\frac{1}{N_{\mathrm N}^{(b,M)}}
\sum_{i\in\mathrm N}
\mathbf 1
\left[
n_{i:\mathrm{Si}}^{(b,M)}\notin\mathcal A_{\mathrm{N:Si}}
\right].
\end{equation}
Unlike the SiO$_2$ case, the nitride coordination screen is not an exact zero-defect rejection condition. Residual Si:N and N:Si defect fractions are refinement-response descriptors. The configured thresholds define a coordination-outlier screen label,
\begin{equation}
A_b^{(M)}
=
\mathbf 1
\left[
f_{\mathrm{Si:N}}^{(b,M)}
\leq
\eta_{\mathrm{Si:N}}^{\mathrm{audit}}
\right]
\mathbf 1
\left[
f_{\mathrm{N:Si}}^{(b,M)}
\leq
\eta_{\mathrm{N:Si}}^{\mathrm{audit}}
\right].
\end{equation}
The ordered audit thresholds are $(\eta_{\mathrm{Si:N}}^{\mathrm{audit}},\eta_{\mathrm{N:Si}}^{\mathrm{audit}})=(0.15,0.20)$.
This screen is configured with \code{exclude=false}. Therefore
\begin{equation}
\mathcal E_{\mathrm{ana}}^{(M)}
=
\{x_b^{(M)}:b=1,\ldots,N_{\mathrm{gen}}\},
\qquad
\mathcal E_{A=0}^{(M)}
=
\{x_b^{(M)}:A_b^{(M)}=0\}.
\end{equation}
The audit-labelled subset is reported as a refinement diagnostic, while
residual defect fractions are reported over the full analysis population:
\begin{equation}
\bar f_{\alpha:\beta}^{(M)}
=
\frac{1}{|\mathcal E_{\mathrm{ana}}^{(M)}|}
\sum_{x_b^{(M)}\in\mathcal E_{\mathrm{ana}}^{(M)}}
f_{\alpha:\beta}^{(b,M)}.
\end{equation}

The Si$_3$N$_4$ descriptor set comprises density, Si--N radial distribution functions, Si--N first-peak position and height, N--Si--N and Si--N--Si angular distributions, analysis-population Si:N and N:Si defect fractions, and primitive rings on the bipartite Si--N graph. Because the Si--N graph is bipartite, chemically meaningful rings have even size. For model $M\in\{\mathrm{PBE},\mathrm{HSE06}\}$ and scalar descriptor $j$ with $\hat\mu_{\mathrm{MG2},j}\ne0$, shifts relative to MG2 are reported as
\begin{equation}
\Delta_j^{M}
=
100
\frac{
\hat\mu_{M,j}-\hat\mu_{\mathrm{MG2},j}
}
{
\hat\mu_{\mathrm{MG2},j}
}
\;\%.
\end{equation}
When $\hat\mu_{\mathrm{MG2},j}=0$, the unnormalised difference $\hat\mu_{M,j}-\hat\mu_{\mathrm{MG2},j}$ is reported instead. This normalisation is used only for reporting relative shifts; convergence is assessed on the original descriptor values and tolerances.

The configured convergence groups separate short-range descriptors, medium-range descriptors, and long-range diagnostics. The short-range group includes density, Si--N first-shell pair descriptors, and coordination-defect fractions. The medium-range group includes angular and bond-graph ring descriptors. Long-range diagnostics include low-$q$ structure-factor features or void/stress fingerprints when configured.

\paragraph{Metric, shape, and topology.}\label{par:si-si3n4-metric-shape-topology}
Let $F_M(r)$ be the equal-parent Si--N pair-distance CDF for branch $M$, let $r_a$ be the registered common fixed anchor, and let $r_{\mathrm{pk}}^{(M)}$ be the branch first-peak position. At radial offset $\Delta l$, the fixed-anchor contrast is $|F_{\mathrm{PBE}}(r_a+\Delta l)-F_{\mathrm{HSE06}}(r_a+\Delta l)|$. With $\widetilde F_M(\Delta l)=F_M(r_{\mathrm{pk}}^{(M)}+\Delta l)$, the peak-referenced contrast is $|\widetilde F_{\mathrm{PBE}}(\Delta l)-\widetilde F_{\mathrm{HSE06}}(\Delta l)|$. The common topology graph uses $r_c^{\mathrm{common}}=2.33$~\AA{}, and its PBE--HSE06 edge-preservation fraction is $|E_{\mathrm{PBE}}\cap E_{\mathrm{HSE06}}|/|E_{\mathrm{PBE}}\cup E_{\mathrm{HSE06}}|$; the species-resolved unchanged-coordination fraction is the fraction of matched atoms whose coordination number is identical in both branches. Bracketed values below are the archived lower and upper matched-parent bounds; the supplied summary does not encode a confidence construction, so no confidence level is assigned here.

\begin{figure}[!htbp]
  \centering
  \includegraphics[width=\textwidth]{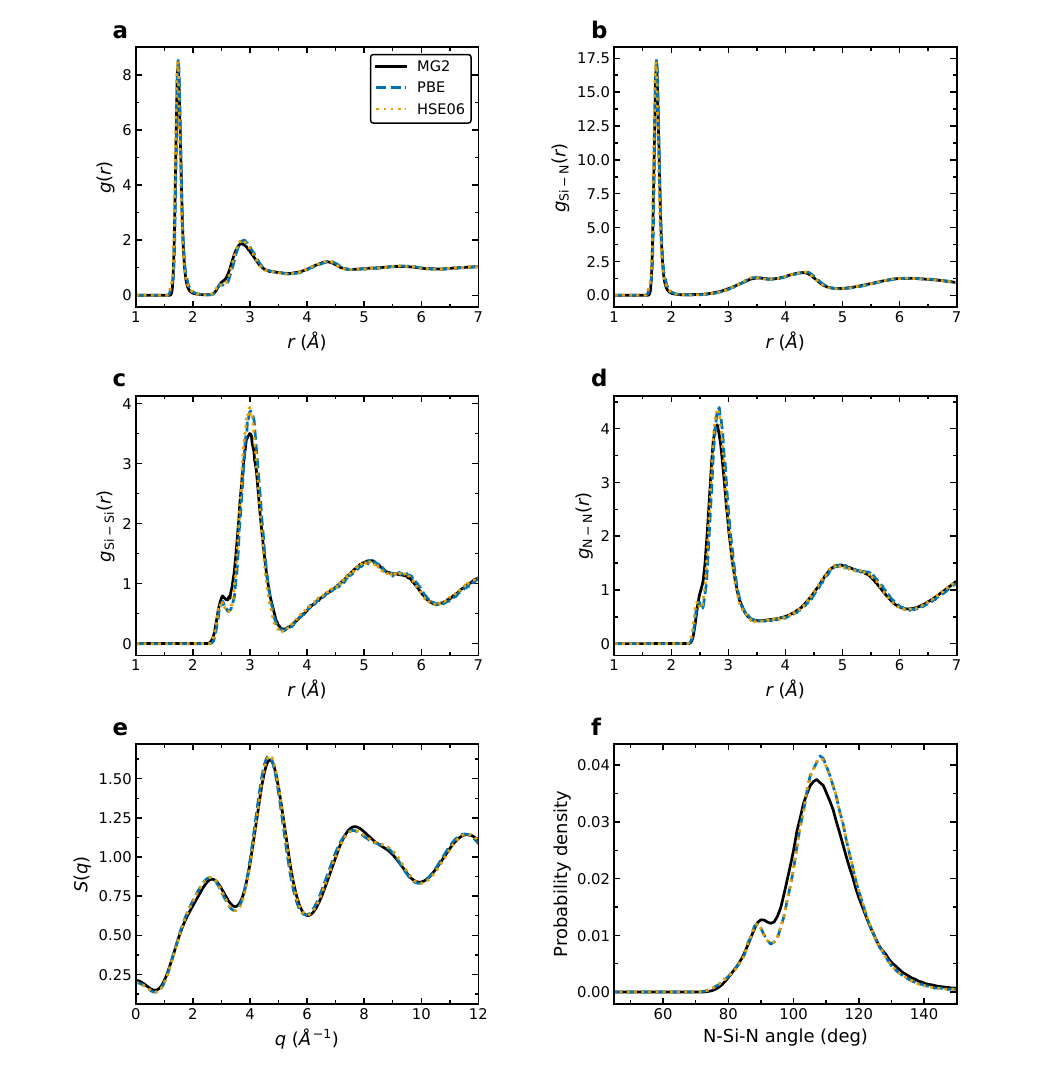}
  \caption{a-Si$_3$N$_4$ MG2/PBE/HSE06 structural comparison. (a) Total radial distribution function. (b) Si--N partial radial distribution function. (c) Si--Si partial radial distribution function. (d) N--N partial radial distribution function. (e) Total structure factor. (f) N--Si--N angular distribution. Radial distribution functions are restricted to 1--7~\AA{}.}
  \label{fig:si-si3n4-structural}
\end{figure}

\begin{figure}[p]
  \centering
  \includegraphics[width=\textwidth]{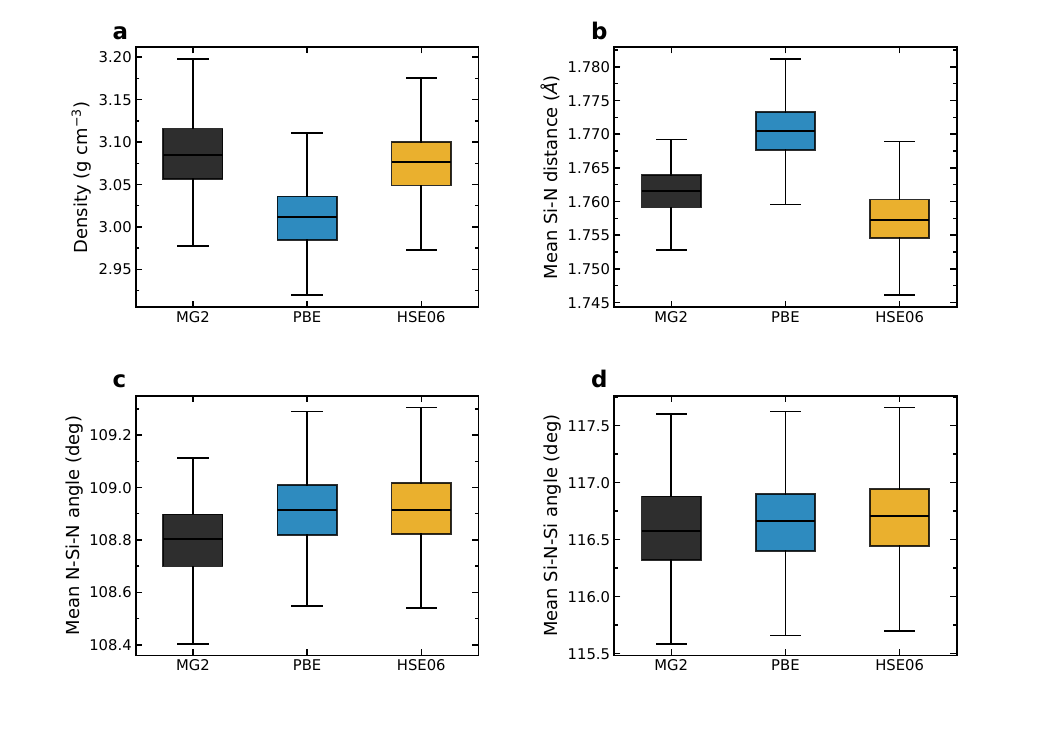}
  \caption{a-Si$_3$N$_4$ scalar descriptor distributions for the MG2, PBE, and HSE06 datasets. (a) Density. (b) Mean Si--N bond length. (c) Mean N--Si--N angle. (d) Mean Si--N--Si angle.}
  \label{fig:si-si3n4-scalars}
\end{figure}

\begin{figure}[p]
\centering
\includegraphics[page=1,width=\textwidth]{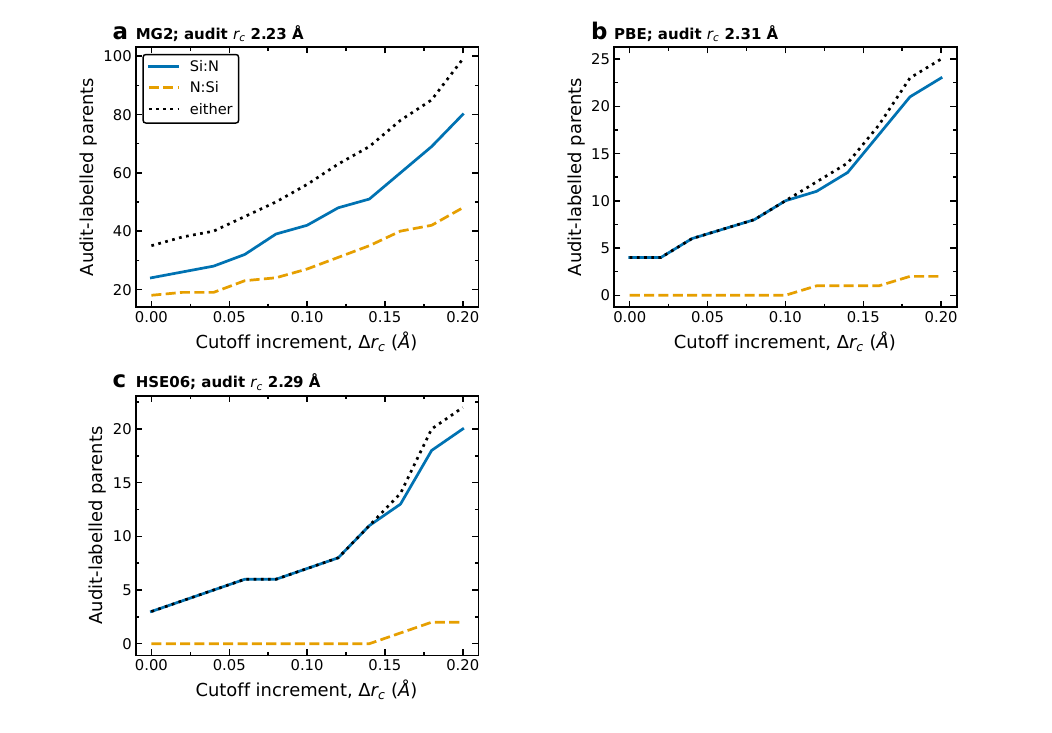}
\caption{Audit-cutoff sensitivity for the 455 matched a-Si$_3$N$_4$ parent branches. All cells remain analysed because the audit action has \texttt{exclude=false}. The registered $r_c^{\mathrm{audit},M}$ values are 2.23, 2.31, and 2.29~\AA{} for MG2, PBE, and HSE06, respectively; the scanned increment is 0--0.20~\AA{}. This audit construction is distinct from the RDF-boundary diagnostic in Fig.~\ref{fig:si-si3n4-boundary} and the common-topology cutoff $r_c^{\mathrm{common}}=2.33$~\AA{}.}
\label{fig:si-si3n4-audit-sensitivity}
\end{figure}

\begin{figure}[p]
\centering
\includegraphics[page=1,width=\textwidth]{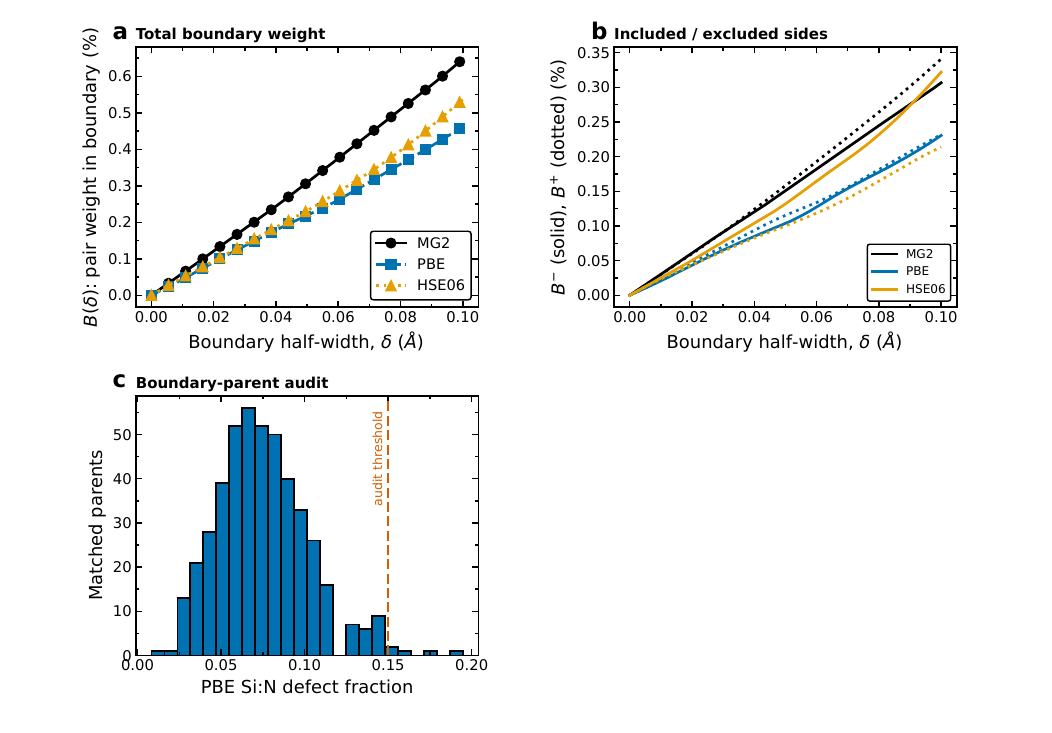}
\caption{RDF-boundary diagnostic for the 455 matched a-Si$_3$N$_4$ parents. For the normalised empirical Si--N pair-distance CDF $F_M^{\rm pair}$ and registered boundary $r_c^{\mathrm{RDF},M}$, the included- and excluded-side weights are $B_M^{-}(\delta)=F_M^{\rm pair}(r_c^{\mathrm{RDF},M})-F_M^{\rm pair}(r_c^{\mathrm{RDF},M}-\delta)$ and $B_M^{+}(\delta)=F_M^{\rm pair}(r_c^{\mathrm{RDF},M}+\delta)-F_M^{\rm pair}(r_c^{\mathrm{RDF},M})$; $B_M(\delta)=B_M^{-}(\delta)+B_M^{+}(\delta)$. The registered $r_c^{\mathrm{RDF},M}$ values are 2.29, 2.31, and 2.23~\AA{} for MG2, PBE, and HSE06. At $\delta=0.05$~\AA{}, the total boundary pair weights are 0.309\%, 0.219\%, and 0.232\%; these are population pair weights, not sitewise ambiguity probabilities. The construction is not merged with the audit-cutoff sensitivity in Fig.~\ref{fig:si-si3n4-audit-sensitivity}.}
\label{fig:si-si3n4-boundary}
\end{figure}

The 455 matched parents give mean densities of 3.08776, 3.01083, and 3.08065~g~cm$^{-3}$ for MG2, PBE, and HSE06, respectively; the PBE--HSE06 method-branch difference is $-0.069815$~g~cm$^{-3}$ ($-2.266\%$). At $\Delta l=0.10$~\AA{}, the fixed-anchor PBE--HSE06 tail contrast is 0.01780 [0.01683, 0.01878], whereas the peak-referenced contrast is 0.00163 [0.00120, 0.00205]. The Kolmogorov--Smirnov distance $D_{\mathrm{KS}}$ decreases from 0.1137 to 0.0020 and the Wasserstein distance $W_1$ from 0.00694~\AA{} to 0.00039~\AA{} after peak referencing. The definition-specific first-neighbour edge union is 99.9664\% [99.9562, 99.9766] preserved between the PBE and HSE06 branches, with 99.8626\% of Si and 99.8970\% of N coordination states unchanged. These are matched method-branch statements; peak-location uncertainty is not propagated in the presented peak-referenced sensitivity, and no universal topology-preservation claim follows. The complete structural and scalar distributions, audit-cutoff sensitivity, and graph-boundary response are shown in Figs.~\ref{fig:si-si3n4-structural}--\ref{fig:si-si3n4-boundary}.

\clearpage

\subsection[a-Sm2O3: mixed coordination and crystal-like screening]{a-Sm$_2$O$_3$: mixed coordination and crystal-like screening}
\label{sec:si-sm2o3-methods}

The a-Sm$_2$O$_3$ benchmark tests the stability and screening layers in a charged mixed-coordination rare-earth oxide. The classical dataset uses a Coulomb--Buckingham interaction model with the tabulated $C^2$ Buckingham--ZBL replacement splice described in Sec.~\ref{sec:si-core}.\cite{Buckingham1938ExpSix,Ziegler1985ZBL} The splice is part of the numerical model used to obtain stable high-temperature dynamics; it is not used as a structural acceptance criterion. The 50~K~ps$^{-1}$ production rate was deliberately selected to explore the order--disorder boundary; the earlier amorphous-only study instead used 100~K~ps$^{-1}$ to suppress crystal-like fragments.\cite{Olsson2019AmorphousSm2O3}

Fixed-coordination screening is not used for Sm$_2$O$_3$ because mixed Sm--O and O--Sm coordination is part of the target amorphous structure. The primary screen targets crystal-like order. For local bond-order analysis, a local spherical-harmonic descriptor may be written as\cite{Steinhardt1983BondOrder,Lechner2008BondOrder}:
\begin{equation}
q_{\ell m}(i)
=
\frac{1}{N_i}
\sum_{j=1}^{N_i}
Y_{\ell m}(\hat{\mathbf r}_{ij}),
\end{equation}
where $\mathcal N(i)$ is the configured neighbour shell, $N_i=|\mathcal N(i)|$, and $\hat{\mathbf r}_{ij}$ is the unit vector from site $i$ to neighbour $j$. The corresponding rotationally invariant magnitude is
\begin{equation}
q_\ell(i)
=
\left[
\frac{4\pi}{2\ell+1}
\sum_{m=-\ell}^{\ell}
|q_{\ell m}(i)|^2
\right]^{1/2}.
\end{equation}
The Lechner--Dellago neighbour average and its invariant magnitude are
\[
\bar q_{\ell m}(i)
=
\frac{q_{\ell m}(i)+\sum_{j\in\mathcal N(i)}q_{\ell m}(j)}{N_i+1},
\qquad
\bar q_\ell(i)
=
\left[
\frac{4\pi}{2\ell+1}
\sum_{m=-\ell}^{\ell}|\bar q_{\ell m}(i)|^2
\right]^{1/2}.
\]
The reported local-order summary uses the configured $\bar q_6$ statistic. Let $\bar{\mathbf q}_{6i}^{(b)}=(\bar q_{6m}^{(b)}(i))_{m=-6}^{6}$, $\widehat{\mathbf q}_{6i}^{(b)}=\bar{\mathbf q}_{6i}^{(b)}/\|\bar{\mathbf q}_{6i}^{(b)}\|$, and $a_{ij}^{(b)}=\min[\bar q_6^{(b)}(i),\bar q_6^{(b)}(j)]$. The implemented crystal-like classifier is
\begin{equation}
\begin{aligned}
s_{ij}^{(b)}&=
\begin{cases}
\operatorname{Re}\!\left[(\widehat{\mathbf q}_{6i}^{(b)})^\dagger\widehat{\mathbf q}_{6j}^{(b)}\right]
\min\!\left(1,\dfrac{a_{ij}^{(b)}}{q_{\mathrm{scale}}}\right),
& a_{ij}^{(b)}\geq q_{\mathrm{floor}},\\
-\infty, & a_{ij}^{(b)}<q_{\mathrm{floor}},
\end{cases}\\
n_{i,\mathrm{solid}}^{(b)}&=
\sum_{j\in\mathcal N(i)}\mathbf 1[s_{ij}^{(b)}\geq s_\star],\\
c_i^{(b)}&=
\mathbf 1[\bar q_6^{(b)}(i)\geq q_{\mathrm{floor}}]\,
\mathbf 1\!\left[n_{i,\mathrm{solid}}^{(b)}\geq
\max\!\left(n_{\min},\left\lceil f_{\min}N_i\right\rceil\right)\right].
\end{aligned}
\end{equation}
Here $q_{\mathrm{scale}}$ and $q_{\mathrm{floor}}$ are the registered local-order amplitude scale and floor, $s_\star$ is the solid-bond threshold, and $n_{\min}$ and $f_{\min}$ are the minimum solid-neighbour count and fraction. The crystalline fraction and largest crystal-like cluster fraction are
\begin{equation}
f_{\mathrm{cryst}}^{(b)}
=
\frac{1}{N_b}
\sum_{i=1}^{N_b}
c_i^{(b)},
\end{equation}
and
\begin{equation}
f_{\mathrm{cluster}}^{(b)}
=
\frac{|\mathcal C_{\max}^{(b)}|}{N_b}.
\end{equation}
The Sm$_2$O$_3$ crystal-like-order screen label is
\begin{equation}
A_b^{\mathrm{Sm_2O_3}}
=
\mathbf 1
\left[
f_{\mathrm{cryst}}^{(b)}
\leq
\eta_{\mathrm{cryst}}
\right]
\mathbf 1
\left[
f_{\mathrm{cluster}}^{(b)}
\leq
\eta_{\mathrm{cluster}}
\right].
\label{eq:si-sm2o3-screen}
\end{equation}
The ordered thresholds are $(\eta_{\mathrm{cryst}},\eta_{\mathrm{cluster}})=(0.18,0.12)$.
This screen is configured with \code{exclude=true}. Therefore $A_b^{\mathrm{Sm_2O_3}}=1$ cells form the retained amorphous analysis population, while $A_b^{\mathrm{Sm_2O_3}}=0$ cells form the excluded crystal-like audit population.

Diffraction-derived order summaries are used as supporting validation of the same classification. For detected peak $i$ and crystalline-reference peak $j$, let $w_i$ and $w_j^{\mathrm{ref}}$ be the prominence weights normalised separately over the detected and reference peaks. Let $\delta q_{ij}=\max(\delta q_0,W_i/2,W_j^{\mathrm{ref}}/2)$ be the registered pairwise matching tolerance and $\gamma_{ij}=\sqrt{\min(W_i,W_j^{\mathrm{ref}})/\max(W_i,W_j^{\mathrm{ref}})}$ the peak-width similarity, with $\gamma_{ij}=1$ when valid widths are unavailable. Define $s_{ij}=\min(w_i,w_j^{\mathrm{ref}})$ and $q_{ij}=\exp[-(q_i-q_j)^2/(2\delta q_{ij}^2)]\gamma_{ij}$. If $\mathcal M^\star$ is the maximum-weight one-to-one assignment under weights $s_{ij}q_{ij}$, the reference-peak overlap is
\begin{equation}
O_{\mathrm{ref}}^{(b)}
=
\min\!\left[
1,
\max\!\left(
0,
\sum_{(i,j)\in\mathcal M^\star}s_{ij}q_{ij}
\right)
\right].
\end{equation}
The Bragg-sharpness score $B^{(b)}$ is the unitless operational scalar returned by the fixed archived structure-factor peak-sharpness evaluator; its exact registered evaluator and parameters are retained in the analysis JSON. These diffraction summaries are reported alongside $f_{\mathrm{cryst}}$ and $f_{\mathrm{cluster}}$ to support the crystal-like label rather than a high-density-amorphous interpretation. The local-order and diffraction summaries do not by themselves establish crystallographic phase identity.

The Sm$_2$O$_3$ reporting descriptor set comprises density, Sm--O and O--Sm coordination distributions, total and partial radial distribution functions, local-order metrics, reference-peak overlap, Bragg sharpness, and angular descriptors when required. The retained-amorphous analysis-population coordination distribution for central species $\alpha$ and neighbour species $\beta$ is
\begin{equation}
p_{\alpha:\beta}(m)
=
\frac{1}{N_{\mathrm{ana}}}
\sum_{x_b\in\mathcal E_{\mathrm{ana}}}
\left[
\frac{1}{N_\alpha^{(b)}}
\sum_{i\in\alpha}
\mathbf 1[n_{i:\beta}^{(b)}=m]
\right].
\end{equation}
This distribution is reported directly, rather than compared with a fixed ideal coordination set.

\begin{figure}[!htbp]
  \centering
  \includegraphics[width=\textwidth]{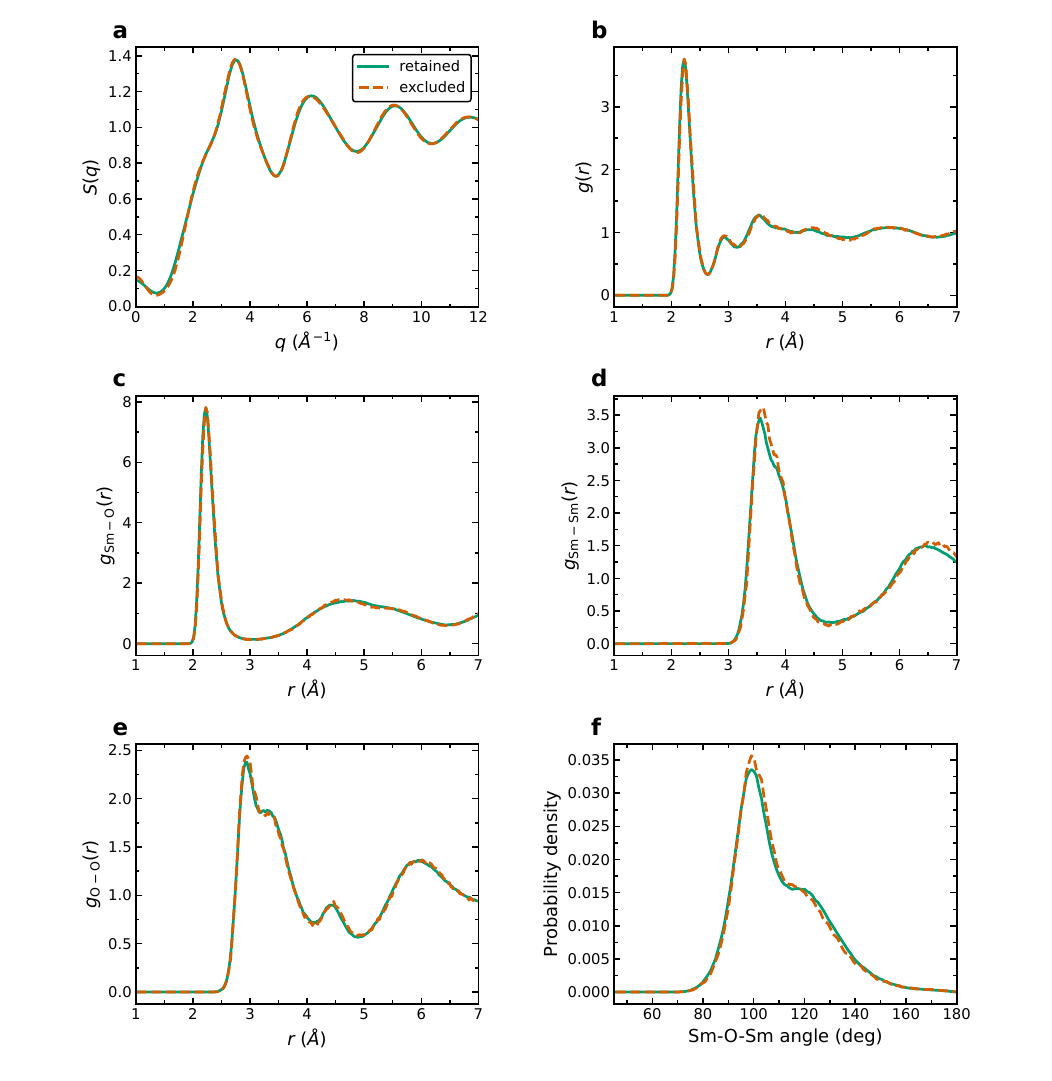}
  \caption{a-Sm$_2$O$_3$ structural descriptors for retained amorphous and excluded crystal-like cells. (a) Total structure factor. (b) Total radial distribution function. (c) Sm--O partial radial distribution function. (d) Sm--Sm partial radial distribution function. (e) O--O partial radial distribution function. (f) Sm--O--Sm angular distribution. Radial distribution functions are restricted to 1--7~\AA{}.}
  \label{fig:si-sm2o3-structural}
\end{figure}

\begin{figure}[p]
  \centering
  \includegraphics[width=\textwidth]{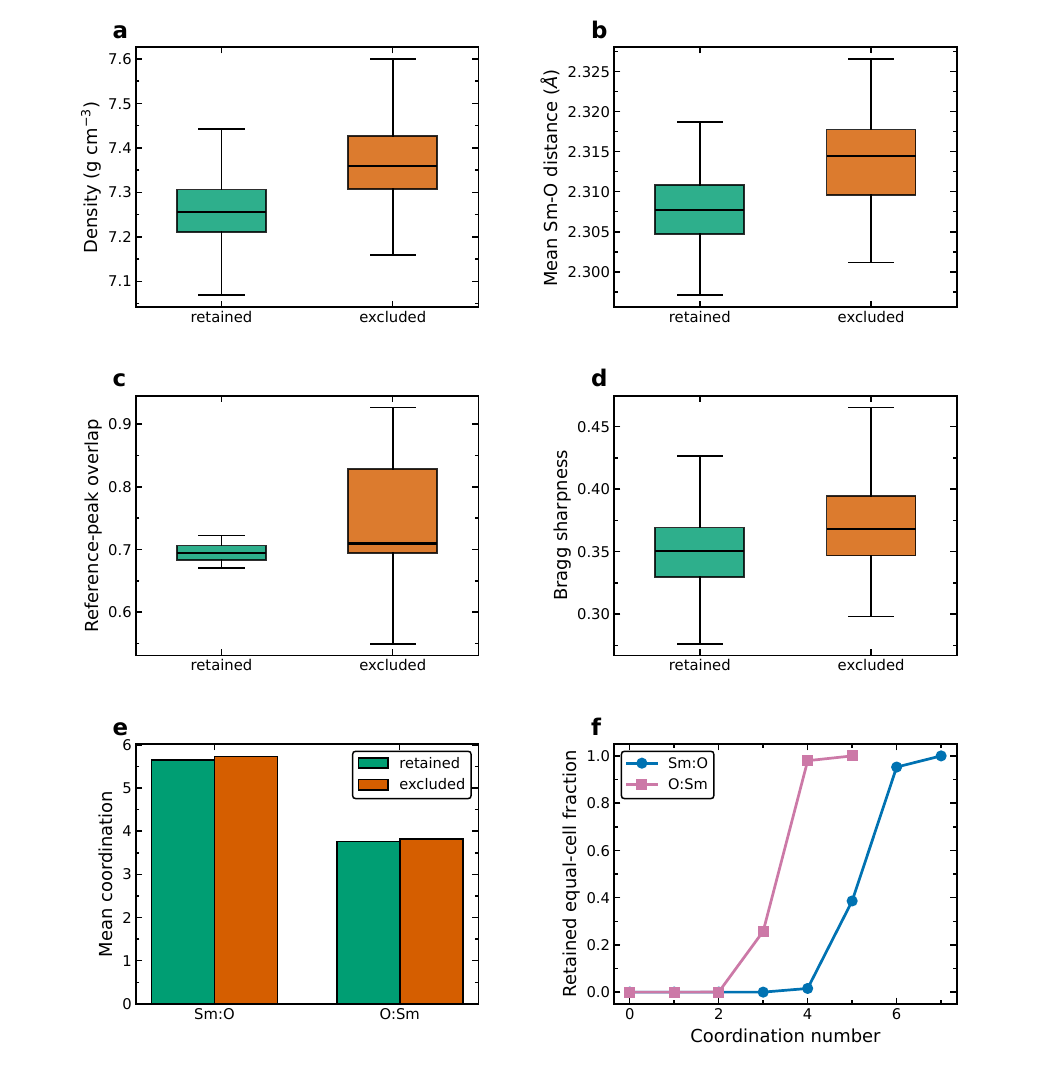}
  \caption{a-Sm$_2$O$_3$ screening descriptors for retained amorphous and excluded crystal-like cells. (a) Density. (b) Mean Sm--O bond length. (c) Reference-peak overlap. (d) Bragg sharpness. (e) Mean Sm:O and O:Sm coordination. (f) Retained amorphous fraction by coordination number for Sm:O and O:Sm environments.}
  \label{fig:si-sm2o3-screening}
\end{figure}

\begin{figure}[p]
\centering
\includegraphics[page=1,width=\textwidth]{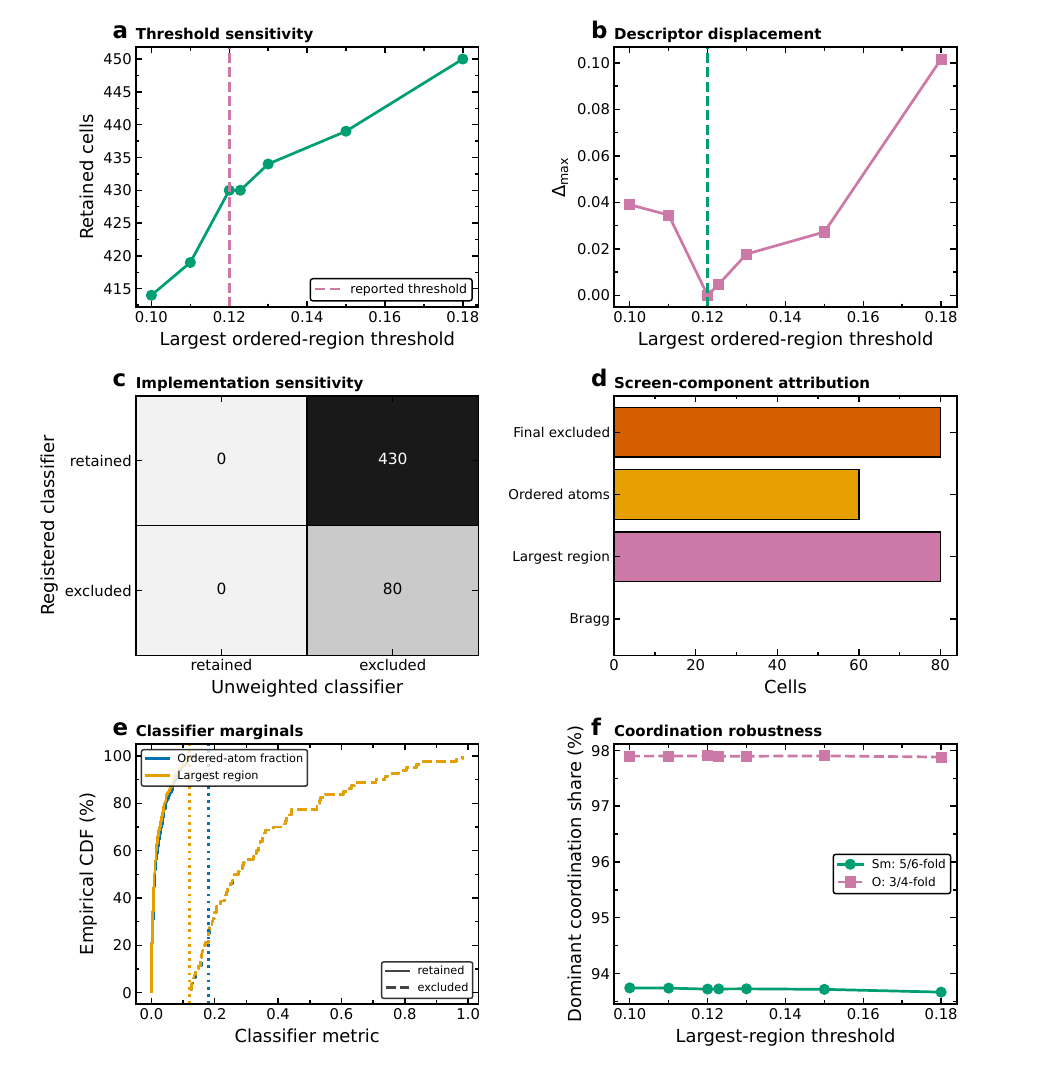}
\caption{Sensitivity of the operational a-Sm$_2$O$_3$ classifier across the tested setting family. The 510-cell source population gives the reported baseline of 430 retained and 80 labelled/excluded cells. Population membership and the monitored descriptor response are reported separately; the sweep is not a phase boundary and does not establish invariance of unmonitored quantities.}
\label{fig:si-sm2o3-classifier-sensitivity}
\end{figure}

\begin{figure}[p]
\centering
\includegraphics[page=1,width=\textwidth]{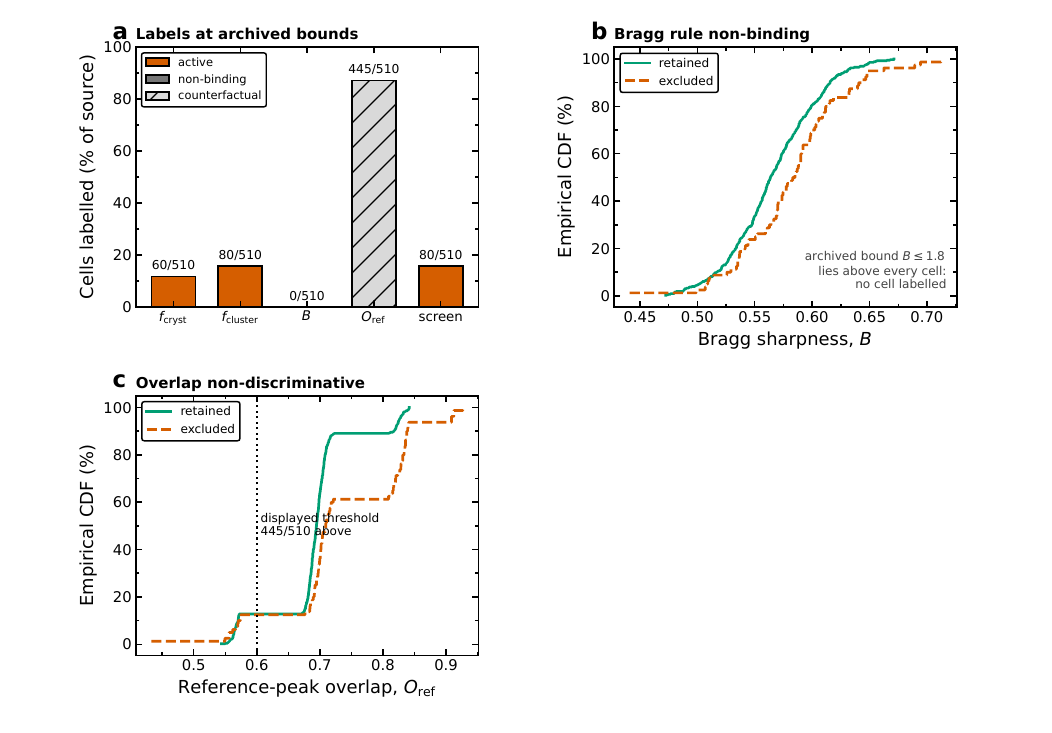}
\caption{Active, non-binding, and counterfactual contributions to the a-Sm$_2$O$_3$ screening diagnostics. The active crystalline-fraction and largest-cluster rules label 60 and 80 of 510 source cells; the combined active screen labels 80. The Bragg-sharpness rule labels none at the archived bound and is non-binding. $O_{\mathrm{ref}}$ is counterfactual/inactive, excluded from the mask, and non-discriminative at the displayed threshold. Connected/crystal-like order is not phase identity.}
\label{fig:si-sm2o3-component-response}
\end{figure}

\paragraph{Classifier membership and structural robustness.}\label{par:si-sm2o3-classifier-robustness}
The archived baseline maps 510 source cells to 430 retained cells, with 80 labelled and excluded. Across 108 declared classifier settings, the retained population spans 182--506 cells. Over that family, the equal-cell five-/sixfold Sm share remains 93.447--94.055\% and the three-/fourfold O share remains 97.846--97.943\%. The connected ordered component has mean Sm:O and O:Sm coordination 5.776 and 3.843, compared with 5.658/3.771 outside and 5.633/3.755 in zero-cluster cells. For each cell containing a connected component, the component--shell contrast is the mean coordination inside that component minus the mean over sites outside it in the same cell; the reported paired cell-level contrasts remain above zero. The supported statement is that the operational connected ordered domain is more highly coordinated. It is not a crystallite, phase-identity, or correlation-length claim. The structural comparison, screening descriptors, classifier sensitivity, and component response are shown in Figs.~\ref{fig:si-sm2o3-structural}--\ref{fig:si-sm2o3-component-response}.

\clearpage

\subsection{Consequences of population declarations}
\label{sec:si-declaration-consequences}

\begin{figure}[!htbp]
\centering
\includegraphics[page=1,width=\textwidth]{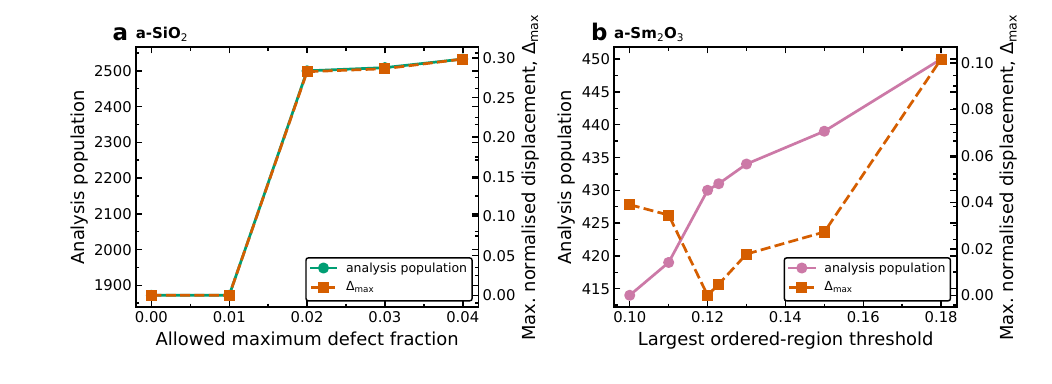}
\caption{Population movement and displacement of the monitored answer under the tested declarations. For a-SiO$_2$, the analysis population spans 1872--2533 cells and the maximum displacement over the declared sensitivity descriptors is 0.299 tolerance units. For a-Sm$_2$O$_3$, the analysis population spans 414--450 cells and the maximum displacement is 0.102 tolerance units. The comparison separates sensitivity of population membership from sensitivity of the monitored answer; it does not establish invariance of unmonitored quantities or equivalence of all declarations.}
\label{fig:si-declaration-consequences}
\end{figure}

The sensitivity of population membership and the corresponding displacement of the monitored material answer are distinct quantities. For declaration $a$ relative to the registered baseline declaration $a_0$, the maximum tolerance-normalised shift is
\[
\delta_{\max}(a)
=
\max_{j\in\mathcal D_{\mathrm{sens}}}
\frac{|\hat\mu_j(a)-\hat\mu_j(a_0)|}
{\tau_j(\hat\mu_j(a_0))},
\]
where $\mathcal D_{\mathrm{sens}}$ is the declared sensitivity-descriptor set. Figure~\ref{fig:si-declaration-consequences} reports both population movement and $\delta_{\max}$ for the tested a-SiO$_2$ and a-Sm$_2$O$_3$ declaration families.

\clearpage

\subsection{Reproducibility outputs}
\label{sec:si-reproducibility}

For each benchmark, the archived \vitriflow{} software output records the numerical preflight candidates, preflight scores, selected numerical setting $\theta^\star$, temperature-scan summaries, selected $T_{\mathrm{high}}$, selected $t_{\mathrm{melt}}^\star$, selected cooling rate $\dot T^\star$, size-scan outcome when applicable, production seed or source-snapshot identifiers, screening functions, screening thresholds, \code{exclude} settings, screen labels, analysis masks, descriptor tables, convergence ratios, and figure-generation inputs. The source-data archive therefore distinguishes three classes of rejected object:
\begin{enumerate}
\item numerically inadmissible settings rejected during preflight;
\item unresolved protocol candidates rejected during calibration;
\item generated final cells screened by material-specific artefact screening.
\end{enumerate}
Only the third class contributes to the $N_{A=0}$ audit-labelled counts reported in the main-text screening table.

\bibliographystyle{elsarticle-num}
\bibliography{refs}